\documentclass[a4paper,twocolumn,11pt,unpublished,noarxiv]{quantumarticle}

\pdfoutput=1

\usepackage[utf8]{inputenc}
\usepackage[english]{babel}
\usepackage[T1]{fontenc}
\usepackage{amsmath}
\usepackage{hyperref}
\usepackage{xurl}
\usepackage{tikz}
\usepackage{lipsum}
\usepackage{amssymb}
\usepackage{bigints}

\newtheorem{theorem}{Theorem}
\newtheorem{Remark}{Remark}

\begin{document}


\newcommand*{\id}{{\normalfont\hbox{1\kern-0.15em \vrule width .8pt depth-.5pt}}}


\title{A thorough introduction to non-relativistic matrix mechanics in multi-qudit systems with a study on quantum entanglement and quantum quantifiers}


\author{Lucas Camponogara Viera}
\affiliation{Institute of Electro-Optical Engineering, National Taiwan Normal University, Taipei 116, Taiwan}
\orcid{0000-0001-6441-2352}
\email{vieracamponogara@gmail.com}
\homepage{https://camponogaraviera.github.io/homepage/}
\author{Shu-Hsien Liao}
\affiliation{Institute of Electro-Optical Engineering, National Taiwan Normal University, Taipei 116, Taiwan}
\orcid{0000-0002-5794-7583}
\maketitle


\begin{abstract}
  Quantum computing is among the most far-reaching technologies of the 21st century, tackling challenges at the cutting edge of physics. This new paradigm in computer science harnesses quantum entanglement, one striking non-intuitive feature of quantum mechanics and a cornerstone of quantum information, to provide computation with a quantum speed-up over the best-known classical algorithms and to enable encrypted data communication against eavesdropping. The bulk of this article is focused on providing a deep and abiding understanding of non-relativistic matrix mechanics by demonstrating the fundamental mathematical identities of the contemporary postulatory approach of quantum mechanics within the state vector and density operator formalism in multipartite systems. In addition to that, we derive and analyze the respective 1-qubit, 1-qutrit, 2-qubit, and 2-qudit coherent and incoherent density operators using Bloch's parametrization for generalized $d$-dimensional $N$-qudit states embedded in the $SU(d)$ Lie group with associate generalized Gell Mann's matrices spanning the $\mathfrak{su}(d)$ Lie algebra. We also address the fundamental concepts of quantum nondemolition measurements, quantum decoherence and, particularly, quantum entanglement providing for the latter a systematic view on its historical development and mathematical description in multipartite systems. We conclude our review by introducing some of the ubiquitous quantum quantifiers required to measure degrees of quantum entanglement and quantum coherence, deriving the $p$-norm quantum coherence measure for a 1-qubit state.
\end{abstract}


\section{Introduction}
The historical development of the so-called old quantum mechanics broadly stems from 1900 with the seminal work of Max Planck \cite{Planck} on the correction of the Rayleigh-Jeans law regarding the ultraviolet catastrophe. Max Planck proposed that the electromagnetic spectrum of black bodies occur according to the emission and absorption of discreet quantities of energy, called quanta, whose separation is proportional to the frequency $\nu$ of the radiation and to a quantum of action $h$ commonly known as Planck constant. Building on Planck's realization, Albert Einstein explained the photoelectric effect, proposing that light waves are quantized by the same amount $h\nu$ carried by a discreet pack of energy termed a photon. The field advanced with a series of atomic models by Ernest Rutherford \cite{Rutherford1911}, Niels Bohr \cite{Bohr1}\cite{Bohr2}, and Arnold Sommerfeld \cite{Sommerfeld}. The first attempt to craft a self-consistent theory of quantum mechanics emerged with the Bohr-Sommerfeld model whose goal was to describe the physics of atomic data from spectroscopy. The data was associated with electronic transitions defined as discontinuous and unpredictable quantum jumps in the atom since its proposal by Niels Bohr in 1913 \cite{Bohr1} and backed by the first experimental observations in 1986 \cite{Bergquist}. Bohr's 1913 model was able to explain the Balmer series in the hydrogen spectral lines. The 1916 Sommerfeld model \cite{Sommerfeld}, a special-relativistic extension of the aforementioned Bohr's planetary model of the atom, correctly described the Stark effect and the fine structure in the emission spectrum of the hydrogen atom. In spite of its theoretical success, the model was inconsistent and unable to accurately explain the ``anomalous'' Zeeman effect, the case where there is a non-zero net spin, for spin had not yet been discovered.

The new quantum mechanics began to soar with the contributions of Werner Heisenberg, Max Born, Pascual Jordan, and Erwin Schrödinger \cite{Mehra1982}. Although the term ``quantum mechanics'' first appeared in Born's 1924 publication \cite{Born1924}, the leapfrog from the classical to the quantum world only happened with Heisenberg’s 1925 breakthrough publication \cite{Heisenberg} entitled \textit{``quantum-theoretical reinterpretation of kinematic and mechanical relations''}. Heisenberg's pivotal idea was to promote the classical physical observables, such as the position coordinate, to analog quantum-theoretical quantities named quantum observables. The realization that Heisenberg’s rule for multiplying the aforementioned quantities was equivalent to a matrix multiplication led to the first formulation of a consistent theory of quantum mechanics by Max Born and Pascual Jordan \cite{Born1925}. The Born-Jordan paper \cite{Born1925} laid a set of postulates, such as the commutation law, and proofs of conservation theorems, while considering quantum observables as Hermitian operators. The sequel ``Zur Quantenmechanik II'' (On quantum mechanics II) by Born, Heisenberg, and Jordan \cite{Born1926} matured Heisenberg's idea into the first complete formalism of the quantum mechanical theory now regarded as matrix mechanics. During the same year of 1926, Erwin Schrödinger \cite{schrodinger} presented his wave theory formalism describing the electron in the atom as an oscillating cloud of probability evolving continuously in space and time according to a second-order differential wave equation. Max Born then proposed that the square of the absolute value of the complex-valued wave-function should represent the probability density associated with a measurement outcome, thus providing a consistent statistical interpretation of quantum mechanics. Being non-relativistic by nature, a feature easily confirmed by the different order of its partial derivatives in space and time, Schrödinger's wave equation provided an analytical solution for the non-relativistic hydrogen atom, and a general form to describe atoms with more than one electron by simply updating its Hamiltonian. However, the shortcomings of spin shared by both models suggested a revamp. Paul Dirac's 1928 relativistic equation \cite{Dirac} naturally yields the spin quantum number, rather than imposing one in an \textit{ad hoc} fashion, and predicted the existence of antimatter. A year earlier, Heisenberg's uncertainty principle \cite{Heisenberg1} stated that conjugate quantum observables have an intrinsic non-simultaneous characteristic: a feature often regarded as the hallmark of the quantum mechanical theory for which quantum mechanics strongly differs from classical mechanics. 

Then, in 1935, Albert Einstein, Boris Podolsky and Nathan Rosen published an article \cite{EPR} describing a thought experiment known today as the EPR paradox. The EPR paper claimed that quantum mechanics was an incomplete theory of nature that should be framed as a deterministic rather than a probabilistic model. To assess the existence of such a deterministic framework, John Bell in 1964 \cite{Bell} proposed a local hidden variable model (LHVM) of quantum mechanics and showed that the expectation value of certain observables, if they were to obey a LHVM, should satisfy some sort of correlation inequality. One such Bell inequality is the ubiquitous CHSH inequality \cite{CHSH} derived by Clauser, Horne, Shimony, and Holt. The fact that some entangled systems violate Bell's inequalities confirmed the incompatibility of quantum mechanics with a LHVM, a result backed by the first loophole-free experiment realized only eighty years later in 2015 \cite{Hensen2015}. Furthermore, the violation of Bell's inequalities has shown that either locality or realism known together as local-realism \cite{Aspect}\cite{Aspect1982} is an incorrect assumption about quantum mechanics, and one must be adopted in place of the other. 

It was only through the intertwined history of quantum mechanics that quantum information science (QIS) was allowed to thrive. QIS is an interdisciplinary field that extends the classical theory of information to explain quantum phenomena. The primary concern regarding the transmission of classical information was how to circumvent the loss of information over noisy communication channels. In 1948, Claude Shannon \cite{Shannon} published two remarkable theorems. The first, Shannon’s \textit{noiseless channel coding theorem}, quantifies the physical resources required to store information. The second, Shannon's \textit{noisy channel coding theorem}, provided an upper bound on how much classical information can be protected from noise using an error-correcting code. A quantum analog to Shannon’s first theorem appeared in 1995 in the work of Benjamin Schumacher \cite{Schumacher}, who also coined the name ``qubit''. The first successful protocol in QIS, named superdense coding, was devised in 1992 by Charles Bennett and Stephen Wiesner \cite{Bennett} with experimental realization in 1996 \cite{Klaus}. The vanilla version of the protocol uses a single qubit (particle or artificial atom) of an entangled qubit pair to realize the transmission of two bits of classical information through a quantum channel. Prior to that was the recognition that quantum mechanics could be used for encrypted communication without the possibility of eavesdropping. This procedure, namely \textit{quantum cryptography} or \textit{quantum key distribution} (QKD), strongly relies on the property known as quantum entanglement. The first consistent quantum cryptographic protocol, the so-called BB84 QKD protocol, was devised by Charles Bennett and Gilles Brassard \cite{Bennett1984} in 1984.

Quantum mechanics truly took a leap towards quantum computing (QC) between 1980 and 1982, when Paul Benioff \cite{Benioff}\cite{Benioff2} spun the idea of a reversible Turing-based machine \cite{Turing1936} operating according to the principles of quantum mechanics. At about the same time, in 1982, Richard Feynman \cite{Feynman1982}, realizing the difficulties in simulating quantum systems on classical computers, proposed a universal quantum simulator. David Deutsch \cite{Deutsch1985} carried out on the previous contributions and in 1985 claimed that quantum computers could outstrip classical computers by efficiently solving certain oracle problems believed to have no efficient solution on a classical Turing machine. The first quantum algorithms to show a quantum speed-up over their classical counterparts were devised by Peter Shor \cite{Shor1994} in 1994 and Lov Grover \cite{Grover} in 1996. Shor’s algorithm is based on the quantum Fourier transform for finding the prime factors of an integer and solving the discrete logarithm problem with an exponential speedup. While Grover’s algorithm performs a quantum searching database with a quadratic speedup in time. 

The QIS community laid the primary ideas \cite{Feynman1982}\cite{Deutsch1985} and requirements \cite{DiVincenzo2000} for the universal QC platform. To date, the Josephson-based superconducting quantum processor architecture is considered to be the best candidate for a large-scale ($N$-qudits), fault-tolerant \cite{Shor1996}\cite{Preskill1997}\cite{Chow2012}, and error-corrected \cite{surfaceCode} QC technology operating within the framework of circuit quantum electrodynamics (QED) \cite{Blais2004}. In the rapidly growing field of circuit QED, Minev et. al \cite{Minev} in 2019 experimentally demonstrated a continuous, coherent, and deterministic evolution in the complete transition from the ground state to an excited state of a superconducting artificial three-level atom (a.k.a qutrit): a remarkable feat that runs counter the original beliefs of discontinuous and unpredictable atomic transitions proposed by Bohr \cite{Bohr1}. In this sense, the authors demonstrated the possibility to predict at which times a transition is about to occur, and reverse it before it happens: a potentially invaluable tool in quantum error correction schemes that can be used to prevent bit-flip errors during gate operations. The experiment also provided a strong support to the modern quantum trajectory theory \cite{Carmichael}, which describes the trajectories of individual particles in open quantum systems, i.e, in systems interacting with their environment. 

Along with Heisenberg's uncertainty principle, Bohr's complementarity principle \cite{BOHR1928} provides another important foundation of quantum mechanics, one whose quantitatively formulation \cite{WoottersZurek}\cite{GreenbergerYasin} open venues for the quantification of the predictability, a feature regarded as a quantum resource just like quantum coherence and quantum entanglement are quantum properties considered as resources for certain tasks in the field of QIS and QC. The purpose of this article is to provide a thorough introduction to the mathematical formalism used in QIS by demonstrating its usage through an example-oriented approach in various quantum information tasks.

The remainder of this article is structured as follows. In \autoref{sec:linearalgebra} we provide a quick review on linear algebra. In \autoref{sec:quantummechanics} we address the Heisenberg uncertainty principle, the postulates of quantum mechanics, and the characteristic of quantum nondemolition measurements. In \autoref{sec:quantuminfo} we introduce the mathematical formalism of coherent quantum superposition within the state vector formalism and derive the Bloch's sphere representation for 1-qubit states. Particular attention is given to the historical development of quantum entanglement as we address the EPR hypothesis and demonstrate the violation of a particular Bell inequality known as the CHSH inequality. In \autoref{sec:densityformalism} we frame the postulates of quantum mechanics within the density operator formalism and derive the criterion to distinguish pure and mixed density operators. In \autoref{sec:partialtrace} we derive the mathematical relations for measurements in the Hilbert space of bipartite systems with the ubiquitous partial trace function for density operators. In \autoref{sec:decoherence} we describe the phenomenon of quantum decoherence. In \autoref{sec:subsecqudit} we derive particular coherent and incoherent density operators from the generalized $N$-qudit density matrix defined by Bloch's parametrization. In \autoref{sec:entangledstates}, the Schmidt decomposition and the Peres-Horodecki criterion are outlined as two important tests used to discriminate between separable and entangled bipartite density operators. In \autoref{sec:entangmeas}, we give the definition of a quantum entanglement quantifier and apply some of the ubiquitous entanglement measures to quantify the degree of entanglement in particular multipartite systems. In \autoref{sec:coherencequantifiers} we introduce the requirements for a quantum coherence quantifier and derive the $p$-norm quantum coherence measure for a 1-qubit state. Finally, in \autoref{sec:teleportation} we provide a worked example of a quantum entanglement application known as quantum teleportation. We demonstrate perfect quantum teleportation with maximally entangled pure states and briefly discuss the necessity for teleportation with nonmaximally multipartite entangled states.


\section{Linear algebra: a primer}
\label{sec:linearalgebra}
The mathematical framework employed in the remainder of this article is the usual matrix mechanics formalism of quantum mechanics initially developed by Werner Heisenberg, Max Born, and Pascual Jordan in a sequel of articles \cite{Heisenberg}\cite{Born1925}\cite{Born1926} between 1925 and 1926. Within this formalism, the physical observables (e.g., spin, electric charge, magnetic flux, position, momentum, etc.) are represented by self-adjoint operators as the central pivots of the theory. Therefore, to master the intricacies of the quantum information theory and the way quantum phenomena are hemmed in, a solid background knowledge is required. This section provides a skill-reaffirming background on elementary linear algebra \cite{Nielsen2010} in the ubiquitous bra-ket notation of Paul Dirac.

\subsection{Hilbert space}
The equivalence between wave mechanics and matrix mechanics is primarily attributed to John Von Neumann \cite{Neumann1932}, who in 1932 proved them to be different realizations of an abstract Hilbert space. Let $\{|\lambda_j\rangle\}_{j=1}^d$ denote any set of $d$ linearly independent orthonormal column vectors $|\lambda_j\rangle$ in a $d$-dimensional complex vector space $V^d$ defined in some finite dimensional inner product space $\mathcal{H}$, known as Hilbert space, over the field (set) $\mathbb{C}$ of the complex numbers $\lambda_j \in \mathbb{C}$ (voiced $\lambda_j$ is an element of $\mathbb{C}$). The number of linearly independent vectors in the set defines the dimension $d$ of the space. Moreover, linear independence states that no vector in the set can be given as a linear combination of the remaining ones and, therefore, the set form a basis for the space $V^d$. As a consequence, any arbitrary state vector (a.k.a ket-vector or column vector) $|\psi\rangle \in V^d$ can be represented as a linear combination of the vectors of the basis set according to: 
\begin{eqnarray}
|\psi\rangle \doteq \sum_{j=1}^d \lambda_j|\lambda_j\rangle = \begin{pmatrix} \lambda_1 \\ \lambda_2 \\ \vdots \\ \lambda_d \end{pmatrix}.
\end{eqnarray}
And the transposed conjugate state vector also known as bra-vector or row vector is defined as:
\begin{eqnarray}
\langle \psi| &=& (|\psi\rangle)^{\dagger}=(|\psi\rangle^T)^* \\ &=& \left( \sum_{j=1}^d \lambda_j | \lambda_j \rangle \right)^{\dagger}
= \sum_{j=1}^d \lambda_j^* \langle\lambda_j| \\
&=& \begin{pmatrix}\lambda_1^* & \lambda_2^* & \cdots & \lambda_d^* \end{pmatrix}.
\end{eqnarray}
The transpose conjugate (a.k.a Hermitian conjugate, self-adjoint or dagger) operation has the following property:
\begin{align}
(A+B)^{\dagger}= A^{\dagger}+B^{\dagger}.
\end{align}
Here, $|\psi\rangle^T$ denotes the transpose of $|\psi\rangle$ swapping rows for columns. The complex scalar quantity $\lambda_j$ is defined in the standard way $\lambda_j\doteq a+bi$ with real scalars $a$ and $b$, and imaginary number $i\doteq \sqrt{-1}$. Whereas $(*)$ denotes the complex conjugate, such that $\lambda_j^*=a-bi$. Orthonormality condition of the basis vectors entails:
\begin{eqnarray}
\label{delta}
\langle\lambda_j|\lambda_k\rangle \doteq \delta_{jk} = \begin{cases}
0, & \mbox{if } j \ne k, \\
1, & \mbox{if } j=k, \end{cases}
\end{eqnarray}
where $\delta_{jk}$ is termed ``Kronecker delta''. In this sense, the basis set $\{|\lambda_j\rangle\}_{j=1}^d$ is termed an orthonormal basis set. In tensorial notation, the $d$x$d$-dimensional unit matrix (a.k.a identity matrix) reads:
\begin{align}
\delta_{\mu\nu}=
    \begin{bmatrix} 
     1 & \dots & 0 \\
    \vdots & \ddots & \vdots \\
    0 & \dots & 1
\end{bmatrix}.
\end{align}

\subsection{Inner product space}
An inner product complex space is a $d$-dimensional complex vector space $V^d$ over the field $\mathbb{C}$ of the complex numbers endowed with an inner product function defined by a map $(\cdot,\cdot) : V^d$ x $V^d \rightarrow \mathbb{C}$. The map is voiced ``a function $(\cdot, \cdot)$, namely inner product, takes as input two vectors from a vector space and produces a scalar quantity, in general a complex number, as output''. The inner product function then satisfies the following conditions for all vectors $x, y, z \in V^d$ and all scalars $\alpha \in \mathbb{C}$:
\begin{itemize}
    \item $(x,\alpha y)= \alpha(x,y)$, (Linearity in the second argument).
    \item $(x,y)=(y,x)^*$, (Conjugate symmetry).
    \item $(x,x) \geq 0$. (Positive semi-definiteness).
\end{itemize}
Within Dirac's bra-ket notation, the inner product $(\cdot,\cdot) \doteq \langle \cdot | \cdot \rangle$ between vectors $|q\rangle = \sum_i q_k |k\rangle$ and $|\phi\rangle = \sum_j \phi_j |j\rangle$ is defined as:
\begin{eqnarray}
(|q\rangle, |\phi\rangle) &\doteq& |q\rangle^{\dagger} |\phi\rangle = \langle q | \phi \rangle \\
&=& \left( \sum_{k=1}^d q_k | k \rangle \right)^{\dagger} \left(\sum_{j=1}^d \phi_j | j \rangle^{\dagger}\right) \\
&=& \sum_{k=1}^d \sum_{j=1}^d  q_k^*\phi_j \langle k|j\rangle) \\
&=& \sum_{k=1}^d \sum_{j=1}^d q_k^* \phi_j \delta_{kj}\\
&=& \sum_{j=1}^d q_j^* \phi_j \\
&=& q_1^* \phi_1 + q_2^* \phi_2 + \cdots + q_d^* \phi_d \\
&=&
\begin{pmatrix} q_1^* & q_2^* & \cdots & q_d^* \end{pmatrix}
\begin{pmatrix} \phi_1 \\ \phi_2 \\ \vdots \\ \phi_d \end{pmatrix}.
\end{eqnarray}
And the corresponding outer product in the space $V^{dxd}$ is defined as:
\begin{eqnarray}
|\phi\rangle|q\rangle^{\dagger} &:=& |\phi\rangle \langle q | \\
&=& \sum_{j=1}^d \sum_{k=1}^d \phi_j q_k^* |j\rangle \langle k | \\
&=&  
\begin{bmatrix}
    \phi_1 q_1^* & \dots & \phi_1 q_d^* \\
    \vdots & \ddots & \vdots \\
    \phi_d q_1^* & \dots & \phi_d q_d^*
\end{bmatrix}.
\end{eqnarray}

\subsection{Completeness relation}
Suppose \{|$j\rangle\}_{j=1}^d$ is any orthonormal basis set for a $d$-dimensional vector space $V^d$ with an arbitrary vector $|v\rangle= \sum_{j=1}^d v_j|j\rangle$, then
\begin{eqnarray}
\left(\sum_{j=1}^d |j\rangle \langle j| \right) |v\rangle &=& \sum_{j=1}^d |j\rangle \langle j|v\rangle \\
&=&\sum_{j=1}^d v_j |j\rangle = |v\rangle.
\end{eqnarray}
And, therefore,
\begin{eqnarray}
\label{completeness}
\sum_{j=1}^d |j\rangle \langle j| = \id_d,
\end{eqnarray}
and known as the completeness relation.

\subsection{Linear operator}
\label{sec:linear}
Suppose $\{|v_j\rangle\}_{j=1}^{d_a}$ and $\{|w_k\rangle\}_{k=1}^{d_b}$ are any orthonormal basis set for vector spaces $V^{d_a}$ and $W^{d_b}$, respectively. A linear operator is any function $\hat{O}$, with a map $\hat{O}$ : $V^{d_a}$ $\rightarrow$ $W^{d_b}$, that is linear in its inputs:
\begin{eqnarray}
\hat{O} \left(\sum_{j=1}^{d_a} v_j |v_j\rangle\right) = \sum_{j=1}^{d_a} v_j \hat{O}(|v_j\rangle).
\end{eqnarray}
The map is voiced ``a function $\hat{O}$, namely linear operator, acts on a vector $V^{d_a}$ and produces a vector $W^{d_b}$''. By definition,
\begin{eqnarray}
\hat{O} \doteq \sum_{j=1}^{d_a} v_j |w_j\rangle \langle v_j|
\end{eqnarray}
is a linear operator which, using the completeness relation, yields the following outer product representation:
\begin{eqnarray}
\hat{O} = I_v\hat{O}I_w &=& \sum_{j=1}^{d_a} | v_j \rangle \langle v_j| \hat{O} \sum_{k=1}^{d_b}|w_k\rangle \langle w_k|\\
&=& \sum_{jk=1}^{d_a,d_b} \langle v_j |\hat{O}|w_k\rangle |v_j\rangle \langle w_k|.
\end{eqnarray}

\subsection{Hermitian Operator}
Suppose $\hat{H}$ is a linear operator on a Hilbert space $\mathcal{H}$. There exists a unique linear operator $\hat{H}^{\dagger}\in \mathcal{H}$ such that for all vectors $|v\rangle, |w\rangle \in \mathcal{H}$:
\begin{eqnarray}
(|v\rangle, \hat{H} |w\rangle)=(\hat{H}^{\dagger}|v\rangle, |w\rangle)=\langle v|\hat{H}| w \rangle.
\end{eqnarray}
Where the operator satisfying $\hat{H}=\hat{H}^{\dagger}$ is dubbed a Hermitian operator.

\subsection{Positive operator}
Positive operators are a special subclass of Hermitian operators. A Positive operator $\hat{J}$ is defined as a linear operator such that, for any vector $|\psi \rangle$, the inner product $(|\psi \rangle, \hat{J}|\psi \rangle)$ is a real and non-negative number. If the inner product $(|\psi \rangle, \hat{J}|\psi \rangle)$ is strictly greater than zero for all $|\psi \rangle \neq 0$ then $\hat{J}$ is termed a positive-definite operator. Within Dirac's bra-ket notation, a positive definite-operator $\hat{J}$ reads
\begin{eqnarray}
\langle \psi |\hat{J}| \psi \rangle > 0,
\end{eqnarray}
for all $|\psi\rangle \neq 0$, while a positive-semidefinite operator $\hat{J}$ reads
\begin{eqnarray}
\langle \psi |\hat{J}| \psi \rangle \geq 0,
\end{eqnarray}
for all $|\psi\rangle$.

\subsection{Unitary operator}
A bounded linear operator $\hat{U}$ in the Hilbert space $\mathcal{H}$ with a map $\hat{U} : \mathcal{H}$ $\rightarrow$ $\mathcal{H}$ is dubbed a Unitary operator if it satisfies:
\begin{eqnarray*}
\hat{U}^{\dagger}\hat{U} \text{ (isometry)}&=& \hat{U}\hat{U}^{\dagger} \text{ (coisometry)}=\id \\ &\implies& \hat{U}^{\dagger}=\hat{U}^{-1} \text{ (inverse)}.
\end{eqnarray*}
Therefore, a Unitary operator is both an isometry and a coisometry, or, equivalently, a surjective isometry.

Let $\hat{U} : V^d$ $\rightarrow$ $W^d$ be a map of a bounded unitary operator $\hat{U}$, and $\{|v_j\rangle\}_{j=1}^d$ any orthonormal basis set for the inner product vector space $V^d$ with an arbitrary vector $|v\rangle= \sum_{j=1}^d v_j|v_j\rangle$. Then $|w_j \rangle = \hat{U}|v_j\rangle$ forms an orthonormal basis since $\hat{U}$ preserves the inner product: 
\begin{eqnarray}
(\hat{U}|v\rangle, \hat{U}|w\rangle) &=& \langle v| \hat{U}^{\dagger}\hat{U}|w\rangle \\
&=&  \langle v|I|w\rangle = \langle v|w\rangle.
\end{eqnarray}
Therefore, one can write the following elegant outer product representation for any Unitary operator $\hat{U}$: 
\begin{eqnarray}
\hat{U}=\sum_{j=1}^d |w_j\rangle \langle v_j|,
\end{eqnarray}
such that
\begin{eqnarray}
\hat{U}|v_j \rangle=\sum_{j=1}^d |w_j\rangle \langle v_j|v_j \rangle=\sum_{j=1}^d|w_j\rangle.
\end{eqnarray}
Moreover, if $\hat{\mathcal{H}}$ $\doteq$ $-ilog(\hat{U})$ is Hermitian for any Unitary $\hat{U}$, than:
\begin{align}
    \hat{U}=e^{i\hat{\mathcal{H}}}, \label{unitexp}
\end{align}
for some Hermitian $\hat{\mathcal{H}}$.

\subsection{Normal operator}
Let $\hat{\mathcal{O}}$ be a linear operator on a finite-dimensional inner product space. Then $\hat{\mathcal{O}}$ is said to be Normal if it satisfies
\begin{eqnarray}
\hat{\mathcal{O}}^{\dagger}\hat{\mathcal{O}} = \hat{\mathcal{O}}\hat{\mathcal{O}}^{\dagger}.
\end{eqnarray}

\begin{theorem}[Spectral decomposition]
    \label{t1}
    Any Normal operator $\hat{\mathcal{O}}$ on a vector space $V$ is diagonal with respect to some orthonormal basis for $V$. Conversely, any diagonalizable operator is normal.
\end{theorem}

From Theorem \eqref{t1}, any Normal operator $\hat{\mathcal{O}}$ can be diagonalized as
\begin{eqnarray}
\label{spectraldec}
\hat{\mathcal{O}} = \sum_{j=1}^d o_j |o_j\rangle \langle o_j|,
\end{eqnarray}
where $\{|o_j \rangle\}$ is any orthonormal set of eigenvectors for $\hat{\mathcal{O}}$ with eigenvalues $o_j$. It is straightforward seeing that Hermitian and Unitary operators are also Normal operators and, therefore, have a espectral decomposition according to Eq.~\eqref{spectraldec}. 

\subsection{Projector operator}
Suppose \{|$j\rangle\}_{j=1}^k$ is any orthonormal basis set of a $k$-dimensional vector subspace $W^k$ of a $d$-dimensional vector space $V^d$ with the completeness relation set by Eq.~\eqref{completeness}. The projection of $V^d$ onto the subspace $W^k$ is given by the projector operator
\begin{eqnarray}
\hat{P}\doteq \sum_{j=1}^k |j\rangle \langle j|,
\end{eqnarray}
It can be shown that such an operator is Hermitian satisfying $P^{\dagger}P=P^2=P$ and, therefore, Normal, hence it can be given a spectral decomposition according to Eq.~\eqref{spectraldec}.
\begin{align*}
\hat{P}^{\dagger} &= \left( \sum_{j=1}^k |j\rangle \langle j| \right)^{\dagger} 
= \left(\sum_{j=1}^k |1\rangle \langle 1| +\cdot\cdot\cdot+  |k\rangle \langle k| \right)^{\dagger} \\
&=\left( |1\rangle \langle 1| \right)^{\dagger} +\cdot\cdot\cdot+ \left( |k\rangle \langle k| \right)^{\dagger} = \sum_{j=1}^k \left( |j\rangle \langle j| \right)^{\dagger}\\
&=\sum_{j=1}^k \langle j|^{\dagger} |j\rangle^{\dagger} = \sum_{j=1}^k |j\rangle \langle j| = \hat{P}.
\end{align*}

\subsection{Eingenvalue-eigenvector equation}
A general operator $\hat{\mathcal{O}}$ with any orthonormal basis set \{|$o_j\rangle\}_{j=1}^d$ obeys the following eingenvalue-eigenvector equation:
\begin{eqnarray}
\hat{\mathcal{O}} |o_j\rangle = o_j |o_j\rangle \implies (\hat{\mathcal{O}}-o_j \id_d)|o_j\rangle=0,
\end{eqnarray}
meaning that the matrix $(o_j \id_d - \hat{\mathcal{O}})$ is not invertible, hence its determinant must be zero. Therefore, the eigenvalues $o_j$ of $\hat{\mathcal{O}}$ are roots of the characteristic polynomial equation 
\begin{align}
   det(\hat{\mathcal{O}}-o_j \id_d) = 0.
   \label{polcharac}
\end{align} 
The fundamental theorem of algebra secures that every polynomial has at least one complex root, so that every operator $\hat{\mathcal{O}}$ has at least one eigenvalue $o_j$ with a corresponding eigenvector $|o_j\rangle$. However, when the operator has only one eigenvalue a diagonal representation cannot be given.

It can be shown that all eigenvalues of a Unitary matrix have modulus 1, that is, they can be written in the form $e^{i\theta}$ for some real number $\theta$. While the eigenvalues of a Projector operator are either 0 or 1. Moreover, a Normal matrix is Hermitian if and only if it has real eigenvalues ($o_j \in \mathbb{R}$). Whereas the eigenvectors of a Hermitian operator with different eigenvalues are necessarily orthogonal. Here, we show that the eigenvalues of a positive operator are all non-negative, as follows:
\begin{align}
\label{positv}
    \langle \phi |\hat{\mathcal{O}}| \phi \rangle &= \sum_{j,k} c_j^*c_k \langle \phi_j |\hat{\mathcal{O}} | \phi_k \rangle \\ &=\sum_{j,k} c_j^*c_k \langle \phi_j | o_k |\phi_k \rangle \\
    &= \sum_{j,k} c_j^*c_k o_k \langle \phi_j |\phi_k \rangle 
     \\&= \sum_{j,k} c_j^*c_k o_k \delta_{jk} 
     \\&= \sum_{j=1}^2 |c_j|^2 o_j \geq 0 \implies o_j \geq 0.
\end{align}

\subsection{Tensor product and Kronecker product}
Suppose $\{|a_j\rangle\}_{j=1}^{d_a}$ and $\{|b_k\rangle\}_{k=1}^{d_b}$ are any orthonormal basis set for vector spaces $V^{d_a}$ and $W^{d_b}$, respectively. Then $\{|a_j\rangle \otimes |b_k\rangle \doteq |a_jb_k\rangle\}_{j,k=1}^{d_a, d_b}$ is a basis set for a $d_a$x$d_b$-dimensional vector space $V^{d_a} \otimes W^{d_b}$, whose elements are given by the tensor product with the property:
\begin{align}
a_j |a_j\rangle \otimes b_k \langle b_k|b_l\rangle&=a_jb_k(|a_j\rangle\otimes \delta_{kl})\\
&= a_jb_l|a_j\rangle.
\end{align}
In matrix representation, the abstract tensor product becomes the Kronecker product. If $A\in M^{m,n}$ and $B\in M^{o,p}$ are $mxn$ and $oxp$ matrices, respectively, then their Kronecker product yelds the following block matrix:
\begin{eqnarray*}
(A \otimes B) = \begin{bmatrix} A_{11} B && \cdots && A_{1n} B \\ 
\vdots  &&\ddots  && \vdots \\ 
A_{m1} B && \cdots && A_{mn} B\end{bmatrix},
\end{eqnarray*}
with dimension $(mo)$ x $(np)$. In general, for a complex number $z \in \mathbb{C}$, matrices $A\in M^{m,n}$, $B\in M^{o,p}$, $C\in M^{q,r}$, $D\in M^{s,t}$ and unit matrix $\id = \delta_{\mu\nu} \in M^{u,u}$, the following identities are satisfied:
\begin{enumerate}
    \item $(A \otimes \id + \id \otimes B)\in M^{mu,nu} = A_{ij} \delta_{\mu\nu} + \delta_{ij} B_{\mu\nu}$.
    \item $(A \otimes \id + \id \otimes B)(|u\rangle \otimes |v\rangle)=(A|u\rangle\otimes |v\rangle)+(|u\rangle \otimes B|v\rangle)$.
    \item $(A \otimes B)(C \otimes D)=AC \otimes BD$.
    \item $(A \otimes B)\sum_j(|a_j\rangle \otimes |b_j\rangle)=\sum_jA|a_j\rangle \otimes B|b_j\rangle$.
    \label{productoftensors}
    \item $(A\otimes B)^{\dagger}=A^{\dagger} \otimes B^{\dagger}$.
    \item $(A\otimes B) \otimes C=A\otimes (B \otimes C)$.
    \item $z(A \otimes B) = (zA) \otimes B = A \otimes (zB)$.
    \item $(A + B) \otimes C = A \otimes C + B \otimes C$.
    \item $A \otimes (B + C) = A \otimes B + A \otimes C$.
\end{enumerate}
The aforementioned identities can be applied to all operators covered so far, as they are all described by matrices. Moreover, the Kronecker product of two Hermitian operators always results in another Hermitian operator. The same applies to Unitary, Positive, and Projector operators. This contrasts with matrix multiplication between operators of the same kind, as it does not always result in an operator of the same kind.

\subsection{Trace}
\label{sec:trace}

Let $\hat{O}$ be a linear operator represented in a basis $|\psi_j\rangle \in \mathcal{H}$ of a $d$-dimensional Hilbert space $\mathcal{H}$. The trace of $\hat{O}$ is a function with a map $tr:\mathcal{L}(\mathcal{H}) \rightarrow \mathbb{C}$ defined as:
\begin{align} \label{tracedef}
    tr(\hat{O})\doteq \sum_{j=1}^d \langle \psi_j|\hat{O}| \psi_j \rangle,
\end{align}
which is the sum of all diagonal elements of the $dxd$ square matrix representation of $\hat{O}$. Another common definition appearing in the literature is:
\begin{align}
    tr(\hat{O})\doteq \sum_{j=1}^d \hat{O}_{jj}.
\end{align}
As a word of caution, one should be wary that the trace of a non-square matrix is undefined. For any arbitrary matrices $A$ and $B$, Unitary matrix $U$, $d$-dimensional identity matrix $\id_d$, and complex number $z$, the following identities are satisfied:
\begin{enumerate}
    \item $tr(\id_d)=d$.
    \item $tr(z)=z$. \label{scalartrace}
    \item $tr(A^{\dagger})=(tr(A))^{\dagger}$.
    \item $tr(A)=tr(A^{T})$.
    \item $tr(zA)=z tr(A)$. \label{tracewithscalar}
    \item $tr(\sum_j A_j) = \sum_j tr(A_j)$. \label{traceofsum}
    \item $tr(AB)=tr(BA)$. \label{traceproduct}
    \item $tr(A^{T} B)=tr(B A^{T})=tr(B^T A)$.
    \item $tr(UAU^{\dagger})=tr(UU^{\dagger}A)=tr(A)$.
    \item $tr(A \otimes B)=tr(B \otimes A)=tr(A)tr(B)$. \label{traceoftensors}
    \item $tr(e^{(A \otimes \id + \id \otimes B)})=tr(e^A)tr(e^B)$.
\end{enumerate}
From the above relations, considering a unit state vector $|\psi\rangle$ and some orthonormal basis set $\{|o_j\rangle\}_{j=1}^d$, one can derive:
\begin{enumerate}
    \item $tr(A|\psi\rangle\langle\psi|)=tr(\langle\psi|A|\psi\rangle)=\langle\psi|A|\psi\rangle$.
    \item $\sum_{j=1}^d tr\left(| \psi_j \rangle\langle \psi_j|\right)=tr\left( \sum_{j=1}^d |\psi_j\rangle\langle\psi_j|\right)$.
    \item $tr(|o_j\rangle\langle o_k|)=tr(\langle o_k | o_j\rangle)=\langle o_k | o_j\rangle = \delta_{kj}$.
    \item $tr(|o_j\rangle\langle o_j|) = 1.$
\end{enumerate}

\section{Quantum mechanics road to quantum information}
\label{sec:quantummechanics}
In this section we introduce the quantum mechanical theory that governs the dynamics of quantized systems, in general. In \autoref{Conjugateobservables} is introduced the commutation relation and the Heisenberg uncertainty principle. In \autoref{sec:postulates} we explore the postulates of quantum mechanics and derive the uncertainty (a.k.a variance) of a measurement given the frequentist probability interpretation. And \autoref{sec:nondemolition} address the concept of quantum nondemolition measurements.

\subsection{Conjugate observables}\label{Conjugateobservables}

To help usher in the field of quantum mechanics, Werner Heisenberg \cite{Heisenberg}, then Bohr's assistant in Copenhagen, promoted the classical observable quantities to analog quantum observables represented by linear operators, such as the position $x \to \hat{x}$ and momentum $p \to \hat{p}$. Max Born \cite{Born1925} then realized that Heisenberg's multiplication rule for quantum observables were equivalent to that of matrix calculus, and that the observables satisfied a non-commutative algebra. In the modern formalism of matrix mechanics, the non-zero commutator of two conjugate observables (a.k.a Fourier transform duals) represented by quantum operators $\hat{A}$ and $\hat{B}$ is defined as: 
\begin{align}
\label{Conjugate}
[\hat{A},\hat{B}]\doteq\left(\hat{A}\hat{B} - \hat{B}\hat{A}\right) \neq 0.
\end{align}
In contrast to classical theory, in quantum mechanics the commutator can be order dependent, i.e, operators representing conjugate observables in quantum mechanics are not defined in the algebra of commutative groups (a.k.a Abelian group) where the commutator would to be zero (see Theorem \eqref{t2}). Moreover, for Hermitian operators, the commutator $[A, B]$ yields an antihermitian operator, while $i[A, B]$ results in a Hermitian operator.

\begin{theorem}[Simultaneous diagonalization]
    \label{t2}
    Suppose $A$ and $B$ are two Hermitian operators. Then $[A, B] = AB-BA$ is equal to the zero matrix (they commute) if and only if there exists an orthonormal basis such that both $A$ and $B$ are diagonal with respect to that basis, i.e, they share the same basis. We say that $A$ and $B$ are simultaneously diagonalizable in this case.
\end{theorem}

The canonical commutation relation is a hallmark of quantum mechanics and has been widely used to classify the different types of superconducting circuit-based qubits \cite{Koch} used in superconducting quantum computing \cite{DiVincenzo2000}\cite{Devoret2004} within the framework of circuit quantum electrodynamics \cite{Blais2004}\cite{blais2020}. In particular, for conjugate observables such as electric charge $Q \to \hat{Q}$ and magnetic flux $\Phi \to \hat{\Phi}$, the commutation relation yields the same result as the prime example of position and momentum:
\begin{align}
[\hat{\Phi}, \hat{Q}] =\left(\hat{\Phi}\hat{Q}-\hat{Q}\hat{\Phi}\right)=i\hbar \id_d.
\end{align}
Here, $\hbar=\frac{h}{2\pi}$ denotes the reduced Planck constant (a.k.a quantum of action), whereas $\id_d$ denotes the $d$-dimensional identity matrix. This result, known as the canonical commutation relation, follows immediately from Heisenberg’s uncertainty principle.

Heisenberg’s uncertainty principle is a consequence of one striking aspect that sets havoc on classical intuition: the intrinsic non-simultaneous nature of conjugate observables. This feature, first described by Werner Heisenberg \cite{Heisenberg1} in the form of epistemological uncertainty relations, states that the expectation values of two conjugate observables cannot be simultaneously measured within the same measurement accuracy/precision, but only up to some characteristic inaccuracy. This is often regarded as the most distinctive feature between classical and quantum mechanics. Heisenberg's prime example considered the measurement of the position and momentum of an electron by a microscope. Historically, the uncertainty relation has been extended to other conjugate observables, and the modern version of the principle is regarded as the Robertson uncertainty relation \cite{Robertson}\cite{Griffiths} stated as the following general inequality for observables $\hat{A}$ and $\hat{B}$:
\begin{align}
\sigma_{\hat{A}}\sigma_{\hat{B}} \geq \frac{|\langle \psi|[A,B]|\psi\rangle|}{2},
\end{align}
where the $\sigma$'s are the corresponding standard deviations of each observable according to Eq.~\eqref{dispersion}. 

\subsection{The fundamental postulates of quantum mechanics}
\label{sec:postulates}

Influenced by Georg Cantor's set theory, German mathematician David Hilbert's perspective on the scientific world is rooted in the axiomatic approach as the basis for any scientific theory to be developed independently of the need for intuition and free from arbitrariness. This realization entails the requirement of a formal and rigorous logical system of mathematical proof based on set theory, in which theorems are derived from a set of axioms via rules of inference conveyed with a symbolic logical language. A formal symbolic logical system of this fashion emerged between 1910-1913 with Bertrand Russell and Alfred Whitehead three-volume ``Principia Mathematica'' \cite{Principia}. In physics, however, Axioms are called Postulates and in the same ways are starting points for any physical theory.

Historically, the postulates of quantum mechanics in matrix notation were first devised by Born and Jordan in 1925 \cite{Born1925}. The original Born–Jordan postulates, as they are commonly known, were devised in close analogy with classical mechanics within a framework where the physical observables are Hermitian operators satisfying a non-commutative algebra (see \autoref{Conjugateobservables}). Nielsen and Chuang \cite{Nielsen2010} provide a systematic view of the contemporary postulatory approach for \textbf{closed systems} within the matrix mechanics formalism, and we now extend.

\subsubsection{Postulate 1 (State space)} 

Any isolated (close) physical system is embedded in a complex $d$-dimensional Hilbert space $\mathcal{H}^d$ known as the \textit{state space} of the system endowed with an inner product $(\cdot,\cdot) \doteq \langle \cdot | \cdot \rangle$. The system is completely described by a normalized \textit{state vector} $|\psi\rangle \in \mathcal{H}^d$ which is a unit vector in the system's state space. In the orthonormal basis set $\{|o_j\rangle\}_{j=1}^d$, the state vector reads
    \begin{eqnarray}
    |\psi\rangle \doteq \sum_{j=1}^d c_j|o_j\rangle,
    \end{eqnarray}
    where $c_j=\langle o_j|\psi\rangle$ denotes the probability amplitude associated with the preparation of an eigenstate $|o_j\rangle$. Normality condition of $|\psi \rangle$ entails
    \begin{eqnarray}
    \label{Normality}
    \langle \psi | \psi \rangle &=&\left(\sum_{j=1}^d c_j^* \langle o_j| \right) 
    \left(\sum_{k=1}^d c_k | o_k\rangle\right) \\
    &=& \sum_{j=1}^d \sum_{k=1}^d c_j^* c_k \langle o_j | o_k \rangle \\
    &=& \sum_{j=1}^d \sum_{k=1}^d c_j^* c_k \delta_{jk} \\
    &=& \sum_{j=1}^d c_j^*c_j \\
    &=&\sum_{j=1}^d |c_j|^2 = 1.
\end{eqnarray}
The state vector is often referred to as a \textit{pure state} to distinguish it from a density operator (see \autoref{sec:densityformalism}).

\subsubsection{Postulate 2 (Evolution)} 

A \textit{closed quantum system} in the initial $|\psi_0\rangle$ evolves, within the \textit{Schrödinger picture formalism}, according to a \textit{unitary transformation}, such that the state of the system after a time $t$ shall be $|\psi_t\rangle=\hat{U}_t|\psi_0\rangle$ given by the action of the propagator $\hat{U}_t$, which is a Unitary operator. Solution for $\hat{U}_t$ comes from the general time-dependent \textit{Schrödinger equation of the time evolution operator}:
\begin{equation} 
\label{timeevolution}
i\hbar \frac{\partial \hat{U}_t}{\partial t}=\hat{H}_t \hat{U}_t,
\end{equation}    
with a time-dependent Hamiltonian operator $\hat{H}_t$ describing the total energy of the system, i.e, the sum of all kinetic and potential energies. For the particular case of closed systems, the Hamiltonian operator is Hermitian. However, open systems (systems interacting with its surroundings), in general, do not allow Hermitian Hamiltonians.

Solution for equation Eq.~\eqref{timeevolution} depends mostly on the characteristic of the Hamiltonian. We shall consider three particular cases:
\begin{enumerate}
    \item Time-independent Hamiltonian. It is possible to construct any Unitary operator by means of a Hermitian observable according to Eq.~\eqref{unitexp}. In this sense, for a \textit{time-independent Hermitian Hamiltonian} $\hat{H}$, the propagator now writes:
\begin{equation} 
\hat{U_t}=e^{-i\hat{H}t/{\hbar}},
\end{equation}  
which satisfies Eq.~\eqref{timeevolution} as can be shown by expanding the exponential function in Taylor series and differentiating term by term with respect to time.
\item Time-dependent Hamiltonian with commuting operators. For a \textit{time dependent Hermitian Hamiltonian}, where the operators describing the Hamiltonian in different moments of time commute $[\hat{H}(t_1), \hat{H}(t_2)]=0$, the solution for the propagator becomes:
\begin{align} 
\hat{U}_t=e^{-\frac{i}{\hbar}\int_{t_0}^{t}dt'\hat{H}(t')}.
\end{align}  
\item Time-dependent Hamiltonian with non-commuting operators. When  the operators describing the Hamiltonian in different moments of time do not commute, the solution for the propagator becomes:
\begin{align} 
\hat{U}_t&=1+\sum_{n=0}^\infty {\frac{(-i)^n}{\hbar}}\left(\prod_{k=1}^n \int_{t_0}^t dt_k\right)\\
&=\int_{t_0}^{t_n-1}dt_n \hat{H}(t_1)\hat{H}(t_2)\cdot\cdot\cdot \hat{H}(t_n),
\end{align}  
known as the Dyson series expansion for unitary operators.
\end{enumerate}
Let $\hat{E}$ denote the Hermitian operator representing the energy observable that commutes with the Hamiltonian, i.e, $[\hat{H}, \hat{E}]=0$. In the basis of the energy eingenvectors, one can write the initial state of the system as $|\psi_{0}\rangle$ $=\id|\psi_{0}\rangle$ $= \sum_{j=1}^d | E_j \rangle \langle E_j|\psi_0\rangle=\sum_{j=1}^d c_j |E_j \rangle$. And for a time-independent hamiltonian the evolution becomes:
\begin{eqnarray*} 
|\psi_t\rangle&=&e^{-i\hat{H}t/{\hbar}}|\psi_0\rangle=e^{-i\hat{H}t/{\hbar}}\sum_{j=1}^d c_j |E_j \rangle \\
&=& \sum_{j=1}^d  c_j e^{-iE_j t/{\hbar}}|E_j \rangle 
= \sum_{j=1}^d  c_j(t) |E_j \rangle,
\end{eqnarray*}  
given the linearity property of linear operators (see \autoref{sec:linear}) and the fact that 
\begin{align}
    f(\hat{\mathcal{O}})|o_j\rangle = f(o_j) |o_j\rangle.
\end{align}

\subsubsection{Postulate 3 (General measurements)} 

Measurements of a quantum system are described by a collection \{$M_m$\} of \textit{measurement operators} acting on the state space of the system with $m$ possible measurement outcomes. If the quantum system is prepared in a general state $|\psi\rangle$, the probability associated with a measurement outcome $o_m$ is:
\begin{equation} 
Pr(o_m)\doteq\langle \psi | M^{\dagger}_{m} M_{m} |\psi \rangle,
\end{equation} 
and the state of the system immediately after the measurement of the eigenvalue $o_m$ will be:
\begin{align} 
    |\psi_{o_m} \rangle= \frac{M_{m} |\psi \rangle}{\sqrt{\langle \psi | M^{\dagger}_{m} M_{m} |\psi \rangle}}. 
\end{align}
The measurement operators satisfy the completeness relation:   
\begin{equation} 
\sum_{m} M_{m}^{\dagger} M_{m} = I,
\end{equation}
meaning that probabilities must sum to one and $|\psi\rangle$ is normalized, i.e, 
\begin{equation} 
\sum_{m} Pr(o_m) = \sum_{m} \langle \psi | M^{\dagger}_{m} M_{m} |\psi \rangle = 1.
\end{equation}  

\subsubsection{Projective measurements}\label{sec:projmeasu}

A special case of postulate 3 is the protective measurement (a.k.a von Neumann measurement) described by a physical observable represented by a Hermitian ($\hat{\mathcal{O}}=\hat{\mathcal{O}}^{\dagger}$) and, therefore, Normal ($ \hat{\mathcal{O}}\hat{\mathcal{O}}^{\dagger}=\hat{\mathcal{O}}^{\dagger}\hat{\mathcal{O}}$) matrix with a diagonal representation (see Theorem \eqref{t1}) of the form:
\begin{eqnarray}
\hat{\mathcal{O}} = \sum_{j=1}^d o_j P_{o_j}= \sum_{j=1}^d o_j |o_j\rangle \langle o_j|.
\end{eqnarray}
Where $P_{o_j}=|o_j\rangle \langle o_j|$ is the projector onto the eigenspace of the observable $\hat{\mathcal{O}}$ in some $d$-dimensional orthonormal basis set $\{|o_j \rangle\}_{j=1}^d$ ($\langle o_j|o_k\rangle=\delta_{jk}$) of eigenvectors $|o_j \rangle$ with eigenvalues $o_j$. A feature of such a linear operator, as it is hermitian, is to have real eigenvalues ($o_j \in \mathbb{R}$). If the system is prepared in the state $|\psi\rangle \doteq \sum_{j=1}^d c_j|o_j\rangle$, a projective measurement (Born's rule) entails the following conditional probability for obtaining an outcome $o_j$ of $\hat{\mathcal{O}}$:
\begin{eqnarray} 
Pr(o_j|\psi\rangle)&=&\langle \psi | P^{\dagger}_{o_j} P_{o_j} |\psi \rangle \\ 
&=& \langle \psi | P_{o_j}^2 |\psi \rangle\\
&=& \langle \psi| (|o_j\rangle \langle o_j|o_j\rangle \langle o_j|)| \psi \rangle\\ \label{eqforexp}
&=&\delta_{jj}\langle\psi|o_j\rangle \langle o_j |\psi\rangle \\ 
&=& |\langle \psi|o_j \rangle|^2=|\langle o_j|\psi \rangle|^2\\
&=& \left|\langle o_j|\sum_{k=1}^d c_k |o_k\rangle\right|^2\\
&=&  \left|\sum_{k=1}^d c_k \langle o_j|o_k\rangle\right|^2\\
&=&\left|\sum_{k=1}^d c_k \delta_{jk} \right|^2 = |c_j|^2.
\label{projective}
\end{eqnarray} 
And the state of the system after the measurement shall collapse to
\begin{eqnarray} 
    \frac{P_{o_j} |\psi \rangle}{\sqrt{Pr(o_j|\psi\rangle)}} &=& \frac{P_{o_j} |\psi \rangle}{|\langle o_j|\psi \rangle|^2}. 
\end{eqnarray}

The aforementioned result of the probability is consistent with the frequentist probability interpretation of statistical mechanics stated as the limit of the frequency of occurrence of a given outcome $o_j$ as the number of measurement trials goes to infinity. In a non-rigorous way, the frequentist probability reads:
\begin{align}
\label{probdef}
   Pr(o_j|\psi \rangle)\doteq lim_{n\rightarrow\infty} \frac{n_j}{n},
\end{align}
where $n_{j}$ denotes the number of outcomes of $o_j$, and $n$ denotes the number of measurement trials. In this formalism, the expectation value (a.k.a mean value or average value) of the observable $ \hat{\mathcal{O}}$ is defined as:
\begin{align}
\langle \hat{\mathcal{O}}\rangle &\doteq lim_{n\rightarrow\infty} \sum_{j=1}^d \frac{n_j}{n} o_j\\
&=\sum_{j=1}^d o_j Pr(o_j|\psi \rangle), 
\end{align}
and using Eq.~\eqref{eqforexp} one then has
\begin{align}
\langle \hat{\mathcal{O}}\rangle &=\sum_{j=1}^d o_j  \langle \psi |o_j \rangle \langle o_j|\psi \rangle \\
&= \langle \psi | \left(\sum_{j=1}^d o_j |o_j \rangle \langle o_j| \right) |\psi \rangle\\
&= \langle \psi | \left(\sum_{j=1}^d o_j P_{o_j} \right) |\psi \rangle\\
&=\langle \psi |\hat{\mathcal{O}}|\psi\rangle.
\end{align}
It is straightforward to extend this result to any function of the observable $\hat{\mathcal{O}}$ as follows:
\begin{align}
    \langle f(\hat{\mathcal{O})}\rangle &= \sum_{j=1}^d Pr(o_j|\psi \rangle)f(o_j) \\
    &= \langle \psi| f(\hat{\mathcal{O}})|\psi \rangle,
\end{align}
such that 
\begin{align}
    \langle \hat{\mathcal{O}}^2 \rangle=\langle \psi | \hat{\mathcal{O}}^2 | \psi \rangle.
\end{align}
Moreover, for any experiment consisting of $N$ measurements, the amount of dispersion of the dataset relative to its mean is quantified by the standard deviation. In quantum mechanics, the more interesting quantity is the uncertainty (a.k.a variance) associated with the observable and defined as the square of the standard deviation. From the definitions of probability and expectation value, the uncertainty can be derived according to:
\begin{align}
&(\Delta\hat{\mathcal{O}})^2 = lim_{n\rightarrow\infty} \sum_{j=1}^d \frac{n_j}{n} (o_j - \langle \hat{\mathcal{O}} \rangle)^2\\
&= \sum_{j=1}^d Pr(o_j|\psi \rangle)(o_j - \langle \hat{\mathcal{O}} \rangle)^2\\
&= \sum_{j=1}^d  \langle \psi |o_j \rangle \langle o_j|\psi \rangle (o_j - \langle \hat{\mathcal{O}} \rangle)^2 \\
&=  \sum_{j=1}^d \langle \psi |o_j \rangle \langle o_j|\psi \rangle (o_j^2 -2o_j \langle \hat{\mathcal{O}}\rangle + \langle\hat{\mathcal{O}}\rangle^2)   \\
&= \sum_{j=1}^d \langle \psi |o_j \rangle \langle o_j|\psi\rangle o_j^2 \label{eigfunc} \\ 
&+ \langle \hat{\mathcal{O}} \rangle^2 \sum_{j=1}^d \langle \psi |o_j \rangle \langle o_j|\psi\rangle\\
&- 2 \langle \hat{\mathcal{O}}\rangle \sum_{j=1}^d \langle \psi |o_j \rangle \langle o_j|\psi\rangle o_j,
\end{align}
which after rewriting Eq.~\eqref{eigfunc} according to $f(\hat{\mathcal{O}})|o_j\rangle = f(o_j) |o_j\rangle$, and bringing the summation sign close to elements of the same index, becomes
\begin{align}
&(\Delta\hat{\mathcal{O}})^2 =  \sum_{j=1}^d \langle \psi| \hat{\mathcal{O}}^2|o_j \rangle \langle o_j|\psi\rangle \\
&+\langle \hat{\mathcal{O}} \rangle^2  \langle \psi| \left( \sum_{j=1}^d|o_j \rangle \langle o_j|\right)|\psi\rangle\\
&- 2 \langle \hat{\mathcal{O}}\rangle \sum_{j=1}^d \langle \psi |\hat{\mathcal{O}}|o_j \rangle \langle o_j|\psi\rangle\\
&=   \langle \psi| \hat{\mathcal{O}}^2 \left( \sum_{j=1}^d|o_j \rangle \langle o_j| \right) \psi\rangle \\
&+\langle \hat{\mathcal{O}} \rangle^2  \langle \psi| \left( \sum_{j=1}^d|o_j \rangle \langle o_j|\right)|\psi\rangle\\
&- 2 \langle \hat{\mathcal{O}}\rangle  \langle \psi |\hat{\mathcal{O}} \left( \sum_{j=1}^d|o_j \rangle \langle o_j|\right)|\psi\rangle\\
&=  \langle \psi | \hat{\mathcal{O}}^2 | \psi \rangle -\langle\hat{\mathcal{O}}\rangle^2-2\langle \hat{\mathcal{O}}\rangle^2\\
&=\langle \hat{\mathcal{O}}^2 \rangle-\langle \hat{\mathcal{O}}\rangle^2.
\label{dispersion}
\end{align}
An equivalent result is obtained from
\begin{align}
(\Delta\hat{\mathcal{O}})^2 &= \sum_j \langle \psi |o_j \rangle \langle o_j|\psi \rangle (o_j - \langle \hat{\mathcal{O}} \rangle)^2 \\
&= \sum_j \langle \psi |o_j \rangle (o_j - \langle \hat{\mathcal{O}} \rangle)^2 \langle o_j|\psi \rangle \\
&= \sum_j \langle \psi | (\hat{\mathcal{O}} - \langle \hat{\mathcal{O}} \rangle)^2 |o_j \rangle \langle o_j|\psi \rangle \\
&=  \langle \psi | (\hat{\mathcal{O}} - \langle \hat{\mathcal{O}} \rangle)^2 \left(\sum_j|o_j \rangle \langle o_j|\right)|\psi \rangle \\
&=  \langle \psi | (\hat{\mathcal{O}} - \langle \hat{\mathcal{O}} \rangle)^2 |\psi \rangle \\
&=\langle (\hat{\mathcal{O}}-\langle\hat{\mathcal{O}}\rangle)^2 \rangle.
\end{align}
Therefore, the uncertainty (variance) in the measurement of the observable $\hat{\mathcal{O}}$ is: 
\begin{align}
\langle (\hat{\mathcal{O}}-\langle\hat{\mathcal{O}}\rangle)^2 \rangle = \langle \hat{\mathcal{O}}^2 \rangle-\langle \hat{\mathcal{O}}\rangle^2.    
\end{align}

A generalization of Born's rule to realistic experiments is another special case of postulate 3, and known as a positive operator valued measure (POVM) \cite{Nielsen2010}. Such a measurement accounts for open systems with intrinsic losses.

\subsubsection{Postulate 4 (Composite Space of multipartite systems)} 
\label{sec:post4}

Consider a quantum system composed of $N$ arbitrary subsystems, where subsystem $s$ is prepared in a $d_s$-dimensional qudit state $|\psi_{s}\rangle_{qd} = \sum_{j=1}^{d_s} c_{j_s} |o_{j}\rangle_s$ of a $d_s$-dimensional Hilbert space $\mathcal{H}_s^{d_s}$ with orthonormal basis set $\{|o_{j}\rangle_s \}_{j=1}^{d_s}$. The Hilbert state space $\mathcal{H}^{d}_{1\cdots N}$ of the composite physical system is given by the tensor product of its constituent Hilbert spaces (the state space of its subsystems a.k.a component physical systems): $\mathcal{H}^{d}_{1\cdots N}=\otimes_{s=1}^N \mathcal{H}_s^{d_s}=\mathcal{H}_1^{d_1} \otimes \mathcal{H}_2^{d_2} \otimes \cdots \otimes \mathcal{H}_N^{d_N}$, where $d=\prod_{s=1}^N d_s$ is the dimension of the composite space. By its turn, the corresponding orthonormal basis set of the composite state space is obtained from the tensor product between the basis of each constituent state space: $\{|o_{j}\rangle \}_{j=1}^{d}=\{|o_{j}\rangle_s \otimes \cdot\cdot\cdot \otimes|o_{k}\rangle_N \}_{j,\cdots,k=1}^{d_s,\cdots,d_N}$. 

\subsection{Quantum nondemolition measurements}
\label{sec:nondemolition}

\begin{figure}[htb!]
  \centering
  \includegraphics[scale=.4]{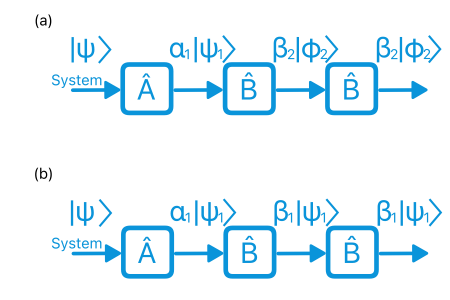}
  \caption{Disturbed and nondisturbed systems. Each square box represents a measure of the indicated observable. The arrows represent the evolution of the system state vector. (a) When the Hamiltonian contains a non-commuting observable $\hat{B}$, which does not commute with a measured observable $\hat{A}$, a measurement of $\hat{A}$ collapses the system initially in the state $|\psi\rangle$ to a state $|\psi_1\rangle$ which is not an eigenstate of $\hat{B}$. As a consequence, the system is disturbed by an increasing uncertainty in the non-commuting observable $\hat{B}$ due to a measurement of $\hat{A}$, such that the prediction of the measurement of $\hat{B}$ is uncertain. (b) When the system Hamiltonian $\hat{H}_s$ is given in terms of a commuting observable $\hat{B}$, which do commute with the measured observable $\hat{A}$, predictability of the measurement outcome of $\hat{B}$ is preserved and certain, and the system is minimally disturbed.}
  \label{fig:figure1}
\end{figure}

In the context of conjugate observables, when a given measured observable $\hat{\mathcal{O}}$ does not commute with the system Hamiltonian $\hat{H_s}$ (see Theorem \eqref{t2}), a measurement of $\hat{\mathcal{O}}$ disturbs the system and, as a consequence, the state of the system after the measurement is no longer constrained to an eigenstate of the Hamiltonian (see Fig.~\ref{fig:figure1}(a)). Conversely, when the Hamiltonian and the measured observable commute, $[\hat{\mathcal{O}}, \hat{H_s}]=0$, the state of the system after the measurement is an eigenstate of the Hamiltonian, and repeated measurements yield the same result (see Fig.~\ref{fig:figure1}(b)) thus minimally disturbing the system. 

A minimally disturbed quantum system, where $[\hat{\mathcal{O}}, \hat{H_s}]=0$, has a sequence of \textit{quantum nondemolition} (QND) measurements \cite{Braginsky1980}\cite{Besse2018} of a continuous QND observable $\hat{\mathcal{O}}$ if and only if $[\hat{\mathcal{O}}(t_1), \hat{\mathcal{O}}(t_n)]=0$ is satisfied for any time $t_n$. The free evolution of the system observables then becomes:
\begin{align}
    i\hbar \frac{d \hat{\mathcal{O}}}{dt} = [\hat{\mathcal{O}}, \hat{H}_s]=0.
\end{align}
A QND measurement enables the realization of multiple measurements on the same physical system without introducing quantum uncertainty into the observable being measured during each subsequent normal evolution that takes place after each measurement trial. In this sense, the predictability of a subsequent measurement outcome of a precisely measured observable is preserved, and any random back-action caused by the quantum uncertainty in a noncommuting observable is avoided. For this reason, a QND measurement is also called a back-action evading (BAE) measurement. One should be wary that a QND measurement, alike any ideal classical measurement, will also cause the wave-function to collapse, i.e, a QND measurement is not a collapse-evading measurement.

\section{Quantum information road to quantum computing}
\label{sec:quantuminfo}

The field of quantum information is primarily concerned with the transmission of quantum data through quantum channels \cite{Shannon} \cite{Bennett} \cite{Klaus} \cite{Bennett1984}. In this section, we focus on the backbones of the quantum information theory as we introduce the state vector formalism of quantum superposition in multipartite systems. The special case of quantum superposition, namely quantum entanglement, is addressed in \autoref{sec:entgsec}. 

\subsection{Quantum superposition}

Consider a quantum system in a $d$-dimensional state space $\mathcal{H}^{d}$. The eigenvectors of some observable $\hat{\mathcal{O}}$ form a set $\{|o_j\rangle\}_{j=1}^{d}$ of orthonormal basis vectors for the space, given that physical observables in quantum mechanics are represented by Hermitian operators. If the quantum system admits more than one independent eigenstate $|o_j\rangle$ of the observable then the global state vector of the system is termed a \textit{coherent state of quantum superposition} \cite{Young} \cite{Broglie}\cite{Gerlach}\cite{Streltsov2017}. According to postulate 1, the system's global state vector is described by a linear combination of the basis vectors as follows:
\begin{eqnarray}
\label{superposition}
|\psi\rangle \doteq \sum_{j=1}^{d} c_j|o_j\rangle,
\end{eqnarray}
where $c_j=\langle o_j|\psi\rangle$ denotes the complex probability amplitude associated with the relative frequency of occurrence (probability density) $|c_j|^2$ of a measurement outcome of $\hat{\mathcal{O}}$. In this sense, quantum superposition represents the uncertainty in the measurement outcome of a physical observable. Henceforth, the information concerning the preparation of the global state $|\psi\rangle$ of the system is then obtained in a non-deterministically way according to a probability distribution. And the associated probabilities ($|c_j|^2$) are given according to the frequentist probability interpretation defined in Eq.~\eqref{probdef}. 

The complex coefficients $c_j$ are also regarded as probability amplitudes of the complex-valued wave-function $\psi(x)$ describing the system. To see this is indeed so, one should recall that the wave-function in the position space is the projection of the system state vector onto the ``position stat'' according to
\begin{align}
\psi(x) = \langle x|\psi\rangle = \sum_{j=1}^{d} c_j \langle x|o_j\rangle = c_j \psi_j(x).
\end{align}
As a word of caution, $|x\rangle$ does not denote a basis in Hilbert space, as it belongs to an uncountable set (a.k.a uncountably infinite set), whereas the inner product Hilbert space is a finite-dimensional space by definition. The probabilistic interpretation of the wave-function is attributed to Max Born, who proposed that the square of the absolute value of the wave-function should represent the probability density associated with a measurement outcome. Furthermore, any ideal classical measurement causes the collapse of the wave-function, consequently updating the global state $|\psi\rangle$ of coherent superposition to one of the eigenstates $|o_j\rangle$ of the measured observable $\hat{\mathcal{O}}$. That is to say, for a quantum system with $N$ interacting 2-level subsystems (a.k.a qubits), the act of measurement yields a single value (a real number) out of $2^N$ possible classical measurement outcomes.

Quantum computing harnesses quantum superposition to perform a computation in all states simultaneously during a time window limited by the decoherence time (see \autoref{sec:decoherence}). To help grasp understanding, consider a system in a superposition of $2^N$ possible configurations (states) from $|000\cdots0\rangle$ to $|111\cdots1 \rangle$ (see postulate 4 in \autoref{sec:postulates} and \autoref{sec:Compositesys}), where each combination in a classical computing is representing a particular function to be computed. A classical computer would require an exponential amount of resources to process each state sequentially or in parallel as the number of bits increases. Quantum computers, however, can encode all $2^N$ combinations in a single state and perform the computation in a single go. This feature known as quantum parallelism backed by quantum superposition does not require a cluster of quantum computers. Superposition leads to a computational speed-up for certain classes of computational problems which, in turn, rely on optimized algorithms that provide speedups over the best classical algorithms for the same computational tasks. In other terms, quantum computers do not uniformly speed up all computational problems.

\subsubsection{The d-level quantum system: qudit} 

A quantum system consisting of $d$ levels is termed a $d$-dimensional qudit. Within the state vector formalism, $d$ denotes both the dimension of the qudit and the dimension of the system's state space, i.e, the number of linearly independent vectors in the orthonormal basis set. Particular quantum systems of $d=2$ and $d=3$ levels are commonly known as qubit and qutrit, respectively. Is a common misconception to suppose that a qudit is a system of $N$ qubits. While the former is a system of $d$ levels, the latter is a composite system of $N$ 2-level subsystems. In this sense, an $N$-qudit system is a quantum system of $N$ particles or artificial atoms each representing a subsystem $s$ of $d_s$ levels. 

In \autoref{sec:subsecqudit}, we describe an $N$-qudit system within the density operator formalism, where the corresponding $N$-qudit state is represented by a density operator (a.k.a density matrix) denoted $\rho_{Nqd}$ that is written in a orthonormal basis formed by $d$x$d$ matrices instead of state vectors (column vectors).

\subsubsection{The 2-level quantum system: qubit}

Generally, in quantum computing, and specially superconducting quantum computing, only systems of $d=2$ levels (qubits) are of interest. The main reason lies in the fact that higher order level systems are hard to control and manipulate. In such a case, to put it bluntly, the main difference between a classical (or conventional) computer and a quantum computer is that the former encodes data into binary digits (a.k.a bits) with their classical information measured by Shannon entropy \cite{Shannon}, whereas the latter encodes data in quantum bits (a.k.a qubits) with quantum information measured by the Von Neumann entropy \cite{Neumann1927}. In the field of quantum information the term ``qubit'', coined by Benjamin Schumacher \cite{Schumacher}, refers to a quantum system in a coherent superposition of two possible states of a single physical observable. Therefore, a qubit is any two-level quantum system in a state of coherent superposition. Those of skill in the art would appreciate that the term qubit can refer to the actual physical device in which information is stored, and it can also refer to the unit of information itself, abstracted away from its physical device. 

According to Postulate 1 (see \autoref{sec:postulates}), the state vector of an arbitrary $d$-level quantum system (qudit) in coherent superposition can be written in the form of Eq.~\eqref{superposition}. For the particular case of a single 2-level quantum system in the orthonormal basis set known as the computational basis (or canonical basis), the 1-qubit state vector becomes:
\begin{align}
|\psi\rangle_{1qb} &= \sum_{j=1}^{d=2} c_j|j\rangle=\sum_{j=0}^{d-1=1}c_j|j\rangle \\&= c_0 |0\rangle + c_1 |1\rangle 
=\begin{pmatrix} c_0 \\ c_1 \end{pmatrix}.
\end{align}
Where $c_0$ and $c_1$ are the alluded complex probability amplitudes, i.e, the coefficients of the complex-valued wave-function. This state is often regarded to be in the $|Z_{\pm}\rangle$ basis:
\begin{eqnarray}
|0\rangle \doteq |Z_+\rangle & \doteq&\begin{pmatrix} 1 \\ 0 \end{pmatrix},
\end{eqnarray}
\begin{eqnarray}
|1\rangle \doteq |Z_-\rangle & \doteq& \begin{pmatrix} 0 \\ 1 \end{pmatrix},
\end{eqnarray}
From Born's rule (see Eq.~\eqref{projective}), a projective measurement yields either outcome $|0 \rangle$ with probability $|c_0|^2$ or $|1 \rangle$ with probability $|c_1|^2$. Normalization constraint is necessary in order to satisfy the law of total probability: $|c_0|^2 + |c_1|^2 = 1$.

It is always possible to write a qubit state vector in a different basis. One possible choice is the following orthonormal basis set: 
\begin{eqnarray}
|+\rangle \doteq |X_+\rangle & \doteq&\frac{1}{\sqrt{2}}(|0\rangle + |1\rangle) \\
&=& \frac{1}{\sqrt{2}}\begin{pmatrix} 1 \\ 1 \end{pmatrix},
\end{eqnarray}
\begin{eqnarray}
|-\rangle \doteq |X_-\rangle & \doteq& \frac{1}{\sqrt{2}}(|0\rangle - |1\rangle) \\
&=& \frac{1}{\sqrt{2}}\begin{pmatrix} 1 \\ -1 \end{pmatrix},
\end{eqnarray}
such that 
\begin{eqnarray}
|0\rangle &=& \frac{|+\rangle + |-\rangle}{\sqrt{2}}, \\
|1\rangle &=& \frac{|+\rangle - |-\rangle}{\sqrt{2}}.
\end{eqnarray}
The qubit state vector in this new basis now reads
\begin{eqnarray}
|\psi\rangle_{1qb} &=& c_0 |0\rangle + c_1 |1\rangle \\
&=& \frac{c_0+c_1}{\sqrt{2}}|+\rangle+\frac{c_0-c_1}{\sqrt{2}}|-\rangle.
\end{eqnarray}
A projective measurement in the $|\pm \rangle$ basis yields a probability
\begin{align} 
Pr(o_{\pm}|\psi\rangle)&=|\langle \pm|\psi \rangle|^2 \\
&= \left|\langle \pm| \left(\frac{c_0+c_1}{\sqrt{2}}|+\rangle+\frac{c_0-c_1}{\sqrt{2}}|-\rangle\right)\right|^2 \\
&=\left| \langle  \pm | \pm \rangle\frac{c_0\pm c_1}{\sqrt{2}} \right|^2 = \left|\frac{c_0\pm c_1}{\sqrt{2}} \right|^2 \\
&= \frac{\left|c_0 \pm c_1\right|^2}{2},
\end{align} 
given the orthonormality condition $\langle \pm | \pm \rangle=1$ as defined in Eq.~\eqref{delta}. That is, measuring with respect to (w.r.t) the $|+\rangle$ basis yields a probability $|c_0+c_1|^2/2$ collapsing the system to the state $|+\rangle$ immediately after the measurement. On the other hand, measuring w.r.t to $|-\rangle$ yields the probability $|c_0-c_1|^2/2$. 

Another commonly basis for a qubit state vector is the following orthonormal basis set:
\begin{align}
|+i\rangle &\doteq |Y_+\rangle  \doteq\frac{1}{\sqrt{2}}(|0\rangle + i|1\rangle) \\
&=  \frac{1}{\sqrt{2}}
\left(
\begin{pmatrix} 1 \\ 0 \end{pmatrix} + i\begin{pmatrix} 0 \\ 1 \end{pmatrix}
\right)=\frac{1}{\sqrt{2}}\begin{pmatrix} 1 \\ i \end{pmatrix},\\
|-i\rangle &\doteq |Y_-\rangle \doteq \frac{1}{\sqrt{2}}(|0\rangle - i|1\rangle) \\
&= 
\frac{1}{\sqrt{2}}\left(\begin{pmatrix} 1 \\ 0 \end{pmatrix} - i\begin{pmatrix} 0 \\ 1 \end{pmatrix}
\right)=\frac{1}{\sqrt{2}}\begin{pmatrix} 1 \\ -i \end{pmatrix}.
\end{align}

\subsubsection{The 3-level quantum system: qutrit}

A single three-level ($d$ $=$ $3$) quantum system known as qutrit has the following state vector representation in the canonical basis:
\begin{align}
|\psi\rangle_{1qt} &= \sum_{j=0}^{d-1=2}c_j|j\rangle = c_0 |0\rangle + c_1 |1\rangle +  c_2 |1\rangle \\
&= c_0 \begin{pmatrix} 1 \\ 0 \\ 0 \end{pmatrix} + c_1 \begin{pmatrix} 0 \\ 1 \\ 0 \end{pmatrix}+ c_2 \begin{pmatrix} 0 \\ 0 \\ 1 \end{pmatrix}=\begin{pmatrix} c_0 \\ c_1 \\ c_2 \end{pmatrix}.
\end{align}

In \autoref{sec:particularstates}, we obtain the 1-qutrit coherent and incoherent density operator from the generalized Bloch's representation for $d$-dimensional $N$-qudit states.

\subsubsection{Composite systems with separable states} \label{sec:Compositesys}

From postulate 4 (see \autoref{sec:post4}), the resulting \textit{separable global state vector} of a composite physical system with $N$ qudit subsystems, where subsystem $s$ is prepared in the qudit state vector $|\psi_s\rangle_{qd}$, is given by the tensor product between the qudit states of each subsystem, as follows:
\begin{align}
\label{separable}
|{\psi_{1,2,\cdots,N}^{sep}}\rangle_{Nqd} &\doteq \otimes_{s=1}^N |\psi_{s}\rangle_{qd}\\
&= |\psi_1\rangle_{qd} \otimes |\psi_2\rangle_{qd} \otimes \cdot \cdot \cdot \otimes |\psi_N\rangle_{qd} \\
&= \bigg(c_{1_1} |o_{1}\rangle_1 + c_{2_1} |o_{2}\rangle_1 \\&+\cdots+ c_{{d}_1} |o_{{d}}\rangle_1\bigg) \\
&  \otimes\bigg(c_{1_2} |o_{1}\rangle_2 + c_{2_2} |o_{2}\rangle_2 \\
&+\cdots+ c_{d_2}|o_{{d}}\rangle_2 \bigg) \\
&\otimes\cdot \cdot \cdot\otimes \bigg( c_{1_N} |o_{1}\rangle_N + c_{2_N} |o_{2}\rangle_N \\
&+ \cdots+ c_{d_N}|o_{{d}}\rangle_N \bigg),
\end{align}
with the normalization condition
\begin{eqnarray}
\sum_{j=1}^{d_s^N} |\gamma_{j}|^2=1.
\end{eqnarray}
Here, the coefficient $\gamma_j$ represents a complex probability amplitude from the sample space of $d_s^N$ possible measurement outcomes associated with each state vector of the composite system. Resulting states of this fashion are called \textit{separable states} (a.k.a \textit{product states}), and they are never entangled. A separable state is one that can be created by quantum local operations and classical communication (LOCC) acting on its subsystems. By local operations \cite{locc} we mean operations that act individually on each subsystem (each single qudit) of a composite system.

For the particular case of a bipartite quantum system ($N=2$ qudits) there is an arbitrary subsystem $s$$=$$1$ in the state $|\psi_1 \rangle$ and subsystem $s$$=$$2$ in the state $|\psi_2 \rangle$ with Hilbert spaces $\mathcal{H}_1^{d_1}$ and $\mathcal{H}_2^{d_2}$, respectively. The Hilbert space of the composite system is denoted by $\mathcal{H}^{d_1 d_2}_{12}\doteq \mathcal{H}_1^{d_1} \otimes \mathcal{H}_2^{d_2}$. Supposing that each subsystem is a $2$-level quantum system (qubit), the resulting composite state vector in the canonical basis is the following separable state (product state) in quantum superposition:
\begin{align}
|\psi_{12}^{sep}\rangle_{2qb}&=|\psi_1 \rangle_{qb} \otimes |\psi_2 \rangle_{qb} \\
&=\left( \alpha_1 |0\rangle_1 + \beta_1 |1\rangle_1 \right)  \\
&\otimes \left(\alpha_2 |0\rangle_2 + \beta_2 |1\rangle_2 \right)\\
&=  \alpha_1  \alpha_2 |0\rangle_1 \otimes |0\rangle_2 +  \alpha_1  \beta_2 |0\rangle_1 \otimes|1\rangle_2 \\
&+ \beta_1 \alpha_2 |1\rangle_1 \otimes |0\rangle_2 + \beta_1 \beta_2 |1\rangle_1 \otimes |1\rangle_2 \\
&= \gamma_{00} |00\rangle_{12} + \gamma_{01} |01\rangle_{12} \\
&+\gamma_{10} |10\rangle_{12} + \gamma_{11}|11\rangle_{12},
\end{align}
with the corresponding matrix representation
\begin{align}
|\psi_{12}^{sep}\rangle_{2qb} &= \gamma_{00} \begin{pmatrix} 1 \\ 0 \\ 0 \\ 0 \end{pmatrix} + \gamma_{01} \begin{pmatrix} 0 \\ 1 \\ 0 \\ 0 \end{pmatrix}  
+ \gamma_{10} \begin{pmatrix} 0 \\ 0 \\ 1 \\ 0 \end{pmatrix} \\
&+ \gamma_{11} \begin{pmatrix} 0 \\ 0 \\ 0 \\ 1 \end{pmatrix}
=
\begin{pmatrix} \gamma_{00} \\ \gamma_{01} \\ \gamma_{10} \\ \gamma_{11} 
\end{pmatrix}.
\end{align}
Here, we have defined: $\gamma_{00}\doteq\alpha_1\alpha_2$, $\gamma_{01}\doteq\alpha_1\beta_2$, $\gamma_{10}\doteq\beta_1\alpha_2$, $\gamma_{11}\doteq\beta_1\beta_2$. Finally, normality condition for state vectors entails
\begin{align}
\sum_{x \in\{0,1\}^N=1}^{4} |\gamma_{x}|^2&=
|\gamma_{00}|^2 + |\gamma_{01}|^2 \\&+ |\gamma_{10}|^2 + |\gamma_{11}|^2 = 1,
\end{align}
where $x$ denotes an element from the set containing all $2^N=4$ possible permutations of strings of length $N=2$. And for a bipartite three-level system ($N$$=$$2$ qutrits), the composite state of the system becomes:
\begin{align}
|\psi_{12}^{sep}\rangle_{2qt}&=|\psi_1 \rangle_{qt} \otimes |\psi_2 \rangle_{qt} \\
&=\left( a_1 |o_{1}\rangle_{1} + b_1 |o_{2}\rangle_{1} + c_1 |o_{3}\rangle_{1} \right) \\
&\otimes \left(a_2 |o_{1}\rangle_{2} + b_2 |o_{2}\rangle_{2} + c_2 |o_{3}\rangle_{2} \right)\\
&=a_1a_2 |o_{1} o_{1}\rangle_{12} + a_1 b_2  |o_{1} o_{2}\rangle_{12} \\&
+ a_1c_2 |o_{1} o_{3}\rangle_{12} + 
b_1a_2 |o_{2} o_{1}\rangle_{12} \\
&+  b_1b_2 |o_{2} o_{2}\rangle_{12} +  b_1c_2 |o_{2} o_{3}\rangle_{12} \\
&+
c_1a_2 |o_{3} o_{1}\rangle_{12} + c_1b_2 |o_{3} o_{2}\rangle_{12} \\
&+ c_1c_2 |o_{3} o_{3}\rangle_{12}.
\end{align}

In the context of computer hardware, a classical register of $N$ qubits has $2^N$ possible configurations, however it can store only one number of information at a time, while a quantum register of $N$ qubits will have a composite system with $2^N$ complex coefficients (amplitudes). Therefore, a quantum register can store a superposition of $2^N$ classical bits of information simultaneously, i.e, it can store all numbers from 1 to $2^{N}$. In this sense a quantum register of $N=11$ qubits will store all numbers from 1 to 2048, whereas a classical register with a sequence 11111011011 will store only the number 2011. Moreover, if each complex amplitude is stored to 128 bits of precision, then each requires 256 bits (or 32 bytes) of information, and $2^N$ qubits will then require $32$x$2^N$ bytes of information for storage.

\subsection{Bloch Sphere}

A generic 1-qubit state vector can be written as
\begin{align}
|\psi\rangle_{1qb} &= c_0 |0\rangle + c_1 |1\rangle \\
&= (\alpha_1 + i \beta_1)  |0\rangle +  (\alpha_2 + i \beta_2) |1\rangle,
\end{align}
where we have expanded the complex coefficients $c_j$'s, with $\alpha$'s and $\beta$'s as real numbers. Normalization condition $\sqrt{c_0^2 + c_1^2}=1$ then implies in the reduction of four degrees of freedom to only three. Choosing the Hopf coordinate system, one can write the complex coefficients as:
\begin{align}
    c_0&= e^{i\omega} cos(\theta/2),\\
    c_1&= e^{i\omega+\phi} sin(\theta/2),
\end{align}
and the 1-qubit state vector becomes
\begin{align}
|\psi\rangle_{1qb} &= e^{i\omega} cos(\theta/2) |0\rangle + e^{i\omega+\phi} sin(\theta/2) |1\rangle \\
&= e^{i\omega} \left( cos(\theta/2) |0\rangle + e^{i\phi} sin(\theta/2) |1\rangle \right).
\end{align}
However, qubit states that differ only by a global factor $e^{i\omega}$ are physically indistinguishable, i.e, a global factor does not change the measurement outcome:
\begin{align}
    |\langle x |(e^{i\omega} | \psi\rangle)|^2 = |e^{i\omega} (\langle x | \psi\rangle)|^2 = |\langle x | \psi\rangle|^2.
\end{align}
With that, one can adopt the following convenient representation for a generic single qubit state vector:
\begin{align}
\label{hopf}
|\psi\rangle_{1qb} &= cos(\theta/2) |0\rangle + e^{i\phi} sin(\theta/2) |1\rangle \\
&=\alpha |0\rangle + \gamma e^{i\phi}|1\rangle,
\end{align}
with $\alpha, \gamma, \theta \in \mathbb{R}$, and $0 \leq \theta \leq \pi$, and $0 \leq \phi \leq 2\pi$. The factor $e^{i\phi}$ is a complex number with absolute value equal to one ($|e^{i\phi}|=1$) that can be decomposed in terms of sines and cosines using Euler's formula $e^{i\phi}=cos(\phi)+isin(\phi)$. Because of that, the real number $\phi$ is regarded as the \textit{relative phase} of the wave-function in analogy to the phase appearing in the argument of sinusoidal functions describing oscillations, even though no time-dependence and, therefore, no oscillation exists in this case. Whereas $\theta$ gives the probability of each possible measurement outcome:
\begin{align}
    p(|0\rangle)&=|\alpha|^2 = cos^2(\theta/2),\\
    p(|1\rangle)&=|\gamma e^{i\phi}|^2 = \gamma e^{i\phi}\gamma e^{-i\phi}=\gamma^2 = sin^2(\theta/2).
\end{align}

\begin{figure}[htb!]
  \centering
  \includegraphics[scale=.2]{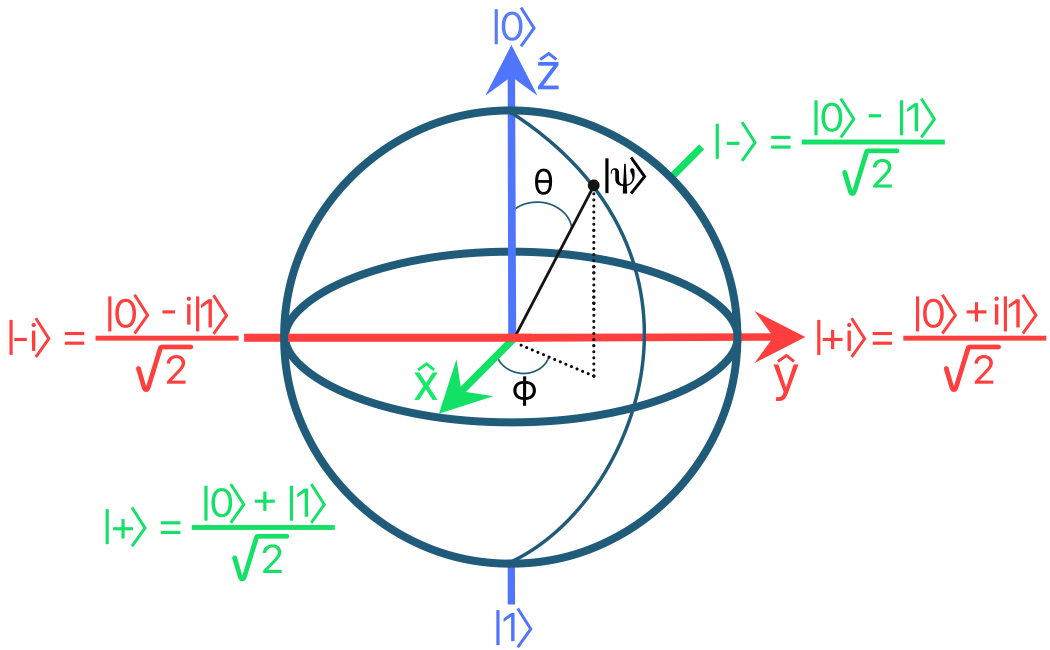}
  \caption{Bloch sphere representation for a generic 1-qubit system. The basis, $|0\rangle$ and $|1\rangle$, are orthonormal column vectors in the Hilbert space, however, they are represented by antiparallel (antipodal) vectors on the Bloch sphere. In this case $\theta/2$ is the actual angle in the Hilbert space, while $\theta$ is the angle in the Bloch sphere. In the context of density operators (see \autoref{sec:densityformalism}), single 1-qubit state vectors whose coordinates of their Bloch's vectors are of points on the surface of the sphere ($||\vec{r}||=1$) have each a corresponding pure density operator, while coordinates of points located inside the sphere ($||\vec{r}||<1$) have corresponding mixed density operators that cannot be associated with a single state vector.}
  \label{fig:figure2}
\end{figure}

Since the magnitude of this normalized 1-qubit state vector is equal to one ($\langle \psi | \psi \rangle=1$), the coordinates $\theta$ and $\phi$ can be regarded as spherical coordinates of a unit sphere of radius $||\vec{r}||=1$ and vector:
\begin{align}
    \vec{r} &= ||\vec{r}||\hat{r} =||\vec{r}||(x\hat{x}+y\hat{y}+z\hat{z})  \\
    &= ||\vec{r}|| sin(\theta)cos(\phi)\hat{x}\\
    &+||\vec{r}|| sin(\theta)sin(\phi)\hat{y} \\
    &+ ||\vec{r}||cos(\theta)\hat{z} \\
    &= ||\vec{r}||\begin{pmatrix} sin(\theta)cos(\phi) \\ sin(\theta)sin(\phi) \\ cos(\theta) \end{pmatrix}.
\end{align}
In this scenario, a generic 1-qubit state vector in the form of Eq.~\eqref{hopf} can be associated with a so-called Bloch's vector whose coordinates are of a point on the surface of a three-dimensional unit sphere known as the Bloch's sphere (see Fig.~\ref{fig:figure2}), even though a 2-level system is embedded in a two-dimensional Hilbert space. For historical reasons, the Bloch's sphere is also referred to as the Poincaré ball when describing classical polarization states. 

In the Bloch's sphere representation, the basis states can be obtained from the following choice of coordinates:
\begin{align*}
\theta &=0, \forall \phi \implies \vec{r}= \begin{pmatrix} 0 \\ 0 \\ 1 \end{pmatrix} \text{and } |\psi\rangle =|0\rangle,\\
\theta &=\pi, \forall \phi \implies \vec{r}= \begin{pmatrix} 0 \\ 0 \\ -1 \end{pmatrix} \text{and } |\psi\rangle = e^{i\phi} |1\rangle,
\end{align*}
and since the factor $e^{i\phi}$ now acts as a global phase it can be neglected. The remaining basis are:
\begin{align*}
\theta &=\frac{\pi}{2}, \phi =0 \implies \vec{r}= \begin{pmatrix} 1 \\ 0 \\ 0 \end{pmatrix} \text{and } |\psi\rangle = |+\rangle,\\
\theta &=\frac{\pi}{2}, \phi =\pi \implies \vec{r}= \begin{pmatrix} -1 \\ 0 \\ 0 \end{pmatrix} \text{and } |\psi\rangle = |-\rangle,\\
\theta &=\frac{\pi}{2}, \phi=\frac{\pi}{2} \implies \vec{r}= \begin{pmatrix} 0 \\ 1 \\ 0 \end{pmatrix} \text{and } |\psi\rangle = |+i\rangle,\\
\theta &=\frac{\pi}{2}, \phi=\frac{3\pi}{2} \implies \vec{r}= \begin{pmatrix} 0 \\ 0 \\ 1 \end{pmatrix} \text{and } |\psi\rangle = |-i\rangle.\\
\end{align*}

In \autoref{sec:particularstates}, we derive the conditions to describe a generic 1-qubit state as a pure or mixed density operator whose Bloch's vectors have coordinates of points located inside the Bloch's sphere or on its surface, respectively.

\subsection{Quantum entanglement}
\label{sec:entgsec}

Quantum entanglement is a special kind of quantum superposition that quantum computers rely on to speed up calculations \cite{Shor1994}\cite{Grover}, protect communication against eavesdropping \cite{Bennett1984}\cite{Morris}, and double the classical capacity of a communication channel \cite{Bennett}. More generally, entanglement is a feature where two or many subsystems of a multipartite quantum system are strongly correlated to one another that the state of each subsystem cannot be described independently of the state of the other without taking into account the global state of the system: a holistic feature of entangled states. Whereas the quantum correlation between two subsystems is such that the state of one subsystem (its physical reality) is immediately influenced by a measurement of the state of the other, regardless of their spacelike separation. This peculiar phenomenon dubbed ``quantum nonlocality'', whereby quantum information appears to travel faster than the speed of light, sparked a heated debate culminating in the famous article co-authored by Einstein, Podolsky, and Rosen (EPR) in 1935 \cite{EPR}. For a more in depth review on quantum entanglement, the reader may resort to Horodecki \textit{et al}. \cite{Horodecki2009}. 

\subsubsection{The EPR hypothesis}

In considering Max Born's probabilistic interpretation of the wave-function, Albert Einstein addressed the mechanism of the wave-function collapse during discussions in the 1927 Solvay conference. Einstein's gedanken (thought experiment) considered the diffraction of a beam of a single electron passing through a pinhole. According to the principle of superposition, before the electron hits a screen, the wave-function in the position space represents the superposition of all the eigenstates associated with the positions available to the electron on that screen. The act of measurement would then cause the wave-function to collapse to a single eingestate of the position observable. Einstein referred to this mechanism by which the wave-function is instantly updated as ``action at a distanc''. The famous phrase ``spooky action at a distance'' is believed to had its origin only twenty years later, in the 1947 letter to Max Born \cite{Forman1970}: \textit{``Ich kann aber deshalb nicht ernsthaft daran glauben, weil die Theorie mit dem Grundsatz unvereinbar ist, daß die Physik eine Wirklichkeit in Zeit und Raum darstellen soll, ohne spukhafte Fernwirkungen''} \textit{(I cannot seriously believe in it (quantum mechanics) because the theory is incompatible with the principle that physics should represent reality in space and time, without spooky action at a distance)}. 

Einstein went further with his hypothesis that quantum mechanics was incomplete and published, among three other physicists, the so-called EPR paper in 1935 \cite{EPR}. The paper devised a thought experiment considering a system of two entangled particles to assess whether a measurement in one particle would have an instantaneous effect on the other. The starting point of the EPR paper adopts the following definitions regarding the quantum mechanical theory:
\begin{itemize}
    \item Realism: every element of reality in nature (the so-called observables) must be predicted within the quantum theory without causing any disturbance to the quantum system. That is, it should be possible to predict the outcome of a measurement with 100\% certainty before the measurement is even realized.
    \item Self-consistency: a quantum theory shall be able to predict and explain experimental data in a consistent manner, i.e, free of contradictions.
    \item Completeness: a complete quantum theory is such that there is a theoretical analog for every element of reality in nature.
    \item Locality: a measurement in one subsystem should not influence the outcome of a measurement in another subsystem.
\end{itemize}
Then, for two conjugate observables $\hat{A}$ and $\hat{B}$ satisfying Eq.~\eqref{Conjugate} (e.g. spin $\hat{\sigma}_x$ and $\hat{\sigma}_z$), the EPR paper states that only one of the following two hypotheses is a correct assumption about quantum mechanics:
\begin{enumerate}
    \item $\hat{A}$ and $\hat{B}$ cannot simultaneously satisfy the criterion of realism;
    \item The theory of quantum mechanics is incomplete.
\end{enumerate}
For the purpose of explaining the thought experiment, consider the following scenario proposed by David Bohm where a subatomic particle of total spin zero decays in two new particles denoted $q_1$ and $q_2$ of subsystems $s=1$ and $s=2$, respectively. (It is worth mentioning that entangled systems have already been produced in the laboratory with spin-1 particles, particularly photons, via a process known as parametric down-conversion \cite{Klaus}, and with Cooper pairs of electrons in a circuit-based superconducting quantum processor driven by a series of microwave pulses.) If the process of decay occurs in an isolated system then the total spin is a conserved quantity, such that the resulting entangled particles have opposite spins and the total spin of the bipartite system adds up to zero as expected. Moreover, if the experiment is controlled in a way that the spins are only allowed to have vertical directions, the two possible configurations for the outcome of the decay are: particle $q_1$ with spin up ($|1\rangle_1$) and particle $q_2$ with spin down ($|0\rangle_2$) or, conversely, particle $q_1$ with spin down and $q_2$ with spin up. The global state of the system is then regarded as a superposition of the aforementioned two possible outcomes of the decay with both particles described by a single wave-function and, therefore, a single Schrödinger equation. Such an entangled state is refereed to as a singlet state or Bell-like state. It is then possible to write the state of the system in two conjugate basis:
\begin{align}
|\psi_{x}^{Bell}\rangle &\doteq \frac{|X+\rangle_{1} \otimes |X-\rangle_{2} - |X-\rangle_{1} \otimes |X+\rangle_{2}}{\sqrt{2}}, \\
|\psi_{z}^{Bell}\rangle &\doteq \frac{|Z+\rangle_{1} \otimes |Z-\rangle_{2} - |Z-\rangle_{1} \otimes |Z+\rangle_{2}}{\sqrt{2}}.
\end{align}
While the spin observables are given according to the spectral decomposition defined in Eq.~\eqref{spectraldec}, as follows:
\begin{align}
\hat{\sigma}_x = \sum_{j=1}^2 o_j P_{o_j}&= 1 P_+ + (-1) P_-\\
         &= |+\rangle \langle +| - |-\rangle \langle -|\\
         &= |0\rangle \langle 1| + |1\rangle \langle 0|,
\end{align}
and
\begin{align}
\hat{\sigma}_z &= \sum_{j=1}^2 o_j P_{o_j}= |0\rangle \langle 0|-|1\rangle \langle 1|.
\end{align}
EPR then assumes locality as a true assumption about quantum mechanics, ensuring both particles to be so far apart that no information traveling at the speed of light can influence the result of a measurement. Within the constraints set by the experiment, in which the spins in each direction are antiparallel to each other, the two particles are regarded as maximally correlated, as are their measurement outcomes. That being so, from the measurement postulate: if a measurement of $\hat{\sigma}_z$ on particle $q_1$ results in spin up, then particle $q_2$ must have spin down with 100\% certainty. This means that particle $q_2$ can be measured without disturbing the system, hence, there should be an element of reality associated with $\hat{\sigma}_z$. The same thought process applies for a measurement of $\hat{\sigma}_x$. The logic inference is that assumption 1 must then be false, i.e, the aforementioned Bell state seems to allow simultaneous reality for $\hat{\sigma}_x$ and $\hat{\sigma}_z$. This led EPR to the erroneous conclusion in which the theory of quantum mechanics is incomplete. Advocating on behalf of EPR, one must recognize the sheer lack of experiments available at the time. As a matter of fact, the first loophole-free experiment demonstrating the non-local phenomenon of quantum entanglement was carried out only 80 years later, in 2015, by Hensen \textit{et al}. \cite{Hensen2015}.

The key conundrum of quantum entanglement lies in the fact that quantum correlations may actually be stronger than non-quantum (classical) correlations. This result becomes evident upon quantification of the degree of correlation of measurement outcomes from conjugate observables, where the prime example is the measurement of the spin in different directions. Once again, the nonlocal quantum correlation of the bipartite system allows the state of one subsystem to be simultaneously determined upon a measurement of the state of the other. Given that such a non-local phenomenon happens seemingly faster than the speed of light, quantum mechanics appears to violate Einstein's principle of special relativity in which the speed of any physical particle (or wave) is upper bounded by the speed of light. However, as measurement outcomes must be compared, and such information travels classically, bounded by the speed of light, Einstein's principle is salved. As a consequence, faster-than-light communication becomes impossible. 

\subsubsection{The CHSH inequality} 

Driven by the misleading conclusion in which the theory of quantum mechanics was incomplete, EPR suggested the existence of a theory of hidden variables to account for the classical ignorance related to the initial conditions of the experiment. A mindset long espoused by Einstein and conveyed through his famous 1926 quote ``God doesn't play dice'', believing that quantum mechanics should not be framed as a probabilistic theory, but rather as a deterministic one. The first deterministic theory of this kind, where the so-called hidden variables represent the position of a given particle, was elaborated in 1952 by David Bohm \cite{Bohm1952} in a similar way to an earlier but less famous theory pioneered by Louis de Broglie in 1927. The theory known as Bohmian mechanics or ``pilot wave interpretation'' was devised as an alternative approach to the standard notion of quantum mechanics formulated by Bohr, Heisenberg, and Schrödinger, namely, ``the Copenhagen interpretation'' whose postulates we have described in \autoref{sec:postulates}. In a nutshell, Bohmian mechanics is based on a so-called guiding field whose definition renders the theory more non-local than the original probabilistic theory of quantum mechanics is. And for that reason, it does not reproduce the same successful predictions as the standard Copenhagen interpretation.

Then, in 1964, John Bell \cite{Bell} proposed a deterministic local hidden variable model (LHVM) to assess the claims put forward by the EPR paper. Bell's approach considered a generic statistical theory (that is, not necessarily related to quantum mechanics) with notions of a probability distribution $p(\lambda)$ of a sample space $\Lambda$ with hidden variables $\lambda$. Restricting the case for a bipartite system, the model allowed the possibility of measuring variables (observables) $A_1$ and $A_2$ in subsystem $A$, and variables $B_1$ and $B_2$ in subsystem $B$ with probability outcomes $a_1, a_2$ and $b_1, b_2$, respectively, sharing the following relation:
\begin{align}
    C(\lambda) &\doteq A_1(\lambda) B_1(\lambda) + A_1(\lambda) B_2(\lambda) \\
    &+ A_2(\lambda) B_1(\lambda) - A_2(\lambda) B_2(\lambda),
\end{align}
where the variables take on two dichotomic measurements whose spectrum of values range between $\pm 1$. Then, for an ensemble of particles in a perfect entangled state (e.g. singlet state), the joint probability of the ensemble statistics yields:
\begin{align}
    p(ab|AB) \neq p(a|A)p(b|B).
\end{align}
If locality is assumed, one then has
\begin{align}
    C(\lambda) &\doteq A_1(\lambda)[B_1(\lambda)+B_2(\lambda)]\\
    &+ A_2(\lambda)[B_1(\lambda)-B_2(\lambda)] = \pm 2.
\end{align}
and also
\begin{align}
p(ab|AB) = \bigints_{\Lambda} p(a|A\lambda)p(b|B\lambda) p(\lambda)d\lambda.
\end{align}
Building on Bell's paper, and assuming locality, Clauser, Horne, Shimony, and Holt devised the following correlation inequality known as the CHSH inequality \cite{CHSH} for bipartite systems:
\begin{align*}
    |\langle C(\lambda) \rangle| &\doteq \bigg|\bigints_{\Lambda} \bigg\{A_1(\lambda)[B_1(\lambda)+B_2(\lambda)]\\
    &+ A_2(\lambda)[B_1(\lambda)-B_2(\lambda)]\bigg\}p(\lambda)d\lambda \bigg|\leq 2,
\end{align*}
corresponding to a bound on any LHVM. The CHSH inequality is regarded as more feasible inequality for an experimental test of the LHVM as an independent hypothesis from quantum formalism. To name a few, Ursin et al \cite{Ursin2006}, in 2006, reported a experimental violation of the CHSH inequality measured by two observers separated by a distance of $144km$. Aspect \textit{et al}. \cite{Aspect}\cite{Aspect1982} were the first to verify experimentally the violation of Bell's inequality. However, to ensure that the state of particle $q_2$ is actually instantly determined once the state of particle $q_1$ is measured, the experiment must be carried out during a time window shorter than the time taken by any possible contaminating information traveling from one particle to the other at the speed of light and which could influence the result of a measurement. The first loophole-free experiment of this kind was only carried out by Hensen \textit{et al}., in 2015 \cite{Hensen2015}, reporting a violation of the CHSH inequality. The experiment consisted of spin-like particles separated by a distance of $1,280m$ setting a $4.27$-$ms$ time window during which two local events are space-like separated. The experimental evidence that certain entangled states do violate Bell's inequality confirms that locality and realism known together as local-realism \cite{Aspect}\cite{Aspect1982} cannot be simultaneously satisfied.

Theoretically, one can show the maximum violation of the CHSH inequality for the following quantum observables:
\begin{align}
    \hat{A}_1 &\doteq R\doteq\vec{r}\cdot \vec{\sigma}, \\
    \hat{A}_2 &\doteq Q\doteq\vec{q}\cdot \vec{\sigma},\\
    \hat{B}_1 &\doteq S\doteq\vec{s}\cdot \vec{\sigma},\\
    \hat{B}_2 &\doteq T\doteq\vec{t}\cdot \vec{\sigma},
\end{align}
where $\vec{r}$, $\vec{q}$, $\vec{s}$, and $\vec{t}$ are 3-dimensional real unit vectors describing the measurements performed by each parties, and $\vec{\sigma} = \sigma_x \hat{x}+  \sigma_y \hat{y} + \sigma_z \hat{z}$ is a 3-dimensional vector with the Pauli-$\sigma$ matrices defined in \autoref{sec:subsecqudit}. One then has
\begin{align}
    \hat{c} &= R \otimes S + R \otimes T + Q \otimes S - Q \otimes T,
\end{align}
so that
\begin{align}
    \langle \hat{c} \rangle &= \langle \psi_{ab} |R \otimes S |  \psi_{ab} \rangle +
     \langle \psi_{ab} |R \otimes T |  \psi_{ab} \rangle \\&+
      \langle \psi_{ab} |Q \otimes S |  \psi_{ab} \rangle -
       \langle \psi_{ab} |Q \otimes T |  \psi_{ab} \rangle.
\end{align}
Then, using Tsirelson's inequality \cite{Tsirelson1980} defined as
\begin{align*}
    4 \id_2 + [Q, R] \otimes [S, T] &\doteq \bigg(Q \otimes S + R \otimes S \\ 
    & +R \otimes T - Q \otimes T \bigg)^2,
\end{align*}
one has
\begin{align}
    \langle \hat{c}^2 \rangle &= \langle 4 \id_2 \rangle + \langle [Q, R] \otimes [S, T] \rangle \\
    &=\langle 4 \id_2 \rangle + \langle [Q, R]  \rangle \langle [S, T] \rangle,
\end{align}
given that $\langle a + b \rangle = \langle a \rangle + \langle b \rangle$. Cauchy-Schwarz inequality then yields
\begin{align}
    |\langle [Q, R] \rangle| \leq 2 \sqrt{\langle Q^2 \rangle  \langle R^2 \rangle} =2.
\end{align}
Moreover,
\begin{align}
    \langle \hat{c}^2 \rangle -  \langle \hat{c} \rangle^2 \geq 0 \implies \langle \hat{c} \rangle \leq \sqrt{\langle \hat{c}^2 \rangle},
\end{align}
and since $\langle 4 \id_2 \rangle=\langle \psi |4\id_2 |\psi \rangle = 4 \langle \psi | \psi \rangle = 4 |\psi|^2 = 4$ one finally has:
\begin{align}
    \langle \hat{c} \rangle \leq \sqrt{4+2\cdot 2} = 2\sqrt{2}.
\end{align}
This result is known as the Tsirelson bound that violates the CHSH inequality. It is worth mentioning that within the density operator formalism, any entangled pure state violates Bell's inequalities, with the exception of mixed entangled states admitting a local model. In the case of entangled pure states, the violation of Bell's inequalities is solely related to the presence of nonlocality.

\subsubsection{Composite systems with entangled states} 

As already mentioned, an entangled state is a special kind of quantum superposition state, so it is reasonable to write the global entangled state vector of a composite system using the superposition principle. Let $\mathcal{H}_{d}=\otimes_{s=1}^N \mathcal{H}_s^{d_s}=\mathcal{H}_1^{d_1} \otimes \mathcal{H}_2^{d_2} \otimes \cdots \otimes \mathcal{H}_N^{d_N}$ denote the composite Hilbert space of a multipartite quantum system with $N$ qudits, where qudit $s$ written in an orthonormal basis set $\{|j_s\rangle \}_{j_s=1}^{d_s}$. From the superposition principle, the $N$-qudit \textit{global entangled state vector} of the multipartite system can be written in the form:
\begin{align}
\label{entangledstate}
|\psi_{1\cdots N}^{entg}\rangle_{Nqd} = \sum_{j_1,\cdots,j_N=1}^{d_1\cdots d_N} c_{j_1\cdots j_N} |j_1\rangle_{1} \otimes \cdots \otimes |j_N\rangle_{N}.
\end{align}
In general, a quantum system represented by a state vector $|\psi\rangle$ is entangled if and only if it is not a product state (separable state) of its subsystems, i.e, if it cannot be written in the form of Eq.~\eqref{separable}. Consequently, an entangled state is one to which it is not possible to assign a single coherent state vector to any one of its subsystems, i.e, there is an intrinsic classical ignorance regarding the state preparation of its subsystems. This inability to write the subsystem of a composite system as a state vector spurred up one of the motivations that led to the development of a quantity known as the density operator (see \autoref{sec:densityformalism}).

Consider the particular case of a bipartite system with some set of orthonormal basis $\{|j_1\rangle\}_{j_1=1}^{d_1}$ $\in \mathcal{H}_1$ of subsystem $s$$=$$1$, and $\{|j_2\rangle\}_{j_2=1}^{d_2}$ $\in \mathcal{H}_2$ of subsystem $s$$=$$2$. From Eq.~\eqref{entangledstate}, one can always write the bipartite global state $|\psi_{12}\rangle \in$ $\mathcal{H}_1 \otimes \mathcal{H}_2$ of the composite system in the form:
\begin{eqnarray}
\label{bipartitestate}
|\psi_{12}^{entg}\rangle_{2qd} = \sum_{j_1=1}^{d_1}\sum_{j_2=1}^{d_2} c_{j_1j_2} |j_1\rangle_{1} \otimes |j_2\rangle_{2},
\end{eqnarray}
with $c_{j_1j_2}= \langle j_1| \otimes \langle j_2|\psi_{12}\rangle$ (we momentarily drop the subscripts to avoid notation overload).
The above global state is separable if and only if the state of its subsystems can be written as state vectors of the form:
\begin{eqnarray}
|\psi_1\rangle_{1qd} = \sum_{j_1=1}^{d_1} c_{j_1} |j_1\rangle_{1},
\end{eqnarray}
and 
\begin{eqnarray}
|\psi_2\rangle_{1qd} = \sum_{j_2=1}^{d_2} c_{j_2} |j_2\rangle_{2},
\end{eqnarray} 
with $c_{j_1j_2}=c_{j_1} c_{j_2}$. Otherwise, the global state is entangled (not separable) and one would not be able to write it as a tensor product of the state vectors of its subsystems such as in Eq.~\eqref{separable}. Special bipartite entangled states of $2$-level subsystems ($d_1=d_2=2$) are the following four maximally entangled two-qubit Bell states \cite{Bell}\cite{Braunstein}, commonly known as EPR states. In the canonical basis and within the state vector formalism, they write:
\begin{align}
|\psi_{12}^{Bell \pm}\rangle &\doteq \frac{1}{\sqrt{2}} \bigg( |0\rangle_{1} \otimes |1\rangle_{2} \pm |1\rangle_{1} \otimes |0\rangle_{2} \bigg) \\
&\doteq \frac{1}{\sqrt{2}}\bigg(|01\rangle_{12} \pm |10\rangle_{12}  \bigg),
\end{align}
and
\begin{align}
|\phi_{12}^{Bell\pm}\rangle &\doteq \frac{1}{\sqrt{2}} \bigg( |0\rangle_{1} \otimes |0\rangle_{2} \pm |1\rangle_{1} \otimes |1\rangle_{2} \bigg) \\
&\doteq \frac{1}{\sqrt{2}}\bigg(|00\rangle_{12} \pm |11\rangle_{12}  \bigg).
\end{align}
Here, $|\psi_{12}^{Bell -}\rangle$ is termed a singlet state, and $|\psi_{12}^{Bell +}\rangle$ a triplet state. Explicitly, for the singlet state:
\begin{align}
\label{singlet}
    |\psi_{12}^{Bell -}\rangle &\doteq  \frac{1}{\sqrt{2}} \Bigg\{ \begin{pmatrix} 1 \\ 0 \end{pmatrix}_1 \otimes \begin{pmatrix} 0 \\ 1 \end{pmatrix}_2 - \begin{pmatrix} 0 \\ 1 \end{pmatrix}_1 \otimes \begin{pmatrix} 1 \\ 0 \end{pmatrix}_2 \Bigg\} \\
    &=\frac{1}{\sqrt{8}} \Bigg\{ \begin{pmatrix} 1 \\ i \end{pmatrix}_1 \otimes \begin{pmatrix} i \\ 1 \end{pmatrix}_2 - \begin{pmatrix} i \\ 1 \end{pmatrix}_1 \otimes \begin{pmatrix} 1 \\ i \end{pmatrix}_2 \Bigg\}. 
\end{align}

Any attempt of writing an entangled state as a tensor product of well defined coherent state vectors yields an absurd. To demonstrate that we equate the Bell state $|\phi_{12}^{Bell +}\rangle$ to a tensor product of state vectors $|\psi_1\rangle\doteq\alpha_1|0\rangle+\beta_1 |1\rangle$ and $|\psi_2\rangle\doteq\alpha_2|0\rangle+\beta_2 |1\rangle$ from subsystems $s=1$ and $s=2$, respectively, as follows:
\begin{align}
    |\psi_{12}^{sep}\rangle &= |\psi_1\rangle \otimes |\psi_2\rangle \\
    &=\alpha_1\alpha_2 |00\rangle+ \alpha_1\beta_2 |01\rangle \\
    &+ \beta_1\alpha_2 |10\rangle+ \beta_1\beta_2 |11\rangle \\
    &= \frac{1}{\sqrt{2}}\left( |00\rangle + |11\rangle  \right) \\
    &\implies \alpha_1\alpha_2=\beta_1\beta_2=\frac{1}{\sqrt{2}} \\
    &\implies \alpha_1\beta_2 \wedge \beta_1\alpha_2 \neq 0,
\end{align}
which is an absurd since the last coefficients must vanish. Therefore, the global state of the composite system is said to be entangled. It is worth mentioning that, for Bell entangled states, while the state preparation of their subsystems is unknown, the global state of the entangled system is completely known. Global states that are completely known are commonly referred to as pure states (whether entangled or not). In section \autoref{sec:densityformalism}, within the density operator formalism, we demonstrate the criteria used to decide whether a generic qudit state is pure or mixed. In \autoref{sec:puritysec}, we explore the characteristic of maximally entangled pure states whose subsystems are described by maximally mixed reduced density operators.

Moreover, it is possible to go from one Bell state to another via simple local operations, which do not change the degree of entanglement. For instance: $\sigma_x |\phi_{12}^{Bell +}\rangle=|\psi_{12}^{Bell +}\rangle$, where $\sigma_x$ is the Pauli-x matrix defined in Eq.~\eqref{paulix}. However, there are multipartite entangled states that, in general, cannot be transformed into one another by stochastic local operations and classical communications (SLOCC) \cite{Bennett2000}, with the exception of tripartite states ($N$$=$$3$ qudits) in the asymptotic regime \cite{Yu2014} such as the 3-qubit Greenberger-Horne-Zeilinger (GHZ)-like \cite{Greenberger} and the W-like \cite{Dur2000} multipartite maximally entangled states:
\begin{align}
|GHZ\rangle_{3qb} &\doteq \frac{1}{\sqrt{2}} \left ( |000\rangle +  |111\rangle \right ), \\
|W\rangle_{3qb} &\doteq \frac{1}{\sqrt{3}} \left ( |001\rangle +  |010\rangle + |100\rangle \right).
\end{align} 
As a word of caution, entanglement cannot be created via LOCC acting on its subsystems, it can either be preserved or destroyed. 

\section{The Density operator formalism} 
\label{sec:densityformalism}

One can rephrase everything that has been said so far in the formalism of the density operator (a.k.a density matrix) developed by John Neumann \cite{Neumann1927} and, independently, by Lev Landau \cite{Landau}, in 1927. The density operator formalism has two main motivations: first, to describe a quantum system whose state is not completely known, i.e, when there is \textit{classical ignorance} regarding the state preparation; and second, to describe entangled systems since their underlying subsystems cannot be assigned to a state vector, i.e, there is classical ignorance about the preparation of the subsystems. One approach considers an ensemble $\{p_j, |\psi_j\rangle\}_{j=1}^n$ corresponding to a statistical mixture of $n$ different state vectors denoted $|\psi_j\rangle$ with probability distribution $\{p_j\}$, meaning that the system can be found in one of the aforementioned state vectors with probability $p_j$. In either case, the global state of the system is fully described by a \textit{mixed density operator} defined as:
\begin{eqnarray}
\label{mixed}
\rho \doteq \sum_{j=1}^n  p_j |\psi_j \rangle \langle \psi_j|.
\end{eqnarray}
On the other hand, whenever the global state of the system is completely known, it is a unique state vector $|\psi\rangle$ with the following corresponding \textit{pure density operator}:
\begin{eqnarray}
\rho \doteq |\psi \rangle \langle \psi|.
\end{eqnarray}
The logical inference is that a pure density operator always corresponds to a state vector. It is worth mentioning that a system whose global state is represented by a pure or a mixed density operator may or may not be entangled, since the nature of entanglement is related to the degree of purity (classical ignorance) of the underlying subsystems, rather than the degree of purity of the global state. For instance, whenever the subsystems of a multipartite system are each described by a pure density operator, it is always possible to assign to each one of them a single state vector, hence the global state of the system is separable and not entangled. The alluded Bell states are examples of global pure states that are maximally entangled for its subsystems are each in a maximally mixed state. There are also global pure states that are partially entangled, and global mixed states with no entanglement at all (see \autoref{sec:entangmeas}). In summary: the degree of purity of the global state is not a metric of entanglement, however, the degree of purity of its subsystems can be regarded as such a metric (see \autoref{sec:puritysec}).

In general, any density operator $\rho$ describing a physical observable satisfies the following conditions:
\begin{enumerate}
\item Unit trace: $tr(\rho)=1$. \label{unittrace}
\item Positiveness: $\langle \phi | \rho| \phi \rangle \geq 0$. \label{Positivity}
\item Hermiticity: $\rho=\rho^{\dagger}$.
\end{enumerate}
Where $|\phi\rangle$ is any arbitrary state vector, and the unit trace condition (a.k.a normalization condition) is the requirement of the law of total probability. It is straightforward to verify conditions no. \eqref{unittrace} and no. \eqref{Positivity} for the generic \textit{mixed density operator} defined in Eq.~\eqref{mixed}. Let $|\psi_j\rangle$ be any orthonormal basis of a $d$-dimensional Hilbert space. Positiveness of the density operator follows directly from 
\begin{align}
\langle \phi | \rho| \phi \rangle &=  \langle \phi| \left( \sum_{j=1}^d  p_j |\psi_j \rangle \langle \psi_j| \right)| \phi \rangle \\
&= \sum_{j=1}^d p_j\langle\phi|\psi_j\rangle\langle\psi_j|\phi\rangle\\
&= \sum_{j=1}^d p_j |\langle \phi | \psi_j \rangle|^2 \geq 0,
\end{align}
given that the probabilities $p_j$ are non-negative real numbers. Then, making use of trace identities no. \eqref{tracewithscalar} and no. \eqref{traceofsum} for the trace of a scalar and trace of sums, respectively, one gets
\begin{align}
tr(\rho) &= tr\left( \sum_{j=1}^d  p_j |\psi_j \rangle \langle \psi_j| \right) \\
&=\sum_{j=1}^d  tr \left(p_j |\psi_j \rangle \langle \psi_j| \right) \\
&= \sum_{j=1}^d  p_j tr \left(\langle \psi_j|\psi_j \rangle \right)\\
&= \sum_{j=1}^d  p_j tr(\delta_{jj})\\
&= \sum_{j=1}^d  p_j = 1,
\end{align}
given the law of total probability and the orthogonality condition of the basis set: $\langle \psi_j|\psi_j\rangle=\delta_{jj}=1$. The same result could have been obtained from the definition of the trace function stated in Eq.~\eqref{tracedef}. 

A bound can then be derived for $tr(\rho^2)$ as follows:
\begin{align}
tr(\rho^2) &= \sum_{j=1}^d \langle \psi_j|\rho^2| \psi_j \rangle,
\end{align}
which after introducing the completeness relation stated in Eq.~\eqref{completeness} becomes
\begin{align}
tr(\rho^2)&= \sum_{j=1}^d \langle \psi_j|\rho \left(\sum_{k=1}^d  |\psi_k \rangle \langle \psi_k | \right)\rho|\psi_j \rangle \\
&=\sum_{j,k=1}^d \langle \psi_j|\rho |\psi_k \rangle  \langle \psi_k|\rho |\psi_j \rangle,
\end{align}
then, on replacing $\rho$ with Eq.~\eqref{mixed} for a mixed density operator, one gets
\begin{align}
tr(\rho^2)&= \sum_{j,k=1}^d \bigg[\langle \psi_j| \left( \sum_{l=1}^d  p_l |\phi_l \rangle \langle \phi_l| \right) |\psi_k \rangle \bigg] \\
&\cdot \bigg[ \langle \psi_k| \left( \sum_{m=1}^d  p_m |\phi_m \rangle \langle \phi_m| \right) |\psi_j \rangle \bigg] \\
&= \sum_{l,m,j,k=1}^d \bigg\{ p_l p_m  \langle \psi_j|\phi_l \rangle \langle \phi_l|\psi_k \rangle \\
&\cdot \langle \psi_k|\phi_m \rangle \langle \phi_m|\psi_j \rangle \bigg\},
\end{align}
then, shifting the summation in $k$ for the completeness relation $\sum_{k=1}^d|\psi_k \rangle \langle \psi_k|=\id_d$ to reappear, and noting that $\{|\phi_l\rangle, |\phi_m\rangle\}$ form a orthonormal basis set ($\langle \phi_l|\phi_m\rangle=\delta_{lm}$), leads to 
\begin{align}
tr(\rho^2)&= \sum_{l,m,j=1}^d \bigg\{p_l p_m \langle \psi_j|\phi_l \rangle\\
&\cdot \langle \phi_l|\bigg(\sum_{k=1}^d|\psi_k \rangle \langle \psi_k|\bigg)|\phi_m \rangle 
\langle \phi_m|\psi_j \rangle \bigg\}\\
&=\sum_{l,m,j=1}^d p_l p_m \langle \psi_j|\phi_l \rangle  \delta_{lm}\langle \phi_m|\psi_j \rangle\\
&= \sum_{l,j=1}^d p_l^2 \langle \psi_j|\phi_l \rangle \langle \phi_l|\psi_j \rangle \\
&=\sum_{l,j=1}^d p_l^2  \langle \phi_l|\psi_j \rangle \langle \psi_j|\phi_l \rangle\\
&= \sum_{l=1}^d p_l^2 \langle \phi_l|\left( \sum_j |\psi_j \rangle \langle \psi_j|\right)|\phi_l \rangle \\
&= \sum_{l=1}^d p_l^2 \langle \phi_l|\phi_l\rangle\\
&= \sum_{l=1}^d p_l^2 \leq \left(\sum_{l=1}^d  p_l\right)^2 =1.
\end{align}
From that, one can devise the following rule-of-thumb for deciding whether a density operator is mixed or pure:
\begin{itemize}
    \item \textit{Mixed density operator}:
    \begin{align}
        tr(\rho^{\dagger}\rho)=tr(\rho^2) < 1,
    \end{align}
    and $\rho^2\neq\rho$.
    \item \textit{Pure density operator}:
    \begin{align}
        tr(\rho^{\dagger}\rho)=tr(\rho^2)=1,
    \end{align}    
    and $\rho^2=\rho$ (idempotent).
\end{itemize}

Furthermore, positiveness condition implies a spectral decomposition (see Theorem \eqref{t1}), so that in the canonical basis the density operator reads:
\begin{align}
    \rho = \sum_{j=1}^d \lambda_j |j\rangle \langle j|,
\end{align}
and
\begin{align}
    \rho^2 = \sum_{j=1}^d \lambda_j^2 |j\rangle \langle j|,
\end{align}
where the $\lambda_j$'s are the corresponding eigenvalues of the density operator. Then, unit trace condition implies
\begin{align}
    tr(\rho) &= \sum_{j=1}^d \lambda_j =1, \\
    tr(\rho^2) &= \sum_{j=1}^d \lambda_j^2.
\end{align}
Thereafter, if only one eigenvalue is non-zero, whose value is equal to one, the matrix represents a pure density operator. Conversely, a diagonalized and normalized (unit trace) pure density operator has only one non-zero eigenvalue, which is always equal to one. To see the converse, let $\{|\xi_j\rangle\}$ denote a set that diagonalize $\rho$, then:
\begin{align}
    \langle \xi_j |\rho^2|\xi_k\rangle=\langle \xi_j |\rho|\xi_k\rangle= \lambda_j\delta_{jk},
\end{align}
and
\begin{align}
    \langle \xi_j |\rho^2|\xi_k\rangle &=\sum_l\langle \xi_j |\rho|\xi_l\rangle \langle \xi_l |\rho|\xi_k\rangle,\\
    &= \lambda_j\delta_{jl}\lambda_l\delta_{lk}=\lambda_j^2\delta_{jk},
\end{align}
so that
\begin{align}
    \lambda_j^2\delta_{jk}=\lambda_j\delta_{jk} \implies \lambda_j(\lambda_j-1)\delta_{jk}=0.
\end{align}
It is straightforward seeing that whenever $j=k$ ($\delta_{jj}=\delta_{kk}=1$) the eigenvalue $\lambda_j$ is either $0$ or $1$, and since the matrix is now diagonal it is sufficient to prove the converse statement.

\subsection{Postulates within the density operator formalism}
\label{sec:densepost}
Within the framework of the density operator it is possible to reformulate the postulates of quantum mechanics stated in \autoref{sec:postulates}, as follows.
\begin{itemize}
    \item Postulate 1: Any isolated physical system (close system) is described within a complex vector space $\mathcal{H}$ (Hilbert space) endowed with an inner product $(\cdot,\cdot) \doteq \langle \cdot | \cdot \rangle$, and known as the \textit{state space} of the system. The system is completely described by a Positive and unit trace operator $\rho = \sum_j  p_j \rho_j$, known as the density operator, which is a statistical mixture of states $\rho_j=\sum_k p_{jk} |\psi_{jk} \rangle \langle \psi_{jk}|$ with probability $p_j$ from some ensemble $\{p_{jk},\psi_{jk}\}$. 
    
    \item Postulate 2: In the \textit{Schrödinger picture formalism}, a \textit{closed quantum system} prepared in the initial state $|\psi_{j,t}\rangle$ evolves according to a unitary transformation $\hat{U}_t|\psi_{j,t}\rangle$, so that the initial density operator $\rho=\sum_j p_j |\psi_{j,t} \rangle \langle \psi_{j,t}|$ evolves to
    \begin{align}
    \rho_t &= \sum_j  p_j (\hat{U}_t|\psi_{j,t} \rangle) (\langle \psi_{j,t}|\hat{U}_t^{\dagger}) \\
    &= \hat{U}_t (\sum_j  p_j|\psi_{j,t} \rangle\langle \psi_{j,t}|)\hat{U}_t^{\dagger} \\
    &=\hat{U}_t\rho \hat{U}_t^{\dagger},
    \end{align}
    given that $(\hat{U}_t|\psi_{j,t}\rangle)^{\dagger}=\langle \psi_{j,t}|\hat{U}_t^{\dagger}$. This equation is valid whenever the Hamiltonian $\hat{H}$ of the system is time-independent, and the unitary operator is $\hat{U}=e^{\frac{-i\hat{H}t}{\hbar}}$.
    
    When the Hamiltonian is time-dependent, the continuous time evolution of this system is given by the Von Neumann equation (a.k.a Liouville–von Neumann equation), as follows:
    \begin{align}
    \frac{\partial\rho}{\partial t}&=\frac{\partial}{\partial t}\left(\sum_j p_j |\psi_{j,t}\rangle \langle \psi_{j,t} |\right) \\
    &=\sum_j p_j \Bigg\{ \left( \frac{\partial}{\partial t}|\psi_{j,t}\rangle \right)\langle\psi_{j,t}|\\
    &+ |\psi_{j,t}\rangle \left(\frac{\partial}{\partial t} \langle\psi_{j,t}|\right)  \Bigg\} \\
    &=\sum_j p_j \left\{ \frac{\hat{H}}{i\hbar}  |\psi_{j,t}\rangle \langle \psi_{j,t} |  -  |\psi_{j,t}\rangle \langle \psi_{j,t} | \frac{\hat{H}}{i\hbar}  \right\} \\
    &= \frac{1}{i\hbar}\left[ \hat{H}, \rho \right],
    \end{align}
    which is true given the product rule for derivatives and the fact that
    \begin{align}
    \frac{\partial}{\partial t}|\psi_{j,t}\rangle=\frac{\hat{H}}{i\hbar}|\psi_{j,t}\rangle,
    \end{align}
    and 
   \begin{align}
   \frac{\partial}{\partial t}\langle\psi_{j,t}|=-\frac{\hat{H}}{i\hbar}\langle\psi_{j,t}|.
   \end{align}
    
    \item Postulate 3:  Measurements of a quantum system are described by a collection \{$M_m$\} of \textit{measurement operators} acting on the state space of the system with $m$ possible measurement outcomes. The measurement operators satisfy the completeness relation 
    \begin{equation} 
    \sum_{m} M_{m}^{\dagger} M_{m} = I.
    \end{equation}
    By the law of total probability, the probability of obtaining an outcome $o_m$ associated to an observable $\hat{\mathcal{O}}$ of a quantum system prepared in the state $|\psi_j \rangle$ with corresponding density operator $\rho$ is
    \begin{align}
    Pr(o_m|\rho)&=\sum_j p_j \langle \psi_j | M^{\dagger}_{m} M_{m} |\psi_j \rangle,
    \end{align} 
    which after applying trace identities no. \eqref{scalartrace} and no. \eqref{traceofsum} for the trace of a scalar and trace of sums, respectively, becomes
    \begin{align}
    Pr(o_m|\rho)&= \sum_j p_j tr(M^{\dagger}_{m} M_{m} |\psi_j \rangle \langle \psi_j |)\\
    &= tr\left(M^{\dagger}_{m} M_{m}\sum_j p_j|\psi_j \rangle \langle \psi_j |\right) \\
    &=tr(M_m^{\dagger}M_m\rho). \label{probtrace}
    \end{align}    
    And the state of the system after the measurement shall collapse to
    \begin{align} 
    |\psi_j^{o_m} \rangle &= \frac{M_{m} |\psi \rangle}{\sqrt{\langle \psi_j | M^{\dagger}_{m} M_{m} |\psi_j \rangle}},
    \end{align}
    while the corresponding new density operator becomes:
    \begin{align} 
    \rho_{o_m} &= \sum_j \frac{p_j \langle \psi_j | M^{\dagger}_{m} M_{m} |\psi_j \rangle}{P_r(o_m)} |\psi_j^{o_m} \rangle \langle \psi_j^{o_m} |   \\
    &= \sum_j \frac{p_j \langle \psi_j | M^{\dagger}_{m} M_{m} |\psi_j \rangle}{P_r(o_m)} \frac{M_m |\psi_j \rangle \langle \psi_j | M_m^{\dagger}}{\langle \psi_j | M^{\dagger}_{m} M_{m} |\psi_j \rangle}\\
    &= \sum_j \frac{p_j}{P_r(o_m)}M_m |\psi_j \rangle \langle \psi_j | M_m^{\dagger}\\
    &=\sum_j p_j \frac{M_m |\psi_j \rangle \langle \psi_j | M_m^{\dagger}}{tr(M_m^{\dagger}M_m\rho)} \\
    &=  \frac{M_{m} \left(\sum_j p_j |\psi_j \rangle \langle \psi_j |\right) M_m^{\dagger}}{tr(M_m^{\dagger}M_m\rho)}\\
    &=\frac{M_{m} \rho M_m^{\dagger}}{tr(M_m^{\dagger}M_m\rho)}.
    \end{align}
Now, consider an observable $\hat{A}$ with an orthonormal basis set $\{a_j\}$, and an observable $\hat{B}$ with an orthonormal basis set $\{b_k\}$. The projective measurement (see \autoref{sec:projmeasu}) for obtaining an outcome $b_k$ of the observable $\hat{B}$ conditioned to a measurement with outcome $a_j$ of the observable $\hat{A}$ in an statistical sub-ensemble given by the \textit{mixed density operator} $\rho$ with probabilities $p_j$ shall be:
\begin{align}
Pr(b_k|\{p_j,|a_j\rangle\})&=\sum_j Pr(b_k|a_j\rangle)p_j \\
&=\sum_j|\langle b_k|a_j\rangle|^2 p_j \\
&=\sum_j\langle b_k|a_j\rangle \langle a_j|b_k\rangle p_j\\
&= \langle b_k \left\{ \sum_j p_j |a_j\rangle\langle a_j| \right\}|b_k\rangle \\
&=\langle b_k |\rho | b_k \rangle,
\end{align}
and with trace identity no. \eqref{scalartrace} of the trace of a scalar, we then have
\begin{align}
Pr(b_k|\{p_j,|a_j\rangle\})&=tr(\langle b_k |\rho | b_k \rangle) \\
&= tr(\rho|b_k\rangle\langle b_k|).
\end{align}
Finally, the expectation value of some arbitrary observable $\hat{\mathcal{O}}$ will be given as
\begin{align}
\label{expvalue}
\langle  \hat{\mathcal{O}}\rangle_{\rho} &= \sum_j p_j \langle \psi_j |\hat{\mathcal{O}}| \psi_j \rangle \\
&= \sum_j p_j tr\left(\langle \psi_j |\hat{\mathcal{O}}| \psi_j \rangle\right)\\
&= \sum_j p_j tr\left(\hat{\mathcal{O}}| \psi_j \rangle\langle \psi_j |\right)\\
&=  tr\left(\hat{\mathcal{O}} \sum_j p_j| \psi_j \rangle \langle \psi_j |\right)\\
&= tr(\rho \hat{\mathcal{O}}).
\end{align}    
    \item Postulate 4: For $N$ arbitrary quantum physical systems, the composite state space of the system is given by the Kronecker product of the state space of its subsystems: $\mathcal{H}_{12\cdots N}=\mathcal{H}_1 \otimes \mathcal{H}_2 \otimes\cdots\otimes \mathcal{H}_N$. And the global state of the composite system will be $\rho_{\otimes^N}=\rho_1 \otimes \rho_2 \otimes \cdot\cdot\cdot \otimes \rho_N$. 
\end{itemize}

\section{The partial trace} 
\label{sec:partialtrace}

The partial trace function \cite{Nielsen2010}\cite{Wilde2013} is one of the many functions used to extract information about a composite quantum system. For instance, in the context of local operations it is expedient to associate to each subsystem a reduced density operator (a.k.a reduced density matrix) via the partial trace function. In \autoref{sec:puritysec}, we demonstrate the use of the partial trace to infer the degree of purity of subsystems from a bipartite system. Here, we derive the analytical expression of the reduced density operator given the definition of the partial trace for bipartite systems. The computation of the partial trace for general multipartite systems requires a numerical approach, and is beyond the scope of this review.

Let $\mathcal{H}_{ab} \doteq \mathcal{H}_a\otimes\mathcal{H}_b$ denote the composite Hilbert space of a bipartite system, and $\{b_j\}_{j=1}^{d_b}$ denote any $d_b$-dimensional orthonormal basis set for $\mathcal{H}_b$. The \textit{partial trace} over subsystem $b$ is a map $tr_b: \mathcal{L} (\mathcal{H}_{ab}) \rightarrow  \mathcal{L}( \mathcal{H}_a)$ defined as \cite{Watrous2018}:
\begin{align}
    tr_b(\mathcal{O})\doteq\sum_{j=1}^{d_b}(\id_a \otimes \langle b_j|)\mathcal{O} (\id_a \otimes | b_j|\rangle). \label{partialtrac}
\end{align}
Conversely, the partial trace over subsytem $a$ is a map $tr_a: \mathcal{L} (\mathcal{H}_{ab}) \rightarrow  \mathcal{L}( \mathcal{H}_b)$. Furthermore, the partial trace is a function with the property:
\begin{align}\label{partialfunc}
tr_a(\hat{A}f(\hat{O}))&=tr_{ab}((\hat{A}\otimes \id_b) \hat{O}),
\end{align}
for generic linear operators $\hat{A} \in \mathcal{L}( \mathcal{H}_a)$ and $\hat{O}\in \mathcal{L}(\mathcal{H}_{ab})$.

One ubiquitous application of the partial trace is to compute the \textit{reduced density matrix} (a.k.a partial state or quantum marginal) of a subsystem. Let $\rho_{ab}\doteq \rho_a \otimes \rho_b \in \mathcal{H}_{ab}$ denote a density operator of a bipartite system. The partial trace of $\rho_{ab}$ over subsystem $b$ yields a \textit{reduced density operator} $\rho_a \in \mathcal{L}( \mathcal{H}_a)$ of subsystem $a$. This assertion can be verified according to \eqref{partialtrac}, as follows:
\begin{align}
   tr_b(\rho_{ab})&=\sum_{j=1}^{d_b} (\id_a \otimes \langle b_j|)(\rho_a \otimes \rho_b)(\id_a \otimes | b_j|\rangle) \\
    &=\sum_{j=1}^{d_b} (\id_a \rho_a \otimes \langle b_j|\rho_b) (\id_a \otimes | b_j|\rangle)\\
    &=\sum_{j=1}^{d_b} \id_a \rho_a \id_a \otimes \langle b_j| \rho_b| b_j \rangle \\
    &= \sum_{j=1}^{d_b} \rho_a \otimes \langle b_j| \rho_b| b_j \rangle,
\end{align}   
and making use of trace identities no. \eqref{tracewithscalar} and no. \eqref{traceproduct}, one gets
\begin{align}
    tr_b(\rho_{ab})&= \sum_{j=1}^{d_b} \rho_a \otimes tr( \langle b_j| \rho_b| b_j \rangle) \\
    &= \sum_{j=1}^{d_b} \rho_a \otimes tr(| b_j \rangle  \langle b_j| \rho_b)\\
    &=  \rho_a \otimes tr\left(\left(\sum_{j=1}^{d_b}| b_j \rangle  \langle b_j|\right) \rho_b\right)\\
    &= \rho_a \otimes tr(\rho_b) = \rho_a \otimes 1 =  \rho_a,
\end{align}
given the completeness relation, and the unit trace condition $tr(\rho_b)=1$. Another equivalent definition of the partial trace for bipartite systems is \cite{Nielsen2010}:
\begin{align}
    tr_b(|a_j\rangle \langle a_k| \otimes |b_l\rangle \langle b_m| )&\doteq|a_j \rangle \langle a_k| \otimes tr_b (|b_l\rangle \langle b_m|) \label{parttrace}\\
    &=|a_j \rangle \langle a_k| \cdot tr (|b_l\rangle \langle b_m|)\\
    &=|a_j \rangle \langle a_k| \cdot tr (\langle b_m|b_l\rangle )\\
    &=|a_j \rangle \langle a_k| \cdot \langle b_m|b_l\rangle, 
\end{align}
for generic state vectors $|a_j\rangle, |a_k\rangle \in \mathcal{H}_a$ and $|b_l\rangle, |b_m\rangle \in \mathcal{H}_b$. 

From the definition of the mixed density operator stated in Eq.~\eqref{mixed}, the bipartite mixed density operator takes the form 
\begin{align}    
\rho_{ab}&=\sum_{i=1}^d p_i |\psi_{i}^{ab}\rangle \langle \psi_i^{ab}|.
\end{align}
The more general expression is obtained by writing the composite state $|\psi_{ab}\rangle$ in the form of Eq.~\eqref{bipartitestate}, such that
\begin{align}    
&\rho_{ab}=\sum_i p_i \Bigg\{ \left(\sum_{j,l} c_{ijl} |a_j\rangle \otimes |b_l\rangle \right) \\
&\cdot \left(\sum_{k,m}  c_{ikm} \langle a_k| \otimes \langle b_m| \right) \Bigg\}\\
&=\sum_{i,j,k,l,m}p_ic_{ijklm}|a_j\rangle\langle a_k|\otimes|b_l\rangle\langle b_m|.
\label{rhobipartite}
\end{align}
Within this new framework, the reduced density operator $\rho_a$ becomes:
\begin{align}
    \rho_a &\doteq tr_b(\rho_{ab})=tr_b\left(\sum_i p_i |\psi_{i}^{ab}\rangle \langle \psi_i^{ab}|\right)\\
    &=tr_b\left(\sum_{i,j,k,l,m}p_ic_{ijklm}|a_j\rangle\langle a_k|\otimes|b_l\rangle\langle b_m| \right)\\
    &=\sum_{i,j,k,l,m}p_ic_{ijklm}tr_b\Bigg(|a_j\rangle\langle a_k|\otimes|b_l\rangle\langle b_m| \Bigg)\\
    &=\sum_{i,j,k,l,m}p_ic_{ijklm}(|a_j\rangle\langle a_k|) \otimes tr_b(|b_l\rangle \langle b_m|),
    \end{align}
which after applying the base independence of the trace function 
\begin{align}
tr_b(|b_l\rangle\langle b_m|)&=tr_b(\langle b_m|b_l\rangle)\\
&=\langle b_m|b_l\rangle\\
&=\langle b_m|\left(\sum_{k=1}^{d_b} |b_k\rangle\langle b_k|\right)|b_l\rangle\\
&=\sum_{k=1}^{d_b} \langle b_m|b_k\rangle\langle b_k|b_l\rangle\\
&=\sum_{k=1}^{d_b}\langle b_k|b_l\rangle\langle b_m|b_k\rangle = \delta_{kl}\delta_{mk}    
\end{align}
one then gets
\begin{align}    
    \rho_a &= \sum_{i,j,k,l,m}p_ic_{ijklm}|a_j\rangle\langle a_k| \delta_{kl}\delta_{mk}\\
           &=\sum_{i,j,k,l,m}p_ic_{ijklm}|a_j\rangle\langle a_k|\delta_{ml}.
\end{align}

\subsection{Measurement in a bipartite system via the partial trace function} \label{sec:bimeas}

A measurement operator in a composite Hilbert space $\mathcal{H}_{ab} \doteq \mathcal{H}_a\otimes\mathcal{H}_b$ of a bipartite system is a local projection operator defined as $M_{ab}\doteq M\otimes\id_d$. For instance, let $M \doteq M_a \doteq |a_j\rangle\langle a_j| \in \mathcal{L}(\mathcal{H}_a)$ denote a projector operator on subsystem $a$, such that $M_{ab}=M_a \otimes \id_d \in \mathcal{L}(\mathcal{H}_{ab})$. Recall that any projector operator satisfies $M_{ab}^{\dagger}M_{ab}=M_{ab}^2=M_{ab}$. From postulate 3 (see \autoref{sec:densepost}), the corresponding probabilities associated to the eigenvalue $a_j$ of an observable $\hat{O}_a \in \mathcal{L}(\mathcal{H}_a)$ on subsystem $a$ are given according to Eq.~\eqref{probtrace}, as follows:
\begin{align}
    Pr(a_j|\rho_{ab})
    &=tr_{ab}((M_{ab}^{\dagger}M_{ab})\rho_{ab})\\
    &=tr_{ab}( M_{ab}\rho_{ab})\\
    &=tr_{ab}\left((M_a\otimes \id_d) p_{ab}\right)\label{probab}.
\end{align} 
From Eq.~\eqref{partialfunc} one has $tr_{ab}((M_a \otimes \id_d)\rho_{ab}) = tr_a(M_a(tr_b(\rho_{ab}))$, such that Eq.~\eqref{probab} becomes 
\begin{align}
    Pr(a_j|\rho_{ab}) &= tr_a(M_a\rho_{a})\\
    &= tr_a(\rho_{a}M_a)\\
    &= tr_a(\rho_{a}(|a_j\rangle \langle a_j|)) \\\label{dropped} 
    &=tr(\rho_{a}(|a_j\rangle \langle a_j|)),
\end{align} 
where we have dropped the sub-index of the trace function in Eq.~\eqref{dropped} since the argument only depends on terms of subsystem $a$, such that the partial trace becomes the ordinary trace defined in \autoref{sec:trace}. 

By its turn, the expectation value according to Eq.~\eqref{expvalue} becomes:
\begin{align}
     \langle \hat{O}_a \rangle_{\rho_{ab}} &= \sum_i p_i\langle \psi_{ab}|(\hat{O}_a \otimes \id_b) |\psi_{ab}\rangle \\
    &= \sum_i p_i tr_{ab}\left( \langle \psi_{ab}|(\hat{O}_a \otimes \id_b) |\psi_{ab}\rangle\right) \\
    &=tr_{ab}\left(\sum_i p_i (\hat{O}_a\otimes \id_b) |\psi_{ab}\rangle \langle \psi_{ab}|\right)\\
    &=tr_{ab}\left((\hat{O}_a\otimes \id_b) \sum_i p_i |\psi_{ab}\rangle \langle \psi_{ab}|\right)\\
    &=tr_{ab}\left((\hat{O}_a\otimes \id_b) p_{ab}\right),
    \end{align}     
and replacing $\rho_{ab}$ with Eq.~\eqref{rhobipartite} yields
\begin{align}
     &=tr_{ab}\Bigg\{(\hat{O}_a\otimes \id_b) \sum_{i,j,k,l,m}p_ic_{ijklm}|a_j\rangle\langle a_k|\otimes|b_l\rangle\langle b_m| \Bigg\}\\
    &=tr_{ab}\Bigg\{\sum_{i,j,k,l,m} p_i c_{ijklm}  \hat{O}_a|a_j\rangle\langle a_k| \otimes \id_b |b_l\rangle\langle b_m|  \Bigg\}\\
    &=\sum_{i,j,k,l,m} tr_{ab}\Bigg\{\hat{O}_a p_i c_{ijklm}|a_j\rangle\langle a_k| \otimes  |b_l\rangle\langle b_m| \Bigg\}\\
    &=\sum_{i,j,k,l,m}tr_{ab}\left(\hat{O}_ap_i c_{ijklm}|a_j\rangle\langle a_k|\right) 
    \cdot tr_{ab}\left(\langle b_m|b_l\rangle \right)\\
    &=tr_{ab}\left(\sum_{i,j,k,l,m}\hat{O}_ap_i c_{ijklm}|a_j\rangle\langle a_k|\right)\delta_{ml}\\
    &=tr_{ab}\left(\hat{O}_a\sum_{i,j,k,l,m}p_i c_{ijklm}|a_j\rangle\langle a_k|\delta_{ml}\right)\\
    &=tr_a\left(\hat{O}_a\rho_a\right).
    \label{exp2}
\end{align}  
Summing up, we proved that $\langle \hat{O}_a \rangle_{\rho_{ab}}=tr_{ab}\left((\hat{O}_a\otimes \id_b) p_{ab}\right)=tr_a\left(\hat{O}_a\rho_a\right)$, which is in accordance with Eq.~\eqref{partialfunc}.

\section{Decoherence} 
\label{sec:decoherence}

As emphasized in previous sections, coherent superposition is an important aspect of quantum mechanics and is widely applied to quantum computing tasks. Coherent superposition is also regarded as the basis for efficient realization of energy and information manipulation tasks \cite{Chenu2015}\cite{Uzdin2016}\cite{Misra2016}\cite{Shi2017}. However, noise effects modeled as interactions between the quantum system and the surrounding environment cause the quantum system to rapidly lose coherent information in a process known as decoherence. 

To help grasp understanding of the phenomenon of quantum decoherence, consider a single generic 2-level quantum system (1-qubit) in the following coherent state of quantum superposition:
\begin{align}
    |\psi\rangle = \alpha |0\rangle + \beta |1\rangle.
\end{align}
The corresponding coherent density matrix is obtained from the outer product, as follows:
\begin{align}
    \rho_{1qb} = |\psi\rangle \langle \psi| = 
\begin{bmatrix} 
\alpha \alpha^* && \alpha \beta^* \\
\beta \alpha^* && \beta \beta^*  
\end{bmatrix},
\end{align}
where the off-diagonal elements are dubbed \textit{coherences}, and the elements in the diagonal entry are the alluded classical probabilities commonly referred to as \textit{populations}. Specifically, for the 1-qubit state vector defined in Eq.~\eqref{hopf} its corresponding density matrix reads:
\begin{align}
    \rho_{1qb}&= \begin{bmatrix} 
cos^2(\frac{\theta}{2}) && cos(\frac{\theta}{2})sin(\frac{\theta}{2})e^{-i\phi} \\
cos(\frac{\theta}{2})sin(\frac{\theta}{2})e^{i\phi} && sin^2(\frac{\theta}{2})    
\label{densqubitmatrix}
\end{bmatrix}.
\end{align}
Upon decoherence, each interaction between the quantum system and the environment randomly changes the relative phase $\phi$ by some amount. In a process of maximum decoherence the average of all phase changes is such that the off-diagonal elements of the coherent density matrix go to zero, and the resulting \textit{decoherent density matrix} becomes:
\begin{align}
\label{denmatrix}
\rho_{1qb} \rightarrow \iota_{1qb} &= 
\begin{bmatrix} 
\alpha \alpha^* && 0 \\
0 && \beta \beta^*
\end{bmatrix}\\
&=
\begin{bmatrix} 
cos^2(\frac{\theta}{2}) &&  0\\
0 && sin^2(\frac{\theta}{2})    
\end{bmatrix}.
\end{align}
When the above density matrix has equal diagonal entries (uniform probability distribution), the associate incoherent 1-qubit state $\iota_{1qb}$ is commonly referred to as a ``maximally mixed state'' (see \autoref{sec:puritysec}). In this particular case, the entropy of the state is maximum (see \autoref{vonentropy}), i.e, one has the least amount of classical knowledge (predictability power) available to predict the state of the system upon measurement. Moreover, after decoherence, the density matrix no longer corresponds to a wave-function, since the wave-function describes a system in a coherent state of quantum superposition, whereas the incoherent density matrix describes the possible classical outcomes from the collapse of the wave-function upon interaction. 

Therefore, a density matrix is dubbed incoherent when it has no nonnull coherence elements in its off-diagonal entry, with its diagonal entry preserving the classical probabilities of a measurement outcome. One then represents an incoherent qudit state $\iota_{qd}$ by a density operator that is diagonal in some $d$-dimensional orthonormal basis set $\{|j\rangle \}_{j=1}^d$, as follows:
\begin{align}
    \iota_{qd} \doteq \sum_{j=1}^d \iota_j |j\rangle \langle j|,
\end{align}
where $\{\iota_j \}_{j=1}^d$ denotes the probability distribution. It is straightforward to verify the particular case of an incoherent 1-qubit state, as follows:
\begin{align}
    \iota_{qb} &= \sum_{j=1}^2 \iota_j |j\rangle \langle j| \\
    &= \iota_1 \begin{pmatrix} 1 \\ 0 \\ \end{pmatrix} \begin{pmatrix} 1 & 0 \end{pmatrix} + \iota_2 \begin{pmatrix} 0 \\ 1 \\ \end{pmatrix} \begin{pmatrix} 0 & 1 \end{pmatrix} \\
    &= \iota_1 \begin{bmatrix} 
1 && 0  \\
0 && 0  
\end{bmatrix} 
+ \iota_2 \begin{bmatrix} 
0 && 0  \\
0 && 1  
\end{bmatrix}
=\begin{bmatrix} 
\iota_1  && 0  \\
0 && \iota_2  
\end{bmatrix},
\end{align}
which is equivalent to the density matrix in Eq.~\eqref{denmatrix}.

\section{The N-qudit coherent and incoherent density operators}
\label{sec:subsecqudit}

The $\rho_{Nqd}$ density matrix of an $N$-qudit system, a multipartite system containing $N$ $d_s$-dimensional qudit subsystems, is embedded in the Special Unitary group of degree $d$ (dimension) denoted $SU(d)$. The $SU(d)$ group is also known as the Lie group or topological space (manifold) of $d$x$d$ Unitary matrices with determinant equal to 1.  

Consider, for instance, a qudit $s$ in $SU(d_s)$, and let $\{\Gamma_{j_s}\}_{j_s=1}^{d_s^2-1} \in SU(d_s)$ denote the group of generators forming an orthonormal basis set with $d_s^2-1$ matrices of dimension $d_s$x$d_s$ satisfying the associated $\mathfrak{su}(d_s)$ Lie algebra. A basis set for the composite system in $SU(d)\doteq \otimes_{s=1}^N SU(d_s)$ with dimension $d=\prod_{s=1}^N d_s$ is obtained by taking the Kronecker product between the basis of each subsystem of $SU(d_s)$. In this basis, one can write a generic density operator (mixed or pure) of an $N$-qudit system embedded in $SU(d)$ according to the following Bloch vector parametrization \cite{Kimura2003}:
\begin{align}
\label{quditstate}
    \rho_{Nqd} &= \sum_{j_1=0}^{d_1^2-1} \cdot\cdot\cdot \sum_{j_N=0}^{d_N^2-1}r_{j_1 \cdot\cdot\cdot j_N}\Gamma_{j_1}^{D,S,A}
    \otimes\cdot\cdot\cdot \otimes \Gamma_{j_N}^{D,S,A} \\
    &\doteq\sum_{j_1,\cdots,j_N=0}^{d_1^2-1,\cdots,d_N^2-1}r_{j_1 \cdot\cdot\cdot j_N}\Gamma_{j_1}\otimes\cdot\cdot\cdot \otimes \Gamma_{j_N}.
\end{align}
For a review on density matrix parametrization, the reader may resort to \cite{Bruning2012}. Here, we have defined $\Gamma_{j_s}^{D,S,A}\doteq\Gamma_{j_s}$ as a shorthand notation, and also:
\begin{align}
    \sum_{j_1=0}^{d_1^2-1} \sum_{j_2=0}^{d_2^2-1} \cdot\cdot\cdot \sum_{j_N=0}^{d_N^2-1} \doteq \sum_{j_1,\cdots,j_N=0}^{d_1^2-1\cdots d_N^2-1}.
\end{align}
The $r_{j_1\cdot\cdot\cdot j_N} = r_{j1}r_{j2}\cdots r_{jN}$ are the components of the so-called \textit{generalized Bloch's vector} (a.k.a \textit{coherence vector}), while $\Gamma_{j_N}^{D,S,A}$ represents all possible combinations of the $d_s$x$d_s$ generalized Gell-Mann matrices (GGM) from the $SU(d_s)$ group of generators satisfying the $SU(d_s)$ Lie algebra. The indices $D, S$ and $A$ denote the matrices from the \textit{diagonal}, \textit{symmetric}, and \textit{antisymmetric} sets, respectively. 

The corresponding incoherent density operator of a generic $N$-qudit state in the computational basis $\otimes_{s=1}^N |j_s\rangle_{j_s=1}^{d_s}$ takes the form
\begin{align}
\iota_{Nqd} &= \sum_{j_{1},\cdots,j_{N}=0}^{d_1-1,\cdots, d_N-1} \iota_{j_{1}j_{2}\cdots j_{N}} \Gamma_{j_1}^{D} \otimes \Gamma_{j_2}^{D}
    \otimes\cdots\otimes \Gamma_{j_N}^{D}, \label{incoherent_qudit}
\end{align}
where $\Gamma_{j_N}^{D}$ denotes the previously mentioned set of diagonal GMM from the $SU(d_s)$ group of generators. And $\iota_{j_{1}j_{2}\cdots j_{N}}$ denotes the probability distribution, with condition $\iota_{j_{1}j_{2}\cdots j_{N}}\ge0$ and $\sum_{j_{1},j_{2},\cdots, j_{N}}\iota_{j_{1}j_{2}\cdots j_{N}}=1$. 

The following are the aforementioned sets of GGM representing the generators for the $SU(d_s)$ manifold \cite{Bertlmann2008}:
\begin{align}
    \Gamma_{j_s}^{D}&= \sqrt{\frac{2}{j_s(j_s+1)}}\sum_{k_s=1}^{j_s+1}(-j_s)^{\delta_{k_s,j_s}+1} |k_s\rangle \langle k_s|, \\
    \Gamma_{(k_s,l_s)}^{S}&= |k_s\rangle \langle l_s| + |l_s\rangle \langle k_s|,\\
    \Gamma_{(k_s,l_s)}^{A} &= -i (|k_s\rangle \langle l_s| - |l_s\rangle \langle k_s|),
\end{align}
for $1 \leq k_s < l_s \leq d_s$, and $1 \leq j_s \leq d_s-1$. It is usual to define $\Gamma_0^D \doteq \id_{d_s}$ as the $d_s$x$d_s$ identity matrix. Using the trace identities (see \autoref{sec:trace}), the GMM satisfy the following $SU(d_s)$ Lie algebra:
\begin{enumerate}
    \item $tr(\Gamma_{j_s})=d_s\delta_{0j_s}$.
    \item $tr(\Gamma_{j_s}\Gamma_{k_s})=d_s^{\delta_{0j_s}}2^{1-\delta_{0j_s}}\delta_{j_sk_s}$. 
    \item $tr(\Gamma_{j_s}\otimes \Gamma_{k_s})=tr(\Gamma_{j_s})tr(\Gamma_{k_s})$.
    \item $tr(\Gamma_{j_s}\otimes \Gamma_{k_s})(\Gamma_{l_s}\otimes \Gamma_{m_s})=\Gamma_{j_s}\Gamma_{l_s}\otimes \Gamma_{k_s}\Gamma_{m_s}$.
    \item $[\Gamma_{j_s},\Gamma_{k_s}]$ $=$ $\Gamma_{j_s}\Gamma_{k_s}$$-$$\Gamma_{k_s}\Gamma_{j_s}$=$2i\sum_{l=1}^{d_s^2-1} f_{jkl}\Gamma_{l_s}$.
    \item $\{\Gamma_{j_s},\Gamma_{k_s}\}=\Gamma_{j_s}\Gamma_{k_s}+\Gamma_{k_s}\Gamma_{j_s}=\frac{4}{d_s}\delta_{jk} \id_{d_s} +2\sum_{l=1}^{d_s^2-1}g_{jkl}\Gamma_{l_s}$.
\end{enumerate}
Where $f_{jkl}$ and $g_{jkl}$ are the structure constants of the $SU(d_s)$ Lie algebra. 

\subsection{SU(2) and SU(3) Lie algebras}
\label{sec:sualgebra}

The GMM reduce to the well known Pauli-$\sigma$ and Gell Mann-$\lambda$ matrices in the particular dimensions $d_s$$=2$ (for qubits) and $d_s$$=3$ (for qutrits), respectively. As a worked example, the $2$x$2$-dimensional identity and Pauli $\sigma$-matrices are obtained as follows:
\begin{align}
\sigma_0&\doteq\Gamma_0^D \doteq \id_2 = |1\rangle \langle 1| + |2\rangle \langle 2| \\
    &=
\begin{bmatrix} 
1 && 0 \\
0 && 1  
\end{bmatrix},\\\label{paulix}
\sigma_1 &\doteq \sigma_x \doteq X \doteq \Gamma_{(1,2)}^{S} =|1\rangle \langle 2| + |2\rangle \langle 1| \\
&= 
\begin{bmatrix} 
0 && 1 \\
1 && 0  
\end{bmatrix},\\
    \sigma_2 &\doteq \sigma_y \doteq Y \doteq \Gamma_{(1,2)}^{A} =-i(|1\rangle \langle 2| - |2\rangle \langle 1|)
\\&=
\begin{bmatrix} 
0 && -i \\
i && 0  
\end{bmatrix},\\
    \sigma_3 &\doteq \sigma_z \doteq Z \doteq \Gamma_{(1)}^{D} = |1\rangle \langle 1| - |2\rangle \langle 2|
\\&=
\begin{bmatrix} 
1 && 0 \\
0 && -1  
\end{bmatrix}.
\end{align}
Where we have defined the basis of reference to be the computational basis:
\begin{align}
    |1\rangle \doteq \begin{bmatrix} 1 \\ 0 \\ \end{bmatrix} \text{ and }
    |2\rangle \doteq \begin{bmatrix} 0 \\ 1 \\ \end{bmatrix}.
\end{align}
The $SU(2)$ group of generators consists of all Unitary, Hermitian, traceless, trace-orthogonal, $2$x$2$-dimensional Pauli $\sigma$-matrices numbered 1 to 3. These matrices represent the observables of spin of the electron (and of the photon in a different linear space) for each direction. Without stressing, the Pauli matrices together with the identify matrix satisfy the following $SU(2)$ Lie algebra, for $j,k,l=1\cdots 3$ and $\mu,\nu=0\cdots3$:
\begin{enumerate}
    \item $[\sigma_j,\sigma_k]=\sigma_j\sigma_k-\sigma_k\sigma_j=2i\varepsilon_{jkl}\sigma_l$.
    \item $\{\sigma_j,\sigma_k\}=\sigma_j\sigma_k+\sigma_k\sigma_j=2\delta_{jk}\id_2$. 
    \item $\sigma_j \sigma_k =\delta_{jk}\id_2+i\varepsilon_{jkl}\sigma_l$.
    \item $tr(\sigma_j\sigma_k\sigma_l)=2i\varepsilon_{jkl}$.
    \item $tr(\sigma_\mu \sigma_\nu)=2\delta_{\mu\nu}$.
    \item $tr(\sigma_\mu)=2\delta_{0\mu}$.
    \item $\sigma_\mu^2=\id_2$.
\end{enumerate}
Where $\varepsilon$ is the Levi-Civita symbol defined as:
\begin{align*}
\varepsilon_{ijk} \doteq
\begin{cases}
+1 & \text{for even permutation of }(i,j,k). \\
-1 & \text{for odd permutation of } (i,j,k).  \\
\;\;\,0 & \text{for any repeated index}.
\end{cases}
\end{align*}

In the case of the $SU(3)$ Lie group, the identity and Gell Mann $\lambda$-matrices reads:
\begin{align} 
\lambda_0&=
    \begin{bmatrix}
    1 && \phantom{-}0 && 0\\
    0 && \phantom{-}1 && 0\\
    0 && \phantom{-}0 && 1  
    \end{bmatrix},\quad\lambda_1 =
    \begin{bmatrix} 
    0 && \phantom{-}1 && 0\\
    1 && \phantom{-}0 && 0\\
    0 && \phantom{-}0 && 0  
    \end{bmatrix},\\
\lambda_2&=
    \begin{bmatrix} 
    0 && -i && 0\\
    i && \phantom{-}0 && 0\\
    0 && \phantom{-}0 && 0  
   \end{bmatrix},\quad\lambda_3 =
    \begin{bmatrix} 
    1 && \phantom{-}0 && 0\\
    0 && -1 && 0\\
    0 && \phantom{-}0 && 0  
    \end{bmatrix},\\
\lambda_4&= 
    \begin{bmatrix} 
    0 && \phantom{-}0 && 1\\
    0 && \phantom{-}0 && 0\\
    1 && \phantom{-}0 && 0  
    \end{bmatrix},\quad\lambda_5=
    \begin{bmatrix} 
    0 && 0 && -i\\
    0 && 0 && 0\\
    i && 0 && 0  
    \end{bmatrix},\\
\lambda_6&= 
    \begin{bmatrix} 
    0 && \phantom{-}0 && 0 \\
    0 && \phantom{-}0 && 1\\
    0 && \phantom{-}1 && 0  
    \end{bmatrix},\quad\lambda_7= 
    \begin{bmatrix} 
    0 && 0 && 0\\
    0 && 0 && -i\\
    0 && i && 0  
    \end{bmatrix},\\
\lambda_8&= \frac{1}{\sqrt{3}}
    \begin{bmatrix} 
    1 && \phantom{-}0 && 0\\
    0 && \phantom{-}1 && 0\\
    0 && \phantom{-}0 && -2  
    \end{bmatrix}.  
\end{align}
The $SU(3)$ group of generators consists of all Unitary, Hermitian, traceless, trace-orthogonal, three-dimensional $\lambda$-matrices numbered 1 to 8. Without stressing, the group of generators satisfy the following $SU(3)$ Lie algebra for $j,k,l=1\cdots 8$ \cite{Simon1987}:
\begin{enumerate}
    \item $[\lambda_j,\lambda_k]=2i\sum_{l=1}^{8} f_{jkl}\lambda_l$.
    \item $\{\lambda_j,\lambda_k\}=\frac{4}{3}\delta_{jk}+2\sum_{l=1}^{8}d_{jkl}\lambda_l$.
    \item $\lambda_j \lambda_k = \frac{2}{3}\delta_{jk}\id_3+d_{jkl} \lambda_l + if_{jkl} \lambda_l$.
    \item $tr(\lambda_j \lambda_k)=2\delta_{jk}$.
    \item $tr(\lambda_j)=0$.
\end{enumerate}
Where $f_{jkl}$ and $d_{jkl}$ are the totally antisymmetric and symmetric expansion coefficients in their indices, respectively. The nonvanishing coefficients are:
\begin{align}
    f_{123}&=f_{458}=f_{678}=\frac{\sqrt{3}}{2},\\
    f_{147}&=f_{246}=f_{257}=f_{345}=f_{516}=f_{637}=\frac{1}{2},\\
    d_{118}&=d_{228}=d_{338}=-d_{888}=\frac{1}{\sqrt{3}},\\
    d_{146}&=d_{157}=-d_{247}=d_{256}=\frac{1}{2},\\
    d_{344}&=d_{355}=-d_{366}=-d_{377}\frac{1}{2},\\
    d_{448}&=d_{558}=d_{668}=d_{778}=\frac{-1}{2\sqrt{3}}.
\end{align}

\subsection{Bloch's vector components}

The $r_{j_1}\cdots r_{j_N}$ components of the generalized Bloch's vector (a.k.a coherence vector) can be obtained as follows:
\begin{align}
    &tr\left(\rho_{Nqd} (\Gamma_{k_1}\otimes\cdots\otimes\Gamma_{k_N})\right)=tr\Bigg\{\bigg(\sum_{j_1,\cdots,j_N=0}^{d_1^2-1,\cdots,d_N^2-1}\\
    &r_{j_1 \cdots j_N}\Gamma_{j_1}\otimes\cdots\otimes \Gamma_{j_N}\bigg)\cdot (\Gamma_{k_1}\otimes\cdots\otimes\Gamma_{k_N})\Bigg\},
\end{align}
which after distributing the Kronecker product according to identity no.~\eqref{productoftensors} becomes
\begin{align}
    &=tr\Bigg(\sum_{j_1,\cdots,j_N=0}^{d_1^2-1,\cdots,d_N^2-1}r_{j_1 \cdots j_N}\Gamma_{j_1}\Gamma_{k_1}\\
    &\otimes\cdots\otimes \Gamma_{j_N}\Gamma_{k_N}\Bigg).
\end{align}    
Trace identity no.~\eqref{traceofsum} then yields
\begin{align}
    &=\sum_{j_1\cdots,j_N=0}^{d_1^2-1,\cdots,d_N^2-1}r_{j_1 \cdots j_N}tr\Bigg(\Gamma_{j_1}\Gamma_{k_1}\\
    &\otimes\cdots\otimes \Gamma_{j_N}\Gamma_{k_N}\Bigg),
\end{align}    
and after applying trace identity no.~\eqref{traceoftensors} one gets
\begin{align}
    &=\sum_{j_1,\cdots,j_N=0}^{d_1^2-1,\cdots,d_N^2-1}r_{j_1 \cdots j_N}tr(\Gamma_{j_1}\Gamma_{k_1})\\
    &\cdots tr(\Gamma_{j_N}\Gamma_{k_N}).
\end{align}
Finally, the $\mathfrak{su}(d)$ Lie algebra yields
\begin{align}
    &= \sum_{j_1,\cdots,j_N=0}^{d_1^2-1,\cdots,d_N^2-1}r_{j_1 \cdots j_N}
    d_1^{\delta_{0j_1}} 2^{(1-\delta_{0j_1})}\delta_{j_1k_1} \\
    &\cdots d_N^{\delta_{0j_N}} 2^{(1-\delta_{0j_N})}\delta_{j_Nk_N} \\
    &=r_{k_1 \cdots k_N}2^{(N-\sum_{s=1}^{N}\delta_{0k_s})}\prod_{s=1}^N d_s^{\delta_{0k_s}},
\end{align}
which implies the components of the generalized Bloch's vector for qudit $s$ to be
\begin{align}
    r_{j_1 \cdots j_N}^s &= \frac{tr(\rho_{Nqd} (\Gamma_{j_1}\otimes\cdots\otimes\Gamma_{j_N}))}{2^{(N-\sum_{s=1}^{N}\delta_{0j_s})}\prod_{s=1}^N d_s^{\delta_{0j_s}}}.
    \label{compbloch}
\end{align}
Unit trace condition \eqref{unittrace} of the density operator then entails:
\begin{align}
    r_{00\cdot\cdot\cdot0}=\frac{1}{d} = \frac{1}{d_1d_2\cdots d_N}.
\end{align}
Note that when the generators $\Gamma_{j_s}^{D,S,A}$ are all skew-Hermitian matrices (square matrices that are Hermitian), we gain a physical significance, the generalization of the expectation value:
\begin{align}
    tr(\rho_{Nqd} (\Gamma_{j_1}\otimes\cdots\otimes\Gamma_{j_N})) = \langle\Gamma_{j_1}\otimes\cdots\otimes\Gamma_{j_N}\rangle_{\rho_{Nqd}}.
\end{align}
Positiveness condition requires all the eigenvalues of $\rho_{Nqd}$ to be non-negatives. In $d$ dimensions, the corresponding characteristic polynomial equation of degree $d$ with roots $\lambda_j$ can be represented in the form:
\begin{align}
    det(\rho_{Nqd}-\lambda \id_d)=\sum_{j=0}^d (-1)^j a_j \lambda^{d-j} = \prod_{j=1}^d (\lambda-\lambda_j),
\end{align}
where $a_j$ are coefficients (polynomials with dependence on $r_{j_1 \cdots j_N}^s$) determined by the generators $\Gamma_{js}$ and the components $r_{j_1 \cdots j_N}^s$ of the generalized Bloch's vector. These coefficients have been calculated explicitly in \cite{Kimura2003} and they satisfy the following relation known as Vieta's formula:
\begin{align}
\label{Npoly}
    a_j = \sum_{1 \leq i_1 < i_2 < \cdots i_j}^d \lambda_{i_1}\lambda_{i_2} \cdots \lambda_{i_j}.
\end{align}

\subsection{Particular coherent and incoherent density operators}
\label{sec:particularstates}

In this section we obtain and analyze particular coherent and incoherent density operators from the generalized $\rho_{Nqd}$ and $\iota_{Nqd}$ density operators of an $N$-qudit system as defined in Eq.~\eqref{quditstate} and Eq.~\eqref{incoherent_qudit}, respectively.

\subsubsection{Coherent states}
Here, we derive the 1-qubit, 1-qutrit, 2-qubit, and 2-qudit coherent density operators from Eq.~\eqref{quditstate}. 
\begin{itemize}
    \item \textbf{1-qubit coherent state ($N=1, d_1=2, d_1^2-1=3$)}. For a 2-level quantum system embedded in $SU(2)$, the GGM $\Gamma_{j_s}$ reduced to the $2$x$2$-dimensional identity and Pauli $\sigma_{j_s}$-matrices. So that the generic density operator of the system reads: 
\begin{align}
    \rho_{1qb} &= \sum_{j_1=0}^{d_1^2-1=3} r_{j_1} \Gamma_{j_1} \\
    &=r_0 \sigma_0 + r_1 \sigma_1+r_2 \sigma_2+r_3 \sigma_3\\
    &= r_0 \sigma_0 + \vec{r}\cdot\vec{\sigma} = r_o \id_2 + \vec{r}\cdot\vec{\sigma}\\
    &=
\begin{bmatrix} 
r_0+r_3 && r_1-ir_2 \\
r_1+ir_2 && r_0-r_3  
\end{bmatrix}, \label{qubitmatrix}
\end{align}
where $\vec{\sigma}$$=$$\sum_{j=1}^3 \sigma_j \hat{e}_j \in \mathbb{R}^3$ is a 3-dimensional vector whose components are the alluded Pauli matrices, whereas $\sigma_0$ $\doteq$ $\id_2$ denotes the identity matrix, and $\vec{r}$$=$$\sum_{j=1}^3r_j\hat{e}_j$ $\in$ $\mathbb{R}^3$ is the 3-dimensional Bloch's vector. Unit trace condition \eqref{unittrace} then implies $r_0$$=$$1/d_1$$=$$1/2$, explicitly
\begin{align}
    tr(\rho_{1qb})&=(r_0+r_3)+(r_0-r_3)\\
    &=2r_0=1 \\
    &\implies r_0=1/2.
\end{align}
So the 1-qubit state can be rewritten as
\begin{align}
\rho_{1qb} &=\frac{1}{2}(\id_2 + 2r_1\sigma_1+2r_2\sigma_2+2r_3\sigma_3)\\
     &=\frac{1}{2}(\id_2 + \vec{v} \cdot \vec{\sigma})\\
&=\frac{1}{2}
\begin{bmatrix} 
1+2r_3 && 2(r_1-ir_2) \\
2(r_1+ir_2) && 1-2r_3  
\end{bmatrix}.\label{rescaledqb}
\end{align}
The 3-dimensional \textit{re-scaled coherence vector} now reads $\vec{v}=\sum_{j=1}^3 v_j \hat{e}_j$, with radius $||\vec{v}||_2 = \sqrt{(2r_1)^2+(2r_2)^2+(2r_3)^2}$. According to Eq.~\eqref{compbloch}, the components of the aforementioned re-scaled vector can be obtained as follows:
\begin{align}
    v_j&=2\cdot 2^{-(1-\delta_{0j})}2^{-\delta_{0j}}tr(\rho_{1qb}\sigma_j)
    \\&=2\cdot 2^{-(1-\delta_{0j})}2^{-\delta_{0j}}\langle \sigma_j \rangle.
\end{align}
Whenever $j$$=$$0,1,2,3$, one has 
\begin{align}
2^{-(1-\delta_{0j})}2^{-\delta_{0j}}=2^{-1},
\end{align}
so that
\begin{align}
     v_j &=(2\cdot2^{-1})\langle \sigma_j \rangle_{\rho_{1qb}}\\
     &=tr(\rho_{1qb}\sigma_j).  
\end{align}
It is straightforward seeing from Eq.~\eqref{qubitmatrix} that the eigenvalues of $\rho_{1qb}$ according to Eq.~\eqref{polcharac} are roots of:
\begin{align}
det(\rho_{1qb}-\lambda \id_2)&=r_0^2-r_0r_3-r_0\lambda\\
&+r_3r_0-r_3^2-r_3\lambda \\
&-\lambda r_0+\lambda r_3+\lambda^2\\
&-r_1^2-r_2^2\\
&=\lambda^2-2r_0\lambda+r_0^2 \\
&-(r_1^2+r_2^2+r_3^2)= 0.
\end{align}
Thus, the eigenvalues are the coefficients
\begin{align}
\lambda^{\pm}&= r_0 \pm \sqrt{r_1^2+r_2^2+r_3^2} \\
&=\frac{1 \pm \sqrt{(2r_1)^2+(2r_2)^2+(2r_3)^2}}{2} \\
             &=\frac{1 \pm||\vec{v}||_2}{2}.
\end{align}
Positiveness of the density operator implies that $\rho_{1qb}$ has non-negative eigenvalues $\lambda_j$, as shown in Eq.~\eqref{positv}. Hence $1-||\vec{v}|| \geq 0 \implies ||\vec{v}|| \leq 1$. Therefore, the $\rho_{1qb}$ state is a mixed density operator (or mixed state) since it is located inside the Bloch sphere (see Fig.~\ref{fig:figure2}). On the other hand, a pure $\rho_{1qb}$ density operator requires $||\vec{v}||=1$.

For the case of the 1-qubit state vector defined in Eq.~\eqref{hopf} with corresponding pure density matrix in Eq.~\eqref{densqubitmatrix}, one identifies the following components of the coherence vector:
\begin{align}
2r_0&=\langle \sigma_0 \rangle=tr(\rho_{1qb}\sigma_0)=1,\\
v_1&=2r_1=\langle \sigma_1 \rangle=tr(\rho_{1qb}\sigma_1)\\&=sin(\theta)cos(\phi),\\
v_2&=2r_2=\langle \sigma_2 \rangle=tr(\rho_{1qb}\sigma_2)\\&=sin(\theta)sin(\phi),\\
v_3&=2r_3=\langle \sigma_3 \rangle=tr(\rho_{1qb}\sigma_3)\\&=cos(\theta).
\end{align}
Where the populations (probabilities) are the diagonal entries:
\begin{align}
    p(0)&=1/2(1+v_3)\\
    &=1/2(1+cos(\theta))
    \\&=cos^2(\theta/2)\\
    p(1)&=1/2(1-v_3)
    \\&=1/2(1-cos(\theta))\\
    &=sin^2(\theta/2).
\end{align}
    \item \textbf{1-qutrit coherent state ($N=1, d_1=3, d_1^2-1=8$)}. For a 3-level quantum system embedded in $SU(3)$, the GGM $\Gamma_{j_s}$ reduced to the $3$x$3$-dimensional identity and Gell-Mann $\lambda$-matrices. So that the generic density operator of the system reads: 
\begin{align}
    \rho_{1qt}&=\left(\sum_{j_1=0}^{d_1^2-1=8} r_{j_1}\Gamma_{j_1}^{D,S,A} \right)\\
              &=\frac{1}{3} \left(\id_3 + \sqrt{3} \sum_{j=1}^8 r_j\lambda_j^{D,S,A} \right)\\
              &=\frac{1}{3}\left(\id_3 +\sqrt{3} \vec{r} \cdot \vec{\lambda} \right).
\end{align}
Where $\vec{r}=\sum_{j=1}^8 r_j \hat{e}_j \in \mathbb{R}^8$ is the coherence vector in $SU(3)$, with components:
\begin{align}
    r_j=\langle \lambda_j^{D,S,A}\rangle=tr(\rho_{1qt}\lambda_j^{D,S,A}),
\end{align}
for $j=1\cdots 8$. These components are given in terms of the symmetric, antisymmetric, and diagonal generators $\lambda_{j}^{D,S,A}$:
\begin{align}
    r^D_{j} &= tr(\lambda_{j}^D \rho_{1qt}) = \langle \lambda_{j}^D \rangle\\
    r^S_{j} &= tr(\lambda_{kl}^S \rho_{1qt}) = \langle \lambda_{kl}^S \rangle \\
    r^A_{j} &= tr(\lambda_{kl}^A \rho_{1qt}) = \langle \lambda_{kl}^A \rangle.
\end{align}
The matrix representation of $\rho_{1qt}$ can then be given as:
\begin{align}
    \rho_{1qt}= \frac{\sqrt{3}}{3}
    {\left(
    \begin{smallmatrix} 
    r_3+\frac{r_8}{\sqrt{3}}+\frac{1}{\sqrt{3}}&&r_1-i r_2&&r_4-i r_5\\
    r_1+i r_2&&-r_3+\frac{r_8}{\sqrt{3}}+\frac{1}{\sqrt{3}}&&r_6-i r_7\\
    r_4+i r_5&&r_6+i r_7&&\frac{1}{\sqrt{3}}-2\frac{r_8}{\sqrt{3}}
    \end{smallmatrix}\right)},
\end{align}
with the populations (probabilities) given by the diagonal entries. Positiveness condition can them be analyzed by the following characteristic equation \cite{Shuming2015}:
\begin{align}
    det(\rho_{1qt} - \lambda \id_3) \doteq \lambda^3 - B_1 \lambda^2 + B_2 \lambda - B_3=0,
\end{align}
with conditions 
\begin{align}
    B_1 &\doteq tr(\rho_{1qt})=1,\\
    B_2 &\doteq \frac{1}{2}(1-tr(\rho_{1qt}^2))=\frac{1}{2}(1-\vec{r}\cdot\vec{r}),\\
    B_3 &\doteq \frac{1}{3}(B_2-tr(\rho_{1qt}^2)+tr(\rho_{1qt}^3)),\\
    &=\frac{1}{27}(1-3\vec{r} \cdot \vec{r}+ 2 \vec{r} * \vec{r} \cdot \vec{r}).
\end{align}
Here, $*$ denotes the star product \cite{Mallesh1997}\cite{Kimura2003} defined as
\begin{align}
    (\vec{r} * \vec{r})_k \doteq \sqrt{3} d_{ijk} r_i r_j.
\end{align}
Positiveness is then satisfied whenever $B_j\geq 0$. First and second conditions translates into the common inequality $\rho_{1qt} \geq 0$, with eigenvalue $\lambda=\frac{1-|\vec{r}|}{3}$. While third condition entails:
\begin{align}
    tr(\rho_{1qt}^2)&=\frac{1}{3}+\frac{2}{3}|\vec{r}|^2,\\
    tr(\rho_{1qt}^3)&=\frac{1}{9}+\frac{2}{9}(\sqrt{3}d_{ijk}r_i r_j\cdot \vec{r}+3|\vec{r}|^2).
\end{align}
    \item \textbf{2-qubit coherent state ($N=2, d_1=d_2=2, d_1^2-1=d_2^2-1=3$)}. For a composite quantum system of two qubits, embedded in $SU(4)\doteq SU(2)\otimes SU(2)$, where the 2-level subsystems are each embedded in $SU(2)$, the $\Gamma_{j_s}$ matrices are identified as the $2$x$2$-dimensional identity and Pauli-$\sigma_{j_s}$ matrices. The generic density operator of the composite system then reads:
\begin{align}
    \rho_{2qb} &= \sum_{j_1=0}^{d_1^2-1=3} \sum_{{j_2}=0}^{d_2^2-1=3} r_{j_1}r_{j_2}\Gamma_{j_1}\otimes \Gamma_{j_2}\\
    &=\sum_{j_2=0}^{3} \Bigg( r_0 r_{{j_2}} \Gamma_0 \otimes \Gamma_{{j_2}} 
    \\&+ r_1 r_{{j_2}} \Gamma_1 \otimes \Gamma_{{j_2}} \\&+ r_2 r_{{j_2}} \Gamma_2 \otimes \Gamma_{{j_2}}
    \\&+ r_3 r_{{j_2}} \Gamma_3 \otimes \Gamma_{{j_2}}\Bigg)\\
    &= r_0 r_{0} \Gamma_0 \otimes \Gamma_{0} + r_0 r_{1} \Gamma_0 \otimes \Gamma_{1} 
    \\&+ r_0 r_{2} \Gamma_0 \otimes \Gamma_{2} + r_0 r_{3} \Gamma_0 \otimes \Gamma_{3} \\
    &+ r_1 r_{0} \Gamma_1 \otimes \Gamma_{0} + r_1 r_{1} \Gamma_1 \otimes \Gamma_{1} 
    \\&+ r_1 r_{2} \Gamma_1 \otimes \Gamma_{2} + r_1 r_{3} \Gamma_1 \otimes \Gamma_{3} \\
    &+ r_2 r_{0} \Gamma_2 \otimes \Gamma_{0} + r_2 r_{1} \Gamma_2 \otimes \Gamma_{1} 
    \\&+ r_2 r_{2} \Gamma_2 \otimes \Gamma_{2} + r_2 r_{3} \Gamma_2 \otimes \Gamma_{3} \\
    &+ r_3 r_{0} \Gamma_3 \otimes \Gamma_{0} + r_3 r_{1} \Gamma_3 \otimes \Gamma_{1} 
    \\&+ r_3 r_{2} \Gamma_3 \otimes \Gamma_{2} + r_3 r_{3} \Gamma_3 \otimes \Gamma_{3}.
\end{align}
Unit trace condition of the density operator then implies $r_{00}=r_{0}r_{0}=1/(d_1d_2)=1/4$. Substituting for the $SU(2)$ matrices, the density matrix can then be written in the Fano form \cite{Fano1983} (a.k.a Hilbert-Schmidt representation) as:
\begin{align}
\label{rho2b}
    \rho_{2qb}&=\frac{1}{4}\bigg\{\id_2^a \otimes \id_2^b \\&+ \sum_{j=1}^{3} r_j^a \sigma_j^a \otimes \id_2^b \\
    &+ \id_2^a \otimes \sum_{k=1}^3 r_k^b \sigma_k^b \\&+ \sum_{j,k=1}^3 c_{jk} \sigma_j^a \otimes \sigma_k^b \bigg\}.
\end{align}
For convenience, we have used indexes $a$ and $b$ to denote subsystems $s=1$ and $s=2$, respectively. The components of the re-scaled coherence vector for each qubit are given according to Eq.~\eqref{compbloch} and also Eq.~\eqref{exp2}, as follows:
\begin{align}
    r_j^a &=\langle \sigma_j^a \rangle_{\rho_{2qb}} \\
    &= tr(\rho_{2qb} (\sigma_j^a \otimes \id_2^b)) \\
    &=tr(\sigma_j^a \rho_{a}), \\
    r_k^b &=\langle \sigma_k^b \rangle_{\rho_{2qb}} \\
    &=tr(\rho_{2qb} (\id_2^a \otimes \sigma_k^b))\\
    &=tr(\sigma_k^b \rho_{b}),
\end{align}
where $\rho_a = tr_b(\rho_{2qb})$ and $\rho_b = tr_a(\rho_{2qb})$ are the reduced density operators of subsystems $s=1$ and $s=2$, respectively:
\begin{align}
    \rho_a &= \frac{1}{2}\id_2^a+\frac{1}{2}\sum_{j=1}^3r_j^a\sigma_j^a,\\
    \rho_b &= \frac{1}{2}\id_2^b+\frac{1}{2}\sum_{k=1}^3r_k^b\sigma_k^b.
\end{align}
And the components of the correlation matrix of the observables (spins) are:
\begin{align}   
    c_{jk}&=4\bigg({2^{-(2-\sum_{s=1}^{2}\delta_{0j_s})}\prod_{s=1}^2 2^{-\delta_{0j_s}}}\bigg)\langle \sigma_j^a \otimes \sigma_k^b \rangle\\
    &=4\bigg({2^{-(2-(\delta_{0j}+\delta_{0k}))}2^{-\delta_{0j}}2^{-\delta_{0k}}}\bigg)\langle \sigma_j^a \otimes \sigma_k^b \rangle,
\end{align}
of which, for any combination of $j,k=0\cdots 3$, yields
\begin{align} 
    c_{jk}&=4\cdot (2^{-2})\langle \sigma_j^a \otimes \sigma_k^b \rangle
    \\&=\langle \sigma_j^a \otimes \sigma_k^b \rangle\\
    &=tr(\rho_{2qb}(\sigma_j^a \otimes \sigma_k^b)).
\end{align}
For the case of the singlet state defined in Eq.~\eqref{singlet}, its density operator is simply \cite{Peres1993}:
\begin{align}
    \rho_{2qb}^{entg} = \frac{1}{4}\left(\id_2^{a}\otimes \id_2^{b}-\sum_{m=1}^3{\sigma_m^{a}\otimes\sigma_m^{b}}\right),
\end{align}
with 
\begin{align}
    \langle \sigma_j \rangle&=0,\\
    \langle \sigma_j \otimes \sigma_k \rangle&=-\frac{1}{4}\delta_{jk},\\
    (\id \otimes \vec{\sigma} + \vec{\sigma} \otimes \id)|\psi_{12}^{Bell -}\rangle&=0.
\end{align}
\item \textbf{2-qudit coherent state}.
The previous results entail the following 2-qudit density operator of a bipartite system:
\begin{align}
    \rho_{2qd}&=\frac{1}{d_a d_b}\bigg\{\id_{d_a}^a \otimes \id_{d_b}^b \\&+ \sum_{j=1}^{d_a^{2}-1} r_j^a \Gamma_j^a \otimes \id_{d_b}^b \\
    &+ \id_{d_a}^a \otimes \sum_{k=1}^{d_b^{2}-1} r_k^b \Gamma_k^b \\&+ \sum_{j=1}^{d_a^{2}-1}\sum_{k=1}^{d_b^{2}-1} c_{jk} \Gamma_j^a \otimes \Gamma_k^b \bigg\}.
\end{align}
Once again, for convenience, indexes $a$ and $b$ denote the subsystems $s=1$ and $s=2$, respectively. Since each subsystem is regarded as an $N=1$ system, the components of each re-scaled coherence vector are obtained according to Eq.~\eqref{compbloch}, as follows:
\begin{align}
    r_j^s&=d_s\cdot 2^{-(1-\delta_{0j})}d_s^{-\delta_{0j}}tr(\rho_{1qb}\Gamma_j)
    \\&=d_s\cdot 2^{-(1-\delta_{0j})}d_s^{-\delta_{0j}}\langle \Gamma_j \rangle,
\end{align}
of which, for $j$$=$$1$$\cdots$$d_s^2-1$, yields
\begin{align}
2^{-(1-\delta_{0j})}d_s^{-\delta_{0j}}=2^{-1}.    
\end{align} 
Therefore,
\begin{align}
    r_j^a &=2^{-1}d_a\langle \Gamma_j^a \rangle_{\rho_{2qb}} \\
    &= 2^{-1}d_a tr(\rho_{2qb} (\Gamma_j^a \otimes \id_2^b)) \\
    &=2^{-1}d_a tr(\Gamma_j^a \rho_{a}), \\
    r_k^b &=2^{-1}d_b\langle \Gamma_k^b \rangle_{\rho_{2qb}} \\
    &=2^{-1}d_b tr(\rho_{2qb} (\id_2^a \otimes \Gamma_k^b))\\
    &=2^{-1}d_b tr(\Gamma_k^b \rho_{b}).     
\end{align}
Whereas the components of the correlation are as follows:
\begin{align}   
    c_{jk}&=d_ad_b\bigg({2^{-(2-\sum_{s=1}^{2}\delta_{0j_s})}\prod_{s=1}^2 d_s^{-\delta_{0j_s}}}\bigg)\langle \Gamma_j^a \otimes \Gamma_k^b \rangle
    \\
    &=2^{-2}d_ad_b\langle \Gamma_j^a \otimes \Gamma_k^b \rangle\\
    &=2^{-2}d_ad_btr(\rho_{2qb}(\Gamma_j^a \otimes \Gamma_k^b)),
\end{align}
for any combination of $j, k$$=$$1$$\cdots$$d_s^2-1$.
\end{itemize}

\subsubsection{Incoherent states}
Particular incoherent density operators are obtained from Eq.~\eqref{incoherent_qudit}. Here, we derive the 1-qubit, 1-qutrit and 2-qubit incoherent density operators.
\begin{itemize}
    \item \textbf{1-qubit incoherent state ($N=1, d_1=2, d_1-1=1$):}
\begin{align}
    \iota_{1qb}&= \sum_{j_{1}=0}^{d_1-1=1}\iota_{j_1}\Gamma_{j_1}^{D}=\iota_0 \Gamma_0+ \iota_1 \Gamma_1\\
    &=\iota_0 \sigma_0+ \iota_1 \sigma_3=\iota_0 \id_2+\iota_1 \sigma_3,
\end{align}    
where $\id_2$ and $\sigma_3$ are the identity and Pauli $\sigma$-diagonal matrices of $SU(2)$. Unit trace condition then implies $tr(\iota_{1qb})=tr(\iota_0 \id_2)+tr(\iota_1\sigma_3)=2\iota_0=1\implies \iota_0=1/2$, so that
\begin{align}
    \iota_{1qb}&= \frac{1}{2}(\id_2 + 2\iota_1 \sigma_3)\\
    &\doteq\frac{1}{2}(\id_2 + \iota_3 \sigma_3)\\
    &=\frac{1}{2}(\id_2 + \vec{\iota} \cdot \vec{\sigma})\\
    &= \frac{1}{2}
\begin{bmatrix} 
1+\iota_3  && 0  \\
0 && 1-\iota_3  
\end{bmatrix},
\end{align}
with the three-dimensional incoherent Bloch's vector $\vec{\iota} = (0,0,\iota_3) \in \mathbb{R}^3$. The components of the aforementioned re-scaled vector according to Eq.~\eqref{compbloch} are:
\begin{align}
    \iota'_j&=2\cdot2^{-1}\langle \Gamma_j^D \rangle=2\cdot2^{-1}tr(\iota_{1qb}\Gamma_j^D). \\
    \iota'_0&=2\iota_0=tr(\iota_{1qb}\sigma_0)=1.\\
    \iota'_1&=2\iota_1=tr(\iota_{1qb}\sigma_3)=\iota_3.
\end{align}
And the populations (probabilities) are the diagonal entries obtained according to Eq.~\eqref{probtrace}, as follows:
\begin{align}
    p_j&= tr(\iota_{1qb}|j\rangle\langle j|)\\
    &=(1+(-1)^{\delta_{j_1}}\langle \Gamma_j^D\rangle)/2\\
    &=(1+(-1)^{\delta_{j_1}}\langle \sigma_3\rangle)/2.\\
    p_0&=\frac{1}{2}(1+\iota_3).\\
    p_1&=1-p_1=\frac{1}{2}(1-\iota_3).
\end{align}
From Eq.~\eqref{rescaledqb}, one realizes that $\iota_3=2r_3=v_3$, and since $||\vec{v}||\leq 1$, one gets the following bound:
\begin{align}
    v_1^2+v_2^2\leq 1-v_3^2&=(1-v_3)(1+v_3)
    \\&=4p_0p_1.
\end{align}
It is straightforward seeing that for the pure density operator defined in Eq.~\eqref{densqubitmatrix} its incoherent counterpart yields $\iota_3=cos(\theta)$.
    \item \textbf{1-qutrit incoherent state ($N=1, d_1=3, d_1-1=2$)}.
\begin{align}
    \iota_{1qt}&=\sum_{j_{1}=0}^{d_1-1=2}\iota_{j_1}\Gamma_{j_1}^{D}\\
    &=\iota_0 \Gamma_0^D+\iota_1 \Gamma_1^D+\iota_2 \Gamma_2^D \\
    &=\iota_0 \lambda_0 +\iota_1 \lambda_3 + \iota_2 \lambda_8,
\end{align}   
where $\lambda_0=\id_3$, $\lambda_3$ and $\lambda_8$ are the identity and Gell-Mann $\lambda$-diagonal matrices of $SU(3)$. Unit trace condition then implies $\iota_0=1/d_1=1/3$, thus
\begin{align}
    \iota_{1qt}&=\frac{1}{3}\bigg(\id_3 +3\iota_1 \lambda_3 + 3\iota_2 \lambda_8\bigg)\\
    &=\frac{1}{3}\bigg(\id_3 + \iota_3\lambda_3+\iota_8\lambda_8\bigg) \\
    &=\frac{1}{3}\bigg(\id_3 + \vec{\iota} \cdot \vec{\lambda}\bigg).
\end{align}    
Where $\vec{\iota}=\sum_{j=1}^8 \iota_j \hat{e}_j \in \mathbb{R}^8$ is the incoherent Blochs's vector in $SU(3)$, with non-zero components:
\begin{align}
    \iota_3&=3\cdot2^{-1}\langle \lambda_3 \rangle=3\cdot2^{-1}tr(\rho_{1qt}\lambda_3),\\
    \iota_8&=3\cdot2^{-1}\langle \lambda_8 \rangle=3\cdot2^{-1}tr(\rho_{1qt}\lambda_8),
\end{align}
    \item \textbf{2-qubit incoherent state ($N=2, d_1=d_2=2, d_1-1=d_2-1=1$)}.
Following the same procedure realized for the 2-qubit coherent state, on substituting for the diagonal Pauli $\sigma$-matrices one gets:
\begin{align}
    \iota_{2qb}&=\sum_{j_{1}=0}^{d_1-1=1}\sum_{j_{2}=0}^{d_2-1=1}\iota_{j_1}\iota_{j_2}\Gamma_{j_1}^{D}\otimes\Gamma_{j_2}^{D}
    \\&=\frac{1}{4}\bigg(\id_2^a \otimes \id_2^b + t_3 \sigma_3^a \otimes \id_2^b \\
    &+ \id_2^a \otimes u_3 \sigma_3^b + v_{33} \sigma_3^a \otimes \sigma_3^b \bigg).
\end{align}    
Where $\id_2$, and $\sigma_3$ are the identity and Pauli $\sigma$-diagonal matrices of each subsystem in $SU(2)$.
\end{itemize}

\section{Separable and entangled states within the density operator formalism}
\label{sec:entangledstates}

In general, a mixed density operator of an $N$-partite composite system represents a separable state if it can be written as a convex combination of product states (probabilistic mixture) in the form \cite{Werner1989}:
\begin{align}
\label{convexcomb}
\rho_{1\cdots N}^{sep}=\sum_j p_j \rho_{j_1} \otimes \cdots \otimes \rho_{j_N}, 
\end{align}
with
\begin{eqnarray}
p_j&\geq 0, \\
\sum_j p_j &= 1.
\end{eqnarray}
Any mixed density operator that cannot be written in the form of Eq.~\eqref{convexcomb} represents an entangled state. For the particular case of a bipartite system, Eq.~{\eqref{convexcomb}} becomes:
\begin{align}
\label{denssep}
\rho_{12}^{sep}=\sum_j p_j \rho_{j_1} \otimes \rho_{j_2},
\end{align}
for any density operators $\rho_{j_1}\in\mathcal{L}(\mathcal{H}_1)$ and $\rho_{j_2}\in\mathcal{L}(\mathcal{H}_2)$. 

One important example of a 2-qubit state is the Werner state defined as
\begin{align}
    \rho_{12}^W (x) = \left(\frac{1-x}{4}\right) \id_2^a \otimes \id_2^b + xS,
\end{align}
where $x$ is a real parameter usually in the range $0 \leq x \leq 1$, and $S$ is the following matrix:
\begin{align}
    S = \frac{1}{2} 
\begin{bmatrix} 
0 && 0 && 0 && 0\\
0 && 1 && -1 && 0\\
0 && -1 && 1 && 0\\
0 && 0 && 0 && 0
\end{bmatrix}.
\end{align}
The Werner state can also be written in the form
\begin{align}
    \rho_{12}^W (x) = \frac{1}{4} \id_2^a \otimes \id_2^b &-\frac{x}{4} \bigg(\sigma_1^a \otimes \sigma_1^b \\&+ \sigma_2^a \otimes \sigma_2^b + \sigma_3^a \otimes \sigma_3^b\bigg).
\end{align}
The Werner state is separable when $-1/3 \leq x \leq 1/3$, and entangled only in the range $1/3 < x \leq 1$.

Determining whether a state is separable or entangled is in general a complex problem. In the following sections we outline two important tests to assess the separability of 2-qubit states.

\subsection{The Schmidt decomposition}

The Schmidt decomposition allows us to determine the existence of entanglement in pure states by looking at the coefficients of the reduced density operator. Let $\mathcal{H}_{ab} \doteq \mathcal{H}_a\otimes\mathcal{H}_b$ denote the composite state space of a bipartite system with some orthonormal basis set $\{|a_j\rangle\otimes|b_k\rangle\}_{j,k=1}^{d_a, d_b}$. Adopting, for convenience, the condition $d_a \leq d_b$, any pure state $|\psi_{ab}\rangle \in \mathcal{H}_{ab}$ can be written in the form of a Schmidt decomposition of Eq.~\eqref{bipartitestate} such that:
\begin{align}
    |\psi_{ab}\rangle_{2qd} = \sum_{j=1}^{d_a} c_j |a_j\rangle \otimes |b_j\rangle,
\end{align}
with coefficients $c_j^2$$\doteq$$\lambda_j$ for some reduced density operator $\rho_a = \sum_j \lambda_j |a_j\rangle \langle a_j |$. The above composite pure state is considered entangled if and only if the Schmidt rank (the number of coefficients strictly greater than zero) is greater than 1, i.e, if there is more than one non-zero eigenvalue $\lambda_j$, otherwise, the state is separable. In this sense, a maximally entangled state is one for which all the eigenvalues $\lambda_j$ are equal, and the composite maximally entangled pure state becomes:
\begin{align}
    |\psi_{ab}^{Mentg}\rangle_{2qd} = \frac{1}{\sqrt{d_a}}\sum_{j=1}^{d_a} |a_j\rangle \otimes |b_j\rangle.
\end{align}

\subsection{The Peres-Horodecki criterion}

Another straightforward way to test whether a bipartite two-qubit system is separable or not is using the Peres-Horodecki criterion also known as positive partial transposition criterion \cite{Horodecki1996} \cite{Peres1996}. This criterion states that a state $\rho_{2qb}$ is separable if the resulting density operator obtained by transposing one of the subsystem's density operator is a positive operator. For the $\rho_{2qb}$ density operator defined in Eq. \eqref{rho2b}, after partial transposition of the second subsystem, the state becomes:
\begin{align}
    \rho_{2qb}^{ppt}&=\frac{1}{4}\bigg\{\id_2^a \otimes \id_2^b \\&+ \sum_{j=1}^{3} r_j^a \sigma_j^a \otimes \id_2^b \\
    &+ \id_2^a \otimes \sum_{k=1}^3 r_k^b (\sigma_k^b)^T \\&+ \sum_{j,k=1}^3 c_{jk} \sigma_j^a \otimes (\sigma_k^b)^T \bigg\}.
\end{align}
And for the Werner state, partially transposition yields
\begin{align}
\rho_{12}^{Wppt} (x) = \frac{1}{4} \id_2^a \otimes \id_2^b &-\frac{x}{4} \bigg(\sigma_1^a \otimes \sigma_1^b \\&- \sigma_2^a \otimes \sigma_2^b + \sigma_3^a \otimes \sigma_3^b\bigg),
\end{align}
with the following matrix
\begin{align}
\rho_{12}^{Wppt} (x)=
\begin{bmatrix} 
1-x && 0 && 0 && -2x\\
0 && 1+x && 0 && 0\\
0 && 0 && 1+x && 0\\
-2x && 0 && 0 && 1-x
\end{bmatrix}.
\end{align}
Since the eigenvalues of a positive operator are all non-negatives, we assess the positiveness of a density operator by looking at the coefficients of its characteristic polynomial equation. In the case of a multipartite system, whose dimension is $d>2$, the aforementioned coefficients are the ones from Eq.~\eqref{Npoly}. In this case, the partially transposed Werner state has non-negative coefficients $a_j$ whenever $-1 \leq x \leq 1/3$, therefore, it is separable when $-1/3 \leq x \leq 1/3$, and entangled in the range $-1 \leq x < -1/3$.

\section{Quantum entanglement quantifiers}
\label{sec:entangmeas}

Several quantum quantifiers have been proposed as entanglement measures to quantify the degree of entanglement of a multipartite quantum system, i.e, to verify if the global state of the system is a maximally entangled state, a partial entangled state, or if it has no entanglement at all. An entanglement measure $\mathcal{E}(\rho)$ is defined by a mapping $\mathcal{E}(\cdot) : \rho \rightarrow \mathbb{R}$, that takes a positive-definite density operator $\rho$ and produces a real number $\lambda \in \mathbb{R}$. At a bare minimum, any entanglement measure must be invariant under similarity transformations, and must not increase under LOCCs. Let $\rho$ be some density operator describing the global state of a quantum system. Without stressing, we outline some of the properties required for an entanglement measure:
\begin{enumerate}
\item $\mathcal{E}(\rho) \in [0, \infty]$ yields a real number and is 0 if the global state of the system is a separable state.
\item Monotonicity: 
\begin{align*}
\mathcal{E}(\Lambda_{\text{LOCC}}(\rho)) \leq \mathcal{E}(\rho),
\end{align*} is monotonically decreasing under LOCC, where $\Lambda_{\text{LOCC}}$ denotes an operation.
\item Convexity: 
\begin{align*}
\mathcal{E}\left(\sum_j p_j |\psi_j\rangle\langle \psi_j|\right)\leq \sum_j p_j \mathcal{E}(|\psi_j\rangle\langle \psi_j|).
\end{align*}
\end{enumerate}

\subsection{Purity and maximally mixed states}\label{sec:puritysec}

One way to quantify the amount of noise (classical ignorance) of a quantum system and, therefore, infer its degree of entanglement, is looking at the purity of the reduced density operator of the global state of that system. Let $\rho$ denote a generic density operator, the \textit{Purity} of $\rho$ is defined as
\begin{align}
    \mathcal{P}(\rho)\doteq tr(\rho^{\dagger}\rho)=tr(\rho^2) \leq 1,
\end{align}
with equality if and only if $\rho$ is a pure density operator, as demonstrated in \autoref{sec:densityformalism}. In addition, there is a lower bound defined by $tr(\rho^2)\geq\frac{1}{2}$ for which $\rho$ is a maximally mixed state. 

Consider, for instance, a bipartite quantum system whose global state is the singlet state defined in Eq.~\eqref{singlet} with associated bipartite pure density operator $\rho_{ab}$. One can show that each subsystem of $\rho_{ab}$ is in a maximally mixed state. To demonstrate that, let us derive the reduced density operator $\rho_a$ of the first qubit in subsystem $a$ as follows:
\begin{align}
\rho_a &= tr_b(\rho_{ab}) =tr_b\left(|\psi_{ab}^{Bell -}\rangle \langle\psi_{ab}^{Bell -}|\right) \\
&=  tr_b \left\{
\left( \frac{|0_a1_b\rangle - |1_a0_b\rangle}{\sqrt{2}} \right)
\left(\frac{\langle0_a1_b| - \langle 1_a0_b|}{\sqrt{2}} \right)
\right\}
\\
&=\frac{1}{2}tr_b
\Bigg\{
|0_a1_b\rangle\langle0_a1_b|-|0_a1_b\rangle\langle1_a0_b|\\
&-|1_a0_b\rangle\langle0_a1_b|+|1_a0_b\rangle\langle1_a0_b|
\Bigg\}\\
&=\frac{1}{2}\Bigg\{tr_b(|0_a\rangle\langle0_a|\otimes|1_b\rangle\langle1_b|\\
&-tr_b(|0_a\rangle\langle1_a|\otimes|1_b\rangle\langle0_b|)\\
&-tr_b(|1_a\rangle\langle0_a|\otimes|0_b\rangle\langle1_b|)\\
&+tr_b(|1_a\rangle\langle1_a|\otimes|0_b\rangle\langle0_b|))\Bigg\},
\end{align}
and with Eq.~\eqref{parttrace} one finally has
\begin{align}
\rho_a&=\frac{1}{2}
\Bigg\{
|0_a\rangle \langle 0_a| \langle 1_b|1_b\rangle  - |0_a\rangle \langle 1_a| \langle 1_b|0_b\rangle \\
&- |1_a\rangle \langle 0_a| \langle 0_b|1_b\rangle + |1_a\rangle \langle 1_a| \langle 0_b|0_b\rangle
\Bigg\}\\
&=\frac{|0\rangle \langle 0|+|1\rangle \langle 1|}{2} = \frac{\id_2}{2} = 
\begin{bmatrix} 
0.5 && 0  \\
0 && 0.5  
\end{bmatrix}.\label{reduced}
\end{align}
Since $tr(\rho_a^2)=tr(\rho_b^2)=tr((\id_2/2)^2)=\frac{1}{4}tr(\id_2^2)=1/2$, such a subsystem is in a maximally mixed state. The matrix representing such a state is dubbed an ``incoherent density matrix'' (see \autoref{sec:decoherence})). Henceforth, we note another hallmark of the maximally entangled bell states: while the global state $|\psi_{ab}\rangle$ of the composite system is known exactly, i.e, there is no classical ignorance associated to its preparation (is a pure state), the state of any of its subsystems, such as the state of the first qubit, is in a maximally mixed state. 

Summing up, when the system is prepared in a maximally entangled pure state:
\begin{align}
    \rho_a \doteq tr_b(\rho_{ab}) = \rho_b \doteq tr_a(\rho_{ab}) = \frac{\id_2}{2}.
\end{align}
In general, any maximally entangled state is necessarily a pure state, and its subsystems are each in a maximally mixed state. However, it is possible to prepare a pure state that is not maximally entangled, as the one defined in Eq.~{\eqref{partentang}}.

\subsection{Concurrence measure} \label{sec:Concurrence}

One example of entanglement measure is the Concurrence measure denoted $\mathcal{E}_c(\rho)$ Concurrence was first introduced in 1996 \cite{Bennett1996} to compute the entanglement of formation for Bell-diagonal two-qubit states, and later established as an entanglement measure for general two-qubit states. The Concurrence metric has the following criteria to determine the degree of entanglement of the global state $\rho$ of a quantum system:
\begin{enumerate}
    \item  If $\mathcal{E}_c(\rho) = 0$, then the global state of the system is separable, i.e, the system has no entanglement.
    \item  If $0 < \mathcal{E}_c(\rho) < 1$, then the global state of the system is partially entangled.
    \item  If $\mathcal{E}_c(\rho) = 1$, then the global state of the system is maximally entangled.
\end{enumerate}
The general definition of Concurrence, valid even for a mixed density operator of two qubits, is:
\begin{align}
    \mathcal{E}_c(\rho)\doteq max\left\{ 0, \sqrt{\lambda_1}-\sqrt{\lambda_2}-\sqrt{\lambda_3}-\sqrt{\lambda_4} \right\},
\end{align}
where $\lambda_1\geq\lambda_2\geq\lambda_3\geq\lambda_4$ are the eigenvalues of the matrix $\Lambda=\rho(\sigma_y\otimes\sigma_y)\rho^*(\sigma_y\otimes\sigma_y)$. 

In section \autoref{sec:puritysec} we have mentioned the possibility of pure states that are not maximally entangled. One such example is the following pure state of a bipartite system:
\begin{align}\label{partentang}
    |\psi_{12}^{entg}\rangle_{2qd}\doteq\frac{1}{\sqrt{3}}\Bigg(|00\rangle + |11\rangle + |01\rangle \Bigg).
\end{align}
The Concurrence measure for such state yields $\mathcal{E}_c(\rho)=2/3<1$ and, therefore, the state is a partially entangled pure state.

\subsection{Von Neumann entropy} \label{vonentropy}
The Von Neumann entropy of a density operator $\rho$ is defined as:
\begin{align}
    S(\rho) \doteq -tr(\rho log(\rho)).
\end{align}
If $\rho$ is diagonal in some orthonormal basis with eigenvalues $\lambda_j$, then
\begin{align}
    S(\rho) = -\sum_j \lambda_j log(\lambda_j).
\end{align}
Let $\{\rho_s\}_{s=1}^N$ be a set of $N$ reduced density operators of $\rho$ for qudit $s$ in subsystem $s$, where $\rho$ defines the global qudit state of the quantum system. The Von Neumann entropy of the reduced density operator is a measure of entanglement (a.k.a entanglement entropy) defined as:
\begin{align}
    \mathcal{E} \rightarrow S(\rho_s) \doteq -tr(\rho_s log_2(\rho_s)).
\end{align}
Here we outline, without stressing, some of the properties of the entanglement entropy:
\begin{enumerate}
    \item $S(\rho_1)=\cdot\cdot\cdot=S(\rho_N)$, whenever $tr(\rho^2)=1$, i.e, when $\rho$ is a pure state its subsystems have all the same entropy.
    \item $S(\rho_s) = 0$, whenever $tr(\rho_s^2)=1$, i.e, when $\rho_s$ is a pure reduced density operator. Hence, the global state of the quantum system represented by the density operator $\rho$ is not entangled. However, its subsystems may have a degree of entanglement of their own, such in the case when the subsystem is the Bell state.
    \item $S(\rho_s) \leq log(d_s)$, with equality (maximum entropy) whenever $\rho_{s}=\id/d_s$, i.e, when $\rho_s$ is a maximally mixed reduced density operator, hence the global state $\rho$ of the system is maximally entangled.
    \item $S(\rho_s) = S(\hat{U}\rho_s\hat{U}^{\dagger})$ is the isometric invariance property.
    \item $S(\rho_1 \otimes \rho_2) = S(\rho_1)+S(\rho_2)$ is the additivity property for independent subsystems.
\end{enumerate}

Consider, for instance, the $|\psi_{12}^{Bell+}\rangle$ entangled state vector with corresponding density matrix
\begin{align}
    \rho_{12} &= |\psi_{12}^{Bell+}\rangle \langle \psi_{12}^{Bell+}| \\
    &=
\begin{bmatrix} 
0.5 && 0 && 0 && 0.5  \\
0 && 0&& 0 && 0\\
0 && 0&&0&&0\\
0.5 && 0&&0&&0.5  
\end{bmatrix},
\end{align}
whose eigenvalues are 0, 0, 0, and 1. The Von Neumann entropy of $\rho_{12}$ yields: $4 \cdot0log(0)+1log(1)=0$, hence it is a pure density operator. While the entanglement entropy of its reduced density operator $\rho_1$ derived in Eq.~\eqref{reduced} yields:
\begin{align}
    - tr(\rho_1 log_2(\rho_1)) = -tr(-\id/2) = log(d_s=2)=1,
\end{align}
with identical result for $\rho_2$. Therefore, the global state $\rho_{12}$ or $|\psi_{12}^{Bell+}\rangle$ is a maximally entangled pure state. 

\section{Quantum coherence quantifiers}
\label{sec:coherencequantifiers}

It is possible to quantify how dissimilar (different) two $N$-qudit states are by computing the distance between their corresponding density operators. Several distance/similarity measures between probability distributions have already been reported \cite{CHA2007}. In general, a distance measure in a normed space (metric space) $\mathcal{D}(\mathcal{H})$ is a function with a map $d: \mathcal{D}(\mathcal{H})$ x $\mathcal{D}(\mathcal{H}) \rightarrow \mathbb{R}[0, \infty)$ induced by the norm (metric) of the space. Without stressing, a good distance measure is a function satisfying the following axioms for all $\rho, \eta, \zeta \in \mathcal{D}(\mathcal{H})$: 
\begin{enumerate}
    \item Positive semi-definiteness: $d(\rho,\eta) \geq 0$.
    \item Identity: $d(\rho,\eta) =0$, iff $\rho$=$\eta$.
    \item Triangle inequality: $d(\rho,\eta) \leq d(\rho,\zeta) - d(\zeta,\eta)$.
    \item Symmetry: $d(\rho,\eta)= d(\eta,\rho)$.
\end{enumerate}
When a distance measure does not satisfy all the above conditions, it is dubbed a pseudometric (if violates no. 2), semi-metric (if violates no. 3) or quasimetric (if violates no. 4). 

In the context of coherence measurements \cite{Baumgratz2014}, the degree of quantum coherence of a generic $N$-qudit state with corresponding density operator $\rho$ may be quantified by the minimum distance between $\rho$ and its incoherent density operator $\iota_{\rho}$. Here, we denote the set of all incoherent states $\iota$ in the form of Eq.~\eqref{incoherent_qudit} by $\mathcal{I}$. Within this formalism, a coherence measure is a nonnegative convex function defined as:
\begin{align}
C(\rho) &\doteq \min_{\iota \in \mathcal{I}}d(\rho,\iota)\\
&= d(\rho,\iota_{\rho}),
\end{align}
which is zero if and only if $\rho\in\mathcal{I}$. In general, a coherence measure should satisfy the following conditions \cite{Baumgratz2014}:
\begin{enumerate}
    \item Positiveness:
    \begin{align}
        C(\rho) \geq 0 \text{ if equality iff } \rho \in \mathcal{I}.
    \end{align}
    \item Monotonicity under incoherent channel $\Lambda_{ICPTP}^{\mathcal{I}}$ a.k.a incoherent completely positive and trace preserving maps:
    \begin{align}
        C(\Lambda_{ICPTP}^{\mathcal{I}} (\rho))\leq C(\rho) ,
    \end{align}
    meaning that quantum coherence does not increase over incoherent operations $\Lambda_{ICPTP}^{\mathcal{I}}$.
    \item Strong monotonicity under incoherent channel:
    \begin{align}
        \sum_j p_j C(\rho_j) \leq C(\rho),
    \end{align}
    meaning that quantum coherence does not increase under selective incoherent operations (measurements) $\Lambda^{\mathcal{I}}$ on average.
    \item Convexity:
    \begin{align}
        \sum_j p_j C(\rho_j) \geq C\left(\sum_j p_j \rho_j\right),
    \end{align}
    for any $p_j\geq 0$ with $\sum_j p_j =1$.
\end{enumerate}
Where $\rho_j \doteq (\kappa_j \rho \kappa_j^{\dagger})/p_j$, with probability $p_j \doteq tr(\kappa_j \rho \kappa_j^{\dagger})$, and $\kappa_j$ are the known $d_j$x$d_j$ incoherent Kraus operators that satisfy $\sum_j \kappa_j \kappa_j^{\dagger} = \id_d$.


\subsection{The p-norm QC}

In the context of quantum information, one ubiquitous norm of a quantum operator $\hat{A}$ is the so-called \textit{$p$-norm} defined as:
\begin{align}
    ||\hat{A}||_{p} \doteq \left( tr \left( \left( \sqrt{\hat{A}^{\dagger}\hat{A}} \right)^p \right) \right)^{1/p},
\end{align}
with $0\leq p < \infty$. If $\hat{A}$ is Hermitian then
\begin{align}
    ||\hat{A}||_{p} = \left( tr \left( \left( \sqrt{\hat{A}^2} \right)^p \right) \right)^{1/p} = \left(\sum_j|a_j|^p\right)^{1/p},
\end{align}
where $|a_j|$ is the absolute value of the real eigenvalue $a_j$ of $\hat{A}$. The $p$-norm induces a dissimilarity measure, dubbed \textit{$p$-norm distance}, between generic $N$-qudit density operators $\rho$ and $\eta$, as follows:
\begin{align}
    d_{p}(\rho,\eta)&\doteq||\rho-\eta||_{p}\\
    &=\left(tr\left(\left(\sqrt{(\rho-\eta)^{\dagger}(\rho-\eta)}\right)^p\right)\right)^{1/p}.
\end{align}
In the limit of $p=1$ and $p=2$ one obtains the \textit{trace-distance} \cite{RANA2016} and the \textit{Hilbert-Schmidt distance}, respectively:
\begin{align}
    d_{1}(\rho,\eta)&=tr\left(\sqrt{(\rho-\eta)^{\dagger}(\rho-\eta)}\right),\\
    d_{2}(\rho,\eta)&=\sqrt{tr\left((\rho-\eta)^{\dagger}(\rho-\eta)\right)}.
\end{align}

Therefore, the $p$-norm QC is the $p$-norm distance between a generic $N$-qudit density operator $\rho$ and its corresponding incoherent density operator $\iota$ is:
\begin{align}
     C_{p}(\rho)&=\min_{\iota}d_{p}(\rho,\iota) = \min_{\iota}||\rho-\iota||_{p}.
\end{align}
\begin{itemize}
    \item \textbf{1-qubit $p$-norm QC:} consider the following particular case of two 1-qubit states $\rho_{qb}=2^{-1}(\id_2 + \vec{r}\cdot \vec{\sigma})$ and $\iota_{qb}=2^{-1}(\id_2 + \vec{\iota} \cdot \vec{\sigma})$ with Bloch's vectors $\vec{r}=||\vec{r}||_2\hat{r} \in \mathbb{R}^3$ and $\vec{\iota}=||\vec{\iota}||_2\hat{\iota}=(0,0,\iota_{3}) \in \mathbb{R}^3$, respectively. We see that $\rho_{qb}-\iota_{qb}=2^{-1}(\vec{r}-\vec{\iota}) \cdot \vec{\sigma}$ such that 
    \begin{align}
        (\rho_{qb}-\iota_{qb})^2&= 2^{-2}\{(\vec{r}-\vec{\iota}) \cdot \vec{\sigma}\}^2\\
        &= 2^{-2}\{(r_1\sigma_1,r_2\sigma_2,(r_3-\iota_3)\sigma_3)\}^2\\
        &= 2^{-2}\{r_1^2\sigma_1^2 + r_2^2\sigma_2^2 + (r_3-\iota_3)^2\sigma_3^2\}\\
        &= 2^{-2}\{r_1^2+ r_2^2+ (r_3-\iota_3)^2\}\sigma_0\\
        &=2^{-2}(||\vec{r}-\vec{\iota}||_2)^2 \id_2,
    \end{align}
    since $\sigma_j^2=\sigma_0=\id_2$. Hence, the $p$-norm distance between these two 1-qubit states yields:
    \begin{align}
    d_{p}(\rho_{qb},\iota_{qb})&= \left(tr\left(\left(\sqrt{2^{-2} (||\vec{r}-\vec{\iota}||_2)^2 \id_2 }\right)^p\right)\right)^{1/p}\\
    &=2^{-1}||\vec{r}-\vec{\iota}||_{2}\left(tr\left(\sqrt{\id_2}\right)^p\right)^{1/p}\\
    &=2^{-1}||\vec{r}-\vec{\iota}||_{2}\bigg(tr(\id_2)\bigg)^{1/p}\\
    &=2^{-1}2^{1/p}||\vec{r}-\vec{\iota}||_{2}\\
    &=2^{(1-p)/p}||\vec{r}-\vec{\iota}||_{2}\\ &=2^{(1-p)/p}\sqrt{r_{1}^{2}+r_{2}^{2}+(r_{3}-\iota_{3})^{2}}.
    \end{align}
    So the $p-$norm QC of an 1-qubit state takes the form
    \begin{align}
    C_{p}(\rho_{qb}) &=\min_{\iota}d_{p}(\rho_{qb},\iota_{qb})\\
    & = \min_{\iota_{3}}2^{(1-p)/p}\sqrt{r_{1}^{2}+r_{2}^{2}+(r_{3}-\iota_{3})^{2}}\\
    & =  2^{(1-p)/p}\min_{\iota_{3}}\sqrt{r_{1}^{2}+r_{2}^{2}+(r_{3}-\iota_{3})^{2}}.
    \end{align}
    Clearly, the minimum of $C_{p}(\rho_{qb})$ is obtained setting $\iota_{3}=r_{3}$. Therefore, the \emph{1-qubit $p$-norm QC} is:
    \begin{align}
    C_{p}(\rho_{qb})=2^{(1-p)/p}\sqrt{r_{1}^{2}+r_{2}^{2}}.
    \end{align}
    \end{itemize}

\section{Quantum teleportation}
\label{sec:teleportation}

The holistic feature of entangled systems is constrained by the no-communication theorem, which states that quantum states, alone, cannot be used for transmission of classical information (bits), or equivalently, that quantum correlations cannot be used for superluminal (faster-than-light) communication. In addition, the no-cloning theorem states that unknown arbitrary quantum states (superpositions) cannot be copied \cite{Wootters1982}. Notwithstanding, quantum entanglement paved the way for the realization of many quantum processes including quantum cryptography \cite{Morris} and quantum communication such as quantum teleportation (QT) \cite{Bennett1993}\cite{Boschi1998} and superdense coding (SC) \cite{Bennett}. Many of which have been realized by optical experiments, in particular, SC that has been reported using parametric down-conversion \cite{Klaus}. Here, we endeavor to explain the quantum teleportation protocol that can be used either as a resource for 1-qubit communication or to build noise-resilient quantum gates.

\begin{figure}[htb!]
  \centering
  \includegraphics[scale=.4]{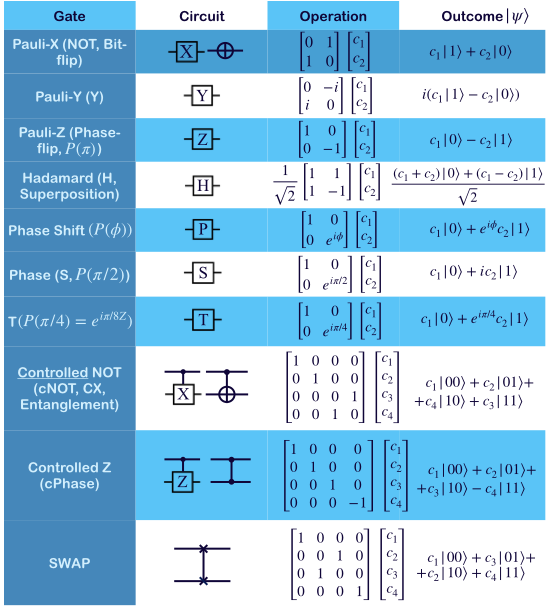}
  \caption{Fundamental single qubit gates and 2-qubit gates with their corresponding circuit symbols, Unitary matrix and post operation states. The set of single qubit and cNOT gates form a universal set of quantum gates to which any possible gate operation can be reduced to a finite sequence of gates from the set. The cNOT, H, S and T gates form a common universal gate set termed Clifford+T gate set.}
  \label{fig:tab2}
\end{figure}

In a nutshell, the vanilla quantum teleportation protocol proposed by Bennett \textit{et al}. \cite{Bennett1993} involves an agent, named Alice, whose aim is to transmit a single random state of 1-qubit denoted $|\psi\rangle=c_1|0\rangle + c_2|1\rangle$ to a second agent, named Bob, by using a small overhead of only two classical bits of information ($c_1$ and $c_2$) and a qubit $q_1$ from a Bell-like pair of entangled qubits. 
\begin{Remark}
As a word of caution, here and hereafter, the term qubit, alone, refers to the actual physical particle or artificial atom in which information is stored. For instance, in a superconducting quantum computer, a qubit is an artificial atom represented by a circuit-based superconducting solid-state device embedded in a coplanar waveguide microwave resonator that can be controlled individually and locally by electromagnetic pulses in the microwave range.
\end{Remark}
Given the no-cloning theorem, a third party, named Telamon, first prepares an entangled pair of particles $q_1$ and $q_2$ in a maximally entangled 2-qubit Bell state $|\psi_{12}^{entg}\rangle$ by applying a quantum logic gate (a local Unitary operator) termed Hadamard gate to qubit $q_1$ followed by a cNOT gate using qubit $q_2$ as target and qubit $q_1$ as control. He then sends qubit $q_1$ to Alice and qubit $q_2$ to Bob. On her turn, Alice applies a cNOT gate to qubit $q_1$ (the target) controlled by a qubit $q_0$ in the state $|\psi\rangle$ that she wants to teleport. She then applies a Hadamard gate to $q_0$ and performs a measurement on both qubits $q_1$ and $q_0$, storing the classical results in classical bits $c_1$ and $c_2$, respectively. Finally, she sends the two classical bits to Bob. On his turn, to recover the original state $|\psi\rangle$, Bob applies one of four single qubit gates ($\id_2$, $X$, $Z$, or $ZX$) on qubit $q_2$, depending on Alice's post-measurement classical state, i.e, on the classical bits he receives from Alice as shown in Table \ref{tab:2}. The fundamental quantum logic gate operations are showcased in Fig.~\ref{fig:tab2}.

From Fig.~\ref{fig:tab2}, single qubit operations entail:
\begin{align}
    H|0\rangle &= |+\rangle,\\
    H|1\rangle &= |-\rangle,\\
    Z|+\rangle &= X|0\rangle = |1\rangle,\\
    Z|-\rangle &= X|1\rangle, = |0\rangle.
\end{align}
Useful gate identities follow the $SU(2)$ Lie algebra (see \autoref{sec:sualgebra}): 
    \begin{align}
        HZH &= X,\\
        HXH &= Z,\\
        HYH &= -Y,\\
        -iXYZ &= \id_2,\\
        ZX &= iY,\\
        XZ &= -iY.
    \end{align}
Moreover, any arbitrary single qubit unitary gate can be written as:
\begin{align}
    U &= e^{i\alpha}R_{\hat{n}}(\theta)=e^{i\alpha} e^{-i\theta \hat{n} \cdot \vec{\sigma}/2}\\
    &=e^{i\alpha} \left(cos\left(\frac{\theta}{2}\right)\id_2-sin\left(\frac{\theta}{2}\right)(\hat{n}\cdot \vec{\sigma})\right),
\end{align}
given $a$ and $\theta \in \mathbb{R}$, $i^2=-1$, 3-dimensional unit vector $\hat{n}=(n_x, n_y, n_z)$, and three component vector $\vec{\sigma}=(X, Y, Z)$. A random 1-qubit state then reads:
\begin{align}
    \left(e^{i(\frac{\pi}{2})}R_{(c_2, 0, c_1)}(\pi)\right)|0\rangle = c_1 |0\rangle+ c_2 |1\rangle = |\psi\rangle.
\end{align}

\begin{figure}[htb!]
  \centering
  \includegraphics[scale=.08]{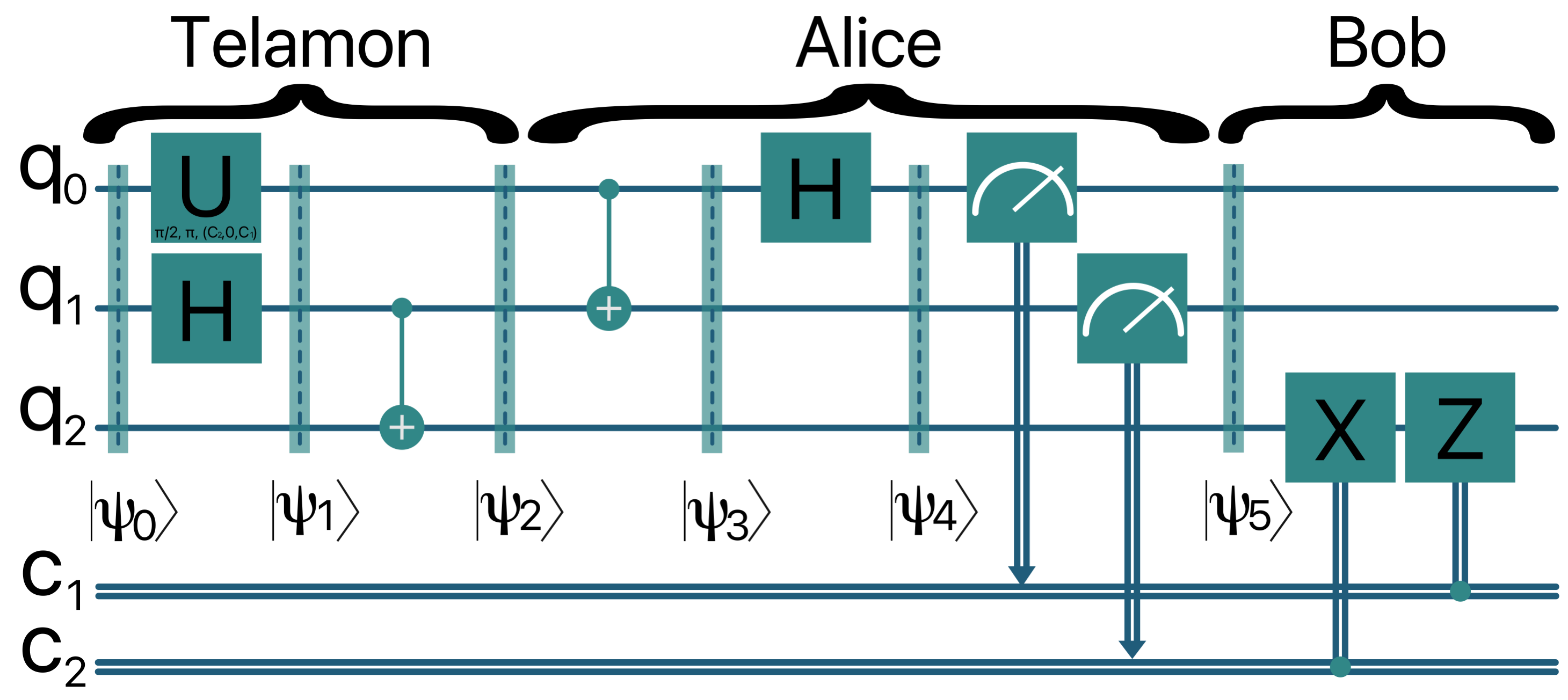}
  \caption{Quantum circuit corresponding to the quantum teleportation protocol. The Meter represents a projective measurement in the computational basis. Each single line represents a qubit register (a physical qubit device) that stores qubit states, whereas double lines represent classical registers for storage of classical bits. The circuit is sliced into five time-steps where the state vector of the global system evolves over time according to a sequence of quantum gates applied on disjoint sets of qubits. Gates of the same time step can in principle be realized simultaneously. To harness quantum superposition the entire circuit must be implemented during a time window smaller than the decoherence time. This requirement imposes the depth of an arbitrary quantum circuit (the length of the qubit register with the longest sequence of gates from the input gate to the output gate) to be as low as possible. In this particular case, the circuit depth is six, corresponding to register $q2$ considering all previously applied gates and discarding any intermediary measurement.}
  \label{fig:circuittele}
\end{figure}

The whole teleportation protocol can be represented by a circuit diagram (see Fig.~\ref{fig:circuittele}) within the \textit{quantum circuit model} of quantum computation. Some authors may adopt the \textit{principle of deferred measurement} and move the measurement operations to the end of the circuit, as real quantum computers may lack support for instructions after measurements.
The evolution of the quantum state of the system throughout the circuit is given as follows:
\begin{itemize}
    \item Each qubit in a real quantum computer is initialized in the ground state $|0\rangle$. The initial global state of the composite tripartite system is, therefore:
    \begin{align}
        |\psi_0\rangle = |0\rangle \otimes |0\rangle \otimes |0\rangle \doteq |000\rangle.
    \end{align}
    \item Telamon then applies a Hadamard gate to qubit $q_1$, while qubit $q_0$ is owned by Alice:
    \begin{align}
    U|0\rangle&= |\psi\rangle=c_1|0\rangle+c_2|1\rangle,\\
    H|0\rangle&=|+\rangle=\frac{1}{\sqrt{2}}(|0\rangle+|1\rangle,\\
        |\psi_1\rangle &=(U\otimes H\otimes \id)|\psi_0\rangle\\
        &=U|0\rangle \otimes H|0\rangle \otimes |0\rangle\\
        &=|\psi\rangle\otimes |+\rangle \otimes |0\rangle\\
        &=|\psi\rangle\otimes\left(\frac{|0\rangle+ |1\rangle}{\sqrt{2}}\otimes |0\rangle\right)\\
        &=|\psi\rangle\otimes\left(\frac{|00\rangle+|10\rangle}{\sqrt{2}}\right)\\
        &= \frac{c_1|000\rangle+c_2|100\rangle+c_1|010\rangle+c_2|110\rangle}{\sqrt{2}}.
    \end{align}
    \item The following cNOT gate applied to qubit $q_2$ (target) controlled by $q_1$ creates an entangled Bell state. The action of the cNOT gate is to flip the target qubit (the rightmost qubit) if the control (the leftmost qubit) is $|1\rangle$:
    \begin{align}
        cNOT\bigg(|+\rangle \otimes |0\rangle\bigg)
        &= cNOT\left(\frac{|00\rangle+|10\rangle}{\sqrt{2}}\right)\\ &=\frac{|00\rangle+|11\rangle}{\sqrt{2}} \\&\doteq |e\rangle,
    \end{align}
    and the global state of the system updates to
    \begin{align}
        |\psi_2\rangle&= (\id_2 \otimes cNOT)|\psi_1\rangle\\
        &=(\id_2 \otimes cNOT)(|\psi\rangle\otimes |+\rangle \otimes |0\rangle)\\
        &= |\psi\rangle \otimes cNOT\bigg(|+\rangle \otimes |0\rangle\bigg)\\
        &=|\psi\rangle \otimes |e\rangle\\
        &= \frac{c_1|000\rangle+c_1|011\rangle+c_2|100\rangle+c_2|111\rangle}{\sqrt{2}}.
    \end{align}
    \item On her turn, Alice applies a cNOT gate on qubit $q_1$ (target) controlled by $q_0$. From left to right, the first qubit is the control and the second is the target, whereas the third remains unchanged:
    \begin{align}
        |\psi_3\rangle &= (cNOT \otimes \id_2)|\psi_2\rangle\\
        &=\frac{c_1|000\rangle+c_1|011\rangle+c_2|110\rangle+c_2|101\rangle}{\sqrt{2}}.
    \end{align}
    \item Alice then applies a Hadamard gate to $q_0$:
    \begin{align}
        |\psi_4\rangle &= (H\otimes\id_2\otimes\id_2)|\psi_3\rangle\\
        &=(H\otimes\id_2\otimes\id_2)(cNOT \otimes \id_2)(|\psi\rangle \otimes |e\rangle)\\
        &=\frac{1}{2}\bigg\{c_1\bigg(|000\rangle+|011\rangle+|100\rangle+|111\rangle\bigg)\\
        &+c_2\bigg(|010\rangle+|001\rangle-|110\rangle-|101\rangle\bigg)\bigg\}\\
        &=\frac{1}{2}\bigg\{
        |00\rangle_a\otimes\bigg(c_1|0\rangle_b+c_2|1\rangle_b\bigg)\\
        &+|01\rangle_a\otimes\bigg(c_1|1\rangle_b+c_2|0\rangle_b\bigg)\\
        &+|10\rangle_a\otimes\bigg(c_1|0\rangle_b-c_2|1\rangle_b\bigg)\\
        &+|11\rangle_a\otimes\bigg(c_1|1\rangle_b-c_2|0\rangle_b\bigg)
        \bigg\}\\
        &=\frac{1}{2}\bigg\{|00\rangle_a \otimes \id_2|\psi\rangle + |01\rangle_a \otimes X|\psi\rangle \\
        &+|10\rangle_a \otimes Z|\psi\rangle + |11\rangle_a \otimes XZ|\psi\rangle\bigg\}.
    \end{align}
    \item Upon measurement in the computational basis $(M_{ja}\otimes M_{ia} \otimes \id_2)|\psi_4\rangle$, Alice's subsystem of two qubits denoted ``$a$'' collapses to one of the four classical states showcased in Table \ref{tab:1}, as does Bob's subsystem. The global state of the tripartite system thus becomes one of the following states, each occurring with probability $1/4$:
    \begin{align}
        |\psi_5\rangle_{ab} = |00\rangle_a\otimes\bigg(c_1|0\rangle_b+c_2|1\rangle_b\bigg),\\
        |\psi_5\rangle_{ab} = |01\rangle_a\otimes\bigg(c_1|1\rangle_b+c_2|0\rangle_b\bigg),\\
        |\psi_5\rangle_{ab} = |10\rangle_a\otimes\bigg(c_1|0\rangle_b-c_2|1\rangle_b\bigg),\\
        |\psi_5\rangle_{ab} = |11\rangle_a\otimes\bigg(c_1|1\rangle_b-c_2|0\rangle_b\bigg).
    \end{align}
    \end{itemize}
\newpage
\begin{center}
\captionsetup{type=table}
\scalebox{1}{
\begin{tabular}{ c c }
 Alice's state ($|\psi_5\rangle_a$) & Bob's state ($|\psi_5\rangle_b$) \\ 
 $|00\rangle$ & $\id_2|\psi\rangle=c_1|0\rangle + c_2|1\rangle$  \\  
 $|01\rangle$ & $X|\psi\rangle=c_1|1\rangle + c_2|0\rangle$ \\
 $|10\rangle$ & $Z|\psi\rangle=c_1|0\rangle - c_2|1\rangle$ \\
 $|11\rangle$ & $XZ|\psi\rangle=c_1|1\rangle - c_2|0\rangle$ 
\end{tabular}}
\captionof{table}{Alice's and Bob's post-measurement classical states.}
\label{tab:1}
\end{center}

At this point, the density operator of the global system becomes:
    \begin{align}
    \rho&=\frac{1}{4}\bigg\{|00\rangle\langle00|\otimes\bigg(c_1|0\rangle+c_2|1\rangle\bigg)\bigg(c_1^*\langle0|+c_2^*\langle 1|\bigg)\\
    &+|01\rangle\langle01|\otimes\bigg(c_1|1\rangle+c_2|0\rangle\bigg)\bigg(c_1^*\langle1|+c_2^*\langle 0|\bigg)\\
    &+|10\rangle\langle10|\otimes\bigg(c_1|0\rangle-c_2|1\rangle\bigg)\bigg(c_1^*\langle0|-c_2^*\langle 1|\bigg)\\
    &+|11\rangle\langle11|\otimes\bigg(c_1|1\rangle-c_2|0\rangle\bigg)\bigg(c_1^*\langle1|-c_2^*\langle 0|\bigg)\bigg\}.
    \end{align}
From that and using Eq.~\eqref{parttrace}, one has the following reduced density operator of Bob's subsystem after Alice's measurement:
    \begin{align}
        \rho_b&=tr_a(\rho)\\
        &=\frac{1}{4}\bigg\{ \langle 00|00\rangle \bigg(c_1|0\rangle+c_2|1\rangle\bigg)\bigg(c_1^*\langle0|+c_2^*\langle 1|\bigg)\\
        &+\langle 01|01\rangle\bigg(c_1|1\rangle+c_2|0\rangle\bigg)\bigg(c_1^*\langle1|+c_2^*\langle 0|\bigg)\\
        &+ \langle 10|10\rangle\bigg(c_1|0\rangle-c_2|1\rangle\bigg)\bigg(c_1^*\langle0|-c_2^*\langle 1|\bigg)\\
        &+\langle 11|11\rangle\bigg(c_1|1\rangle-c_2|0\rangle\bigg)\bigg(c_1^*\langle1|-c_2^*\langle 0|\bigg)
        \bigg\}\\
        &=\frac{1}{4}\bigg\{2(|c_1|^2+|c_2|^2)|0\rangle\langle0|\\
        &+2(|c_1|^2+|c_2|^2)|1\rangle\langle1|\bigg\}\\
        &=\frac{|0\rangle\langle0|+|1\rangle\langle1|}{2}=\frac{\id_2}{2}.
    \end{align}
This last result entails that faster-than-light communication is impossible since Bob's measurement alone does not yield any information about the teleported state without prior knowledge of the classical bits sent by Alice.
    
Finally, to recover the teleported state $|\psi\rangle$, Bob applies on his qubit $q_2$ the single qubit gates corresponding to the classical bits he receives from Alice, as shown in Table \ref{tab:2}.
\begin{center}
\captionsetup{type=table}
\scalebox{0.9}{
\begin{tabular}{c c c}
 Bit sent & Gate applied on $|\psi_5\rangle_b$ & Final state ($|\psi\rangle$) \\ 
 $00$ & $\id_2$ & $\id_2(\id_2|\psi\rangle)$ \\
 $01$ & $X$ & $X(X|\psi\rangle)$\\
 $10$ & $Z$ & $Z(Z|\psi\rangle)$\\
 $11$ & $ZX$ & $ZX(XZ|\psi\rangle)$
\end{tabular}}
\captionof{table}{Classical bits sent by Alice and corresponding single qubit gate operations on qubit $q_2$ ($|\psi_5\rangle_b$) for recovering the teleported state $|\psi\rangle$.}
\label{tab:2}
\end{center}

Bennet \textit{et al}. \cite{Bennett1993} proved that perfect quantum communication requires at least two maximally entangled qubits. However, open quantum systems are prone to noise processes such as thermalization and dephasing which causes the quantum system to rapidly lose coherent information (see \autoref{sec:decoherence}). In this scenario, 2-qubit maximally entangled pure states (Bell states) are often unavailable, while non-maximal Bell states are insufficient for realization of perfect quantum teleportation \cite{Pati2004} and perfect superdense coding \cite{Pati2005}. A workaround is to use a noise-resilient class of states such as the allured special class of $GHZ$-like and $W$-like maximally entangled tripartite states \cite{Dur2000}. Particularly, $W$-like states form a more robust class that can be used even after particle loss \cite{Dur2001}. In spite of that, the requirement of non-local operations in quantum communication protocols restricts the aforementioned $w$-like state to only imperfect quantum teleportation and imperfect superdense coding schemes. Then, in 2006, Agrawal \textit{et al.} \cite{Agrawal2006} showed that a class of 3-qubit $W$-like nonmaximally entangled states meet the criteria for both perfect teleportation and superdense coding. For a generalization of the scheme to higher dimensions, the reader may resort to Li \textit{et al.} \cite{Lvzhou2007}.


\section{Conclusion}

In this letter, we focused on deriving the fundamental mathematical identities of the postulates of quantum mechanics in two different formalisms, in addition to obtaining and analyzing particular multipartite states using Bloch's parametrization for generalized multi-qudit density operators and deriving the $p$-norm quantum coherence quantifier for a generic 1-qubit state. We also addressed the mathematical description of quantum entanglement, providing worked examples of entanglement quantification and quantum communication. Although the results obtained here are primarily due to the field of quantum information science, they span a wide range of applications throughout areas such as quantum optics, and quantum computing: in the preparation and quantification of maximally and nonmaximally entangled states for quantum cryptography and perfect quantum communication; in the estimation of quantum decoherence in classical computer simulations of quantum systems; in the calculation of correlation functions for phase transitions; and in quantum state tomography used to reconstruct the coherent density matrix of a qudit state by computing the components (expectation values) of its Bloch's vector through a series of tomographic measurements on the observables.


\section{Version}
This is quantum article version v\quantumarticleversion.


\bibliographystyle{unsrt}
\bibliography{paper}

\end{document}